\documentclass{article}

\pdfoutput=1

\usepackage{arxiv}

\usepackage[utf8]{inputenc} 
\usepackage[T1]{fontenc}    
\usepackage{hyperref}       
\usepackage{url}            
\usepackage{booktabs}       
\usepackage{amsfonts}       
\usepackage{nicefrac}       
\usepackage{microtype}      
\usepackage{lipsum}		
\usepackage{graphicx}
\usepackage[numbers]{natbib}
\usepackage{doi}

\usepackage{amsmath,amssymb,amsfonts,amsthm}%

\usepackage{rotating}

\title{Basic Mechanisms for Understanding Container-Content Interactions : a fundamental perspective and application to polymer materials}

\author{ \href{https://orcid.org/0000-0003-1837-4601}{\includegraphics[scale=0.06]{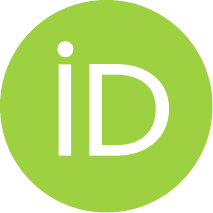}\hspace{1mm}Emmanuelle Feschet-Chassot}\\
	  Universit\'e Clermont Auvergne\\
        Clermont Auvergne INP, CNRS, ICCF\\	
	Clermont-Ferrand, F-63000 \\
	\texttt{emmanuelle.feschet@uca.fr} \\
	\And
	\href{https://orcid.org/0000-0001-8086-7517}{\includegraphics[scale=0.06]{orcid.pdf}\hspace{1mm}Philip Chennell} \\
	Universit\'e Clermont Auvergne\\
        CHU Clermont-Ferrand, Clermont Auvergne INP, CNRS, ICCF\\
	Clermont-Ferrand, F-63000 \\
	\texttt{philip.chennell@uca.fr} \\
}

\renewcommand{\shorttitle}{\textit{arXiv} Template}

\hypersetup{
pdftitle={Basic Mechanisms for Understanding Container-Content Interactions : a fundamental perspective and application to polymer materials},
pdfsubject={cond-mat.soft,physics.chem-ph},
pdfauthor={Emmanuelle Feschet-Chassot, Philip Chennell},
pdfkeywords={Adsorption, Absorption, Leaching, Permeation, Polymers},
}

\begin{document}
\maketitle

\begin{abstract}
	This article provides a comprehensive understanding of the interactions that can occur at the interface between liquids and materials. It describes the phenomena of sorption (adsorption and absorption), permeation and leaching from a physicochemical perspective. In addition to examining these interactions, the historical context, detailing how these phenomena have been studied and demonstrated over time, it provides an understanding of the evolution of research in this field and the methodologies used to study these interactions, with a specific focus on polymer materials. The mechanisms underlying these interactions are presented as well as the equations that describe the processes involved, thus providing a scientific basis for understanding the complexities of container-content interactions.
\end{abstract}


\section{Introduction}

During their life (including their manufacture, storage and administration to the patients), pharmaceutical compounds and medications come into contact with multiple materials, such as those from the primary packaging materials and medical devices used for their administration. This contact is rarely devoid of possible interactions, especially (but not exclusively) for liquid formulations. These interactions are grouped into the generic term "container-content interactions", and correspond to the physicochemical phenomena that can occur at an interface, in this case a solid-liquid (and sometimes solid-gaseous) interface. For the purpose of this review, we will provide hereafter a brief historical context of the subject of surface interactions (namely sorption, permeation and leaching), and the readers are invited to consult the corresponding section for the definitions of the terms used and for more detailed explanations. Then, for each type of interaction, the description of the phenomenon and its mathematical models will be presented.

From a historical point of view, it seems that the principles of interface interactions were first noticed a very long time ago in practical life, mainly by the use of activated charcoal in medical applications or for the purification of water (the first references appear around 1550 BC with the Egyptians, then around 460 BC with Hippocrate and Pliny). Much latter on, the Abb\'e Fontana (1777) carried out the first experiments on charcoal in order to study this phenomenon and Von Saussure then observed the exothermic aspect of this phenomenon \cite{Saussure1814}. In 1843, Mitscherling introduced the notions of exposed surface area and porous volume, notions that are nowadays routinely employed \cite{Mitscherlich1843}. The term \textit{adsorption} was first individualised by E. du Bois-Reymond and introduced into the literature by \citet{Kayser1881} in 1881. Prior that time both phenomena (\textit{adsorption} and \textit{absorption}) were named \textit{absorption}, as it was not possible to distinguish these two phenomena and define them precisely.

A few years after the introduction of the term \textit{adsorption}, Kayser developed some theoretical concepts which became the basis for the theory of monomolecular adsorption \cite{Kayser1881}. The notation \textit{isothermal adsorption} seems to be introduced by Ostwald in 1885. McBain used the term \textit{absorption} in 1909 in order to explain the much slower interaction of hydrogen gas with a carbon surface than what would be expected from just an adsorption phenomenon. He proposed the term \textit{sorption} for adsorption and absorption when it is not easy or meaningful to make a difference between the two. The first theoretical analysis of adsorption was due to Irving Langmuir in 1914, it described the adsorption of a monolayer of adsorbate on a homogeneous surface in the form of an equation which formed the now well known \textit{Langmuir isotherm} \cite{Langmuir1916}.

Nowadays, the term \textit{absorption}, as opposed to that of desorption (reverse of sorption), is generally used to describe any process integrating the penetration and then the dispersion of a diffusing agent into a matrix, whereas the less specific \textit{sorption} is a physical and chemical process by which one substance becomes attached to another. It can be stated that:

\begin{itemize}
\item Adsorption is the physical adherence or bonding of ions and molecules onto the surface of another phase;
\item Absorption correspond to the transfer of a substance in one state into another substance of a different state.
\end{itemize}

This phenomenon of absorption can lead after a sufficient amount of time to the phenomenon of permeation. The permeation process involves the diffusion of molecules that form the permeate, most commonly across a membrane or interface. Permeation is due to diffusion; the permeate will move from the medium of high concentration to that of low concentration across the interface. The history of material permeability is quite old. Indeed, the first study relating the permeability of membranes dates back to 1748. It was carried out by the French Abb\'e Nollet \cite{Nollet1743} who had filled wine containers closed by animal bladders and had immersed them in pure water. Because of their higher permeability to water than to wine, the bladders swelled and some even burst. In 1829, Thomas Graham observed the swelling of a pig's bladder in the presence of $\rm CO_2$ \cite{Graham1834}. The publication of this work also marked the beginning of studies of the permeability of polymers to gases. In 1855, Fick published his broadcasting laws which are still used today \cite{Fick1855}. He saw the analogy that applies between mass transfer and heat transfer. A year later, Darcy published his treatise on \textit{the public fountains of the city of Dijon} \cite{Darcy1856} in which the formula which bears his name appears, inspired by the laws of Jean Louis Poiseuille. 

\section{Adsorption}\label{adsorption-1}

\subsection{Definition}\label{adsorption}

Adsorption is essentially a surface phenomenon linked to surface energy. The molecular species or substance, which concentrates or accumulates at the surface is termed \textit{adsorbate} and the material on the surface of which the adsorption takes place is called \textit{adsorbent}. In a bulk material, all of the bonding requirements (whether ionic, covalent, or metallic) of the constituent atoms of the material are met by other atoms nearby. Thus, inside the adsorbent, all the forces acting between the particles are mutually balanced, but on the surface the particles are no longer surrounded by other atoms or molecules of the same type, and therefore they have unbalanced or residual attractive forces. The molecules on the surface therefore have a higher energy than those of the bulk. This extra energy per unit surface area is called \textit{surface energy}. The residual attractive forces of the adsorbent are responsible for attracting the adsorbate particles onto its surface. The extent of adsorption increases with the increase of surface area per unit mass of the adsorbent at a given temperature and pressure.

The exact nature of the bond depends on the details of the species involved, but the adsorption process is generally classified as \textit{physisorption} (characteristic of weak van der Waals forces) or \textit{chemisorption} (characteristic of covalent bond). It can also occur due to electrostatic attractions. Physical adsorption, which will be described in detail further, is a reversible process, i.e.~the physisorbed layer can be removed by reducing the pressure or increasing the temperature \cite{West1945}. Chemisorption requires more drastic conditions to break and create new chemical bonds. Another important factor explaining adsorption is the enthalpy of adsorption as mentioned before by Von Saussure \cite{Saussure1814} who observed the exothermic aspect of this phenomenon. When adsorption occurs, there is always a decrease in residual forces of the surface, i.e., there is decrease in surface energy which is freed as heat. Adsorption, therefore, is invariably an exothermic process.

Improved comprehension of quantum mechanics and chemistry have greatly helped understanding the characteristics of the interactions in the physical and chemical adsorption processes. Excellent reviews on the development of fundamental adsorption forces have been presented in for example by \citet{Young1962} and \citet{Israelachvili2011}.

\subsection{Mechanisms of adsorption}\label{mechanism-of-adsorption}

\subsubsection{Physical adsorption}\label{physical-adsorption}

As previously mentioned, there are two types of adsorption : physisorption and chemisorption. Sometimes these two processes occur simultaneously or evolve from one to another and it is not easy to distinguish between the two. For example, a physical adsorption at low temperature may become a chemisorption process as the temperature is increased.

The forces responsible for physical adsorption are the Van der Waals or London dispersing forces. Hydrogen bonds can also be a cause of physical adsorption. The theoretical approach to this phenomenon is based on the works of London in the 1930s, who calculated the forces of attraction between molecules, and those of Born and Mayer who calculated the repulsive forces (1932). The different contributions of the Van der Waals forces are as follows:

\begin{itemize}
 
\item
  Keesom's forces, which are electrostatic interactions between two permanent multipoles depending on their orientations (orientation effects) \cite{Keesom1915}. These forces are caused by the interaction between two polar molecules. The dipole-dipole interaction is much weaker than an ion-dipole interaction since the interaction occurs between partial charges.
\item
  Debye's forces: an attractive interaction between a permanent multipole and an induced multipole (induction effects also known as polarization). Most often, this force would be caused by the interaction between a polar molecule and an induced dipole.
\item
  London forces: this dominant contribution consists of an attractive electrostatic interaction between two induced multipoles, like to induced dipoles (dispersion effects) \cite{London1930}.
\end{itemize}

Physical adsorption can occurs with any container-content system provided only that the conditions of temperature and pressure are appropriate. Under various temperature and pressure conditions, polymolecular layers of physically adsorbed molecules can be found.

\subsubsection{Chemical adsorption}\label{chemical-adsorption}

Chemical adsorption or chemisorption is a process that results from a chemical reaction with the formation of chemical bonds between the adsorbate molecules and the adsorbent surface. The chemical bonding can be of two sorts:

\begin{itemize}
 
\item
  Covalent bonding which is a chemical bond in which two atoms share two electrons of one of their external layers. There must be an electronegativity difference of less than 1.7 on the Pauling scale for this sort of bond to be created.
\item
  Ionic bonding which is a type of chemical bonding that involves the electrostatic attraction between oppositely charged ions, and is the primary interaction occurring in ionic compounds. It can be formed by a pair of atoms having a large difference in electronegativity (by convention, higher than 1.7)
\end{itemize}

\subsubsection{Differences between physisorption and chemisorption}\label{difference-between-physisorption-and-chemisorption}

\begin{figure}[!ht]
\centering
\includegraphics[scale=0.3]{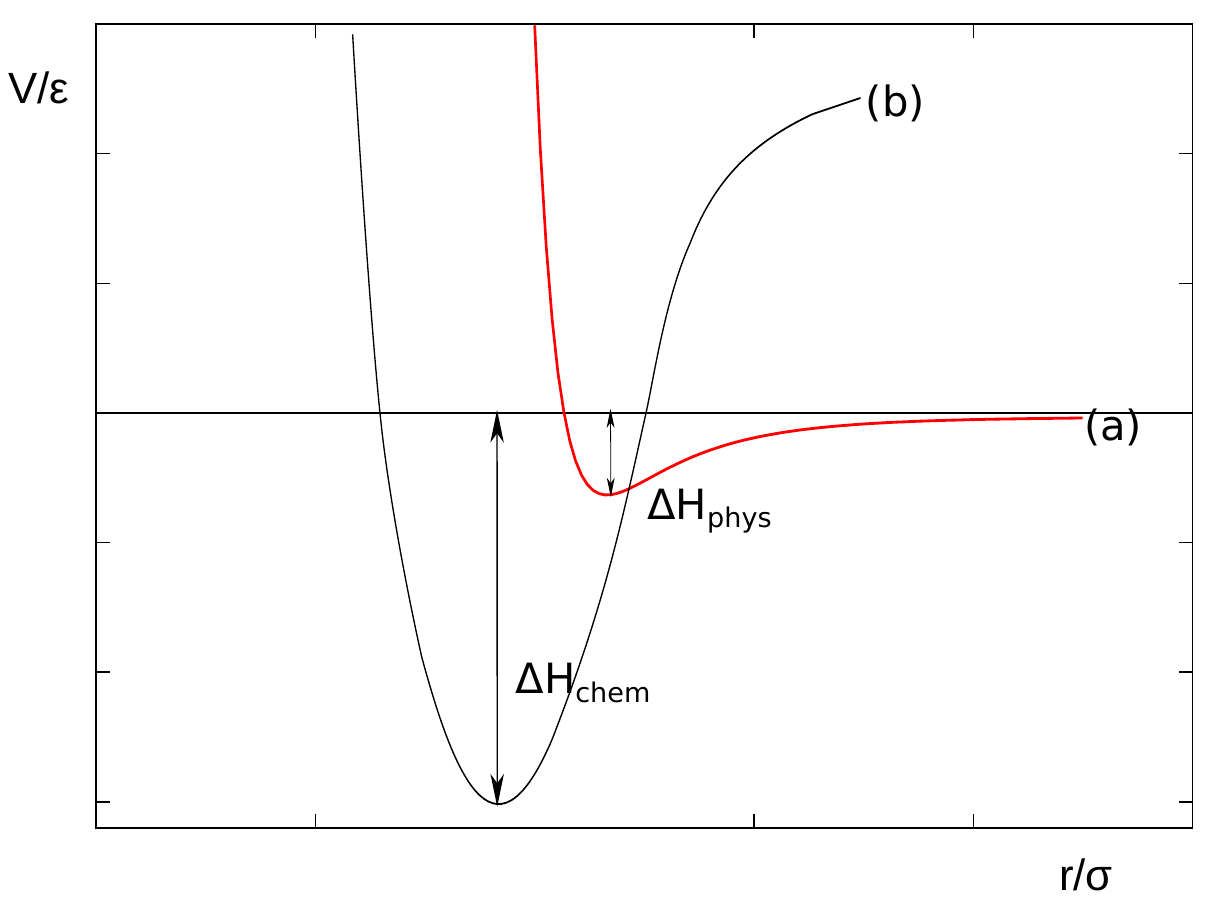}
\caption{Lennard-Jones potential energy near a surface for adsorption: a) for chemisorption and b) for physisorption}\label{fig:lennard}
\end{figure}

The process of adsorption can be described using the Lennard-Jones potential energy diagram \cite{Lennard-Jones1932}. This diagram show the potential energy of a molecule approaching a surface. In physical adsorption, as we mentioned above, the forces involved are weak, inducing energy exchanges that can be for example between 4 and 40 kJ/mol. In the case of chemical adsorption we have strong chemical bonds, hence higher exchanged energies (usually between 40 and 200 kJ/mol). The Lennard-Jones potential (L-J potential) is a mathematically simple model that approximates the interaction between a pair of neutral atoms or molecules (Figure \ref{fig:lennard}).

A form of this interatomic potential was first proposed in 1924 by John Lennard-Jones \cite{Jones1924}. The most common expressions of the L-J potential are:
\begin{equation}
V_{LJ} = \varepsilon\left[\left(\frac{r_m}{r}\right)^{12} - 2 \left(\frac{r_m}{r}\right)^6\right] = 4 \varepsilon \left[\left(\frac{\sigma}{r}\right)^{12} - \left(\frac{\sigma}{r}\right)^6\right]
\label{eq:lennard}
\end{equation}
where $\varepsilon$ is the potential well depth, $\sigma$ is the finite distance at which the inter-particle potential is zero, $r$ is the distance between the particles, and $r_m$ is the distance at which the potential reaches its minimum. The distances are related as $r_m = 2^{1/6} \sigma \approx 1.122 \sigma$. The $r^{12}$ term, which is the repulsive term, describes the Pauli repulsion at short ranges due to overlapping electron orbitals, and the $r^6$ term, which is the attractive long-range term, describes the attractions at long ranges (Van der Waals force, or dispersion force).

The other differences between physisorption and chemisorption are cited in (Table \ref{tbl:table1}).

\begin{sidewaystable*}
\caption{Main differences between physisorption and chemisorption \label{tbl:table1}}
\begin{tabular}{p{2cm}|p{5cm}|p{7cm}}
\toprule%
Properties & Physorption & Chemisorption\\
\midrule
Bonding & Weak, Long range forces, Van der Walls interactions & Strong, Short range forces, Chemical bonding involved\\

Surface specificity & No, physisorption takes place between all molecules on any surface providing the temperature is low enough & Yes, chemisorption is highly specific and it will only occur if there is some possibility of chemical bonding between adsorbent and adsorbate\\

Surface area of adsorbent & \multicolumn{2}{c}{Increases with the increase of surface area of the adsorbent}.\\

Nature & Reversible, physical adsorption occurs readily at low temperature and decreases with increasing temperature (Le Chatelier's principle \cite{Lechatelier1884}) & Irreversible, as chemisorption involves new compound formation, it is usually irreversible by nature\\

Saturation & Multi-layer & Mono-layer\\

Effect of temperature & Occurs at low temperature and decreases with increase in temperature & Even though chemical adsorption is an exothermic process, it does not occur slowly at lower temperature due to high kinetic energy barrier. Hence, like most chemical changes, the extent of chemisorption increases with temperature up to a certain limit, then after that it starts decreasing.\\

Effect of pressure (gas) or concentration (liquid) & Increases with increase in pressure / concentration of adsorbate (Le Chatelier's principle) & Not appreciably affected by small changes in pressure/concentration. However, very high pressures/concentrations are favorable.\\

Activation energy & Not needed, non activated with equilibrium achieved relatively quickly & High activation energy is needed, can be activated, in which case equilibrium can be slow.\\
\end{tabular}
\end{sidewaystable*}

Chemisorption requires more drastic conditions to break the chemical bonds. Furthermore, the heat of physisorption is much lower than the heat of chemisorption and is comparable to the liquefaction of the adsorbate. The heat released by chemisorption is of the same order as the heat of the relevant chemical reaction. The forces responsible for any type of adsorption are named adsorption forces. The theory of adsorption forces has been developed by London \cite{London1930, London1930a}, de Boer and Custers \cite{DeBoer1934}, Lenel \cite{Lenel1933} and de Boer \cite{DeBoer1956}.

\subsection{Adsorption isotherm}\label{adsorption-isotherm}

The adsorption process can be characterized by determining how many ions or molecules are adsorbed onto the surface. Generally this is managed by measuring the correlation between the concentration of adsorbate in the fluid phase and the amount of adsorbate that is trapped by the surface at a given temperature. If the adsorbent and adsorbate are in contact for long enough, an equilibrium will be established between the amount of adsorbate adsorbed and the amount of adsorbate in solution. The representation of these results can take several forms, the most common of which is the measurement of \textit{adsorption isotherms}. Shortly after having introduced the concept of \textit{adsorption}, the terms isotherm and isothermal curve were applied to the results of the adsorption measurements carried out at constant temperature by Kayser \cite{Kayser1881, Kayser1881a}. The first liquid solute-solid adsorption equilibria were reported by Friedrich Stohmann and Wilhelm Henneberg in 1858 (ammonium adsorption on soil) \cite{Henneberg1858} and the first recorded isotherms of adsorption from solution were probably those reported by Van Bemmelen in 1881 \cite{Bemmelen1910}.

In 1909, Freundlich expressed an empirical equation for representing the isothermal variation of adsorption of a quantity of gas adsorbed by unit mass of solid adsorbent with pressure. This equation (known as the Freundlich Isotherm \cite{Freundlich1907}) can be expressed in two different ways :
\begin{eqnarray}
q_e & = K_f~ P^{1/n}\\
q_e & = K_f~  C_e^{1/n}
\label{eq:freundlich}
\end{eqnarray}
$q_e$ is the quantity of adsorbate fixed per unit of adsorbent surface at the equilibrium, $C_e$ the concentration of adsorbate in the equilibrium solution, $P$ the equilibrium pressure of the gaseous adsorbate in case of experiments made using a gas phase, $K_f$ is the Freundlich constant which represents the adhesion capacity of the adsorbate on the adsorbent and characterizes the interactions between the adsorbate and the adsorbent and gives information on the heterogeneity of the surface. Though the Freundlich isotherm correctly established the relationship of adsorption with pressure at lower values, it failed to predict the value of adsorption at higher pressures. Indeed, the complexity of the isotherms obtained experimentally rarely agrees with this simple power law over a large pressure range. In addition, this equation does not correlate with Henry's Law when the pressure P tends to 0. Indeed, Henry's Law \cite{Henry1803}, initially defined to study the quantity of gas dissolved in a liquid, covers a more general principle that can be applied to adsorption: when the concentration of adsorbate tends towards 0, the adsorbed quantity also tends to 0.

\begin{figure}[!ht]
\centering 
\includegraphics[scale=0.4]{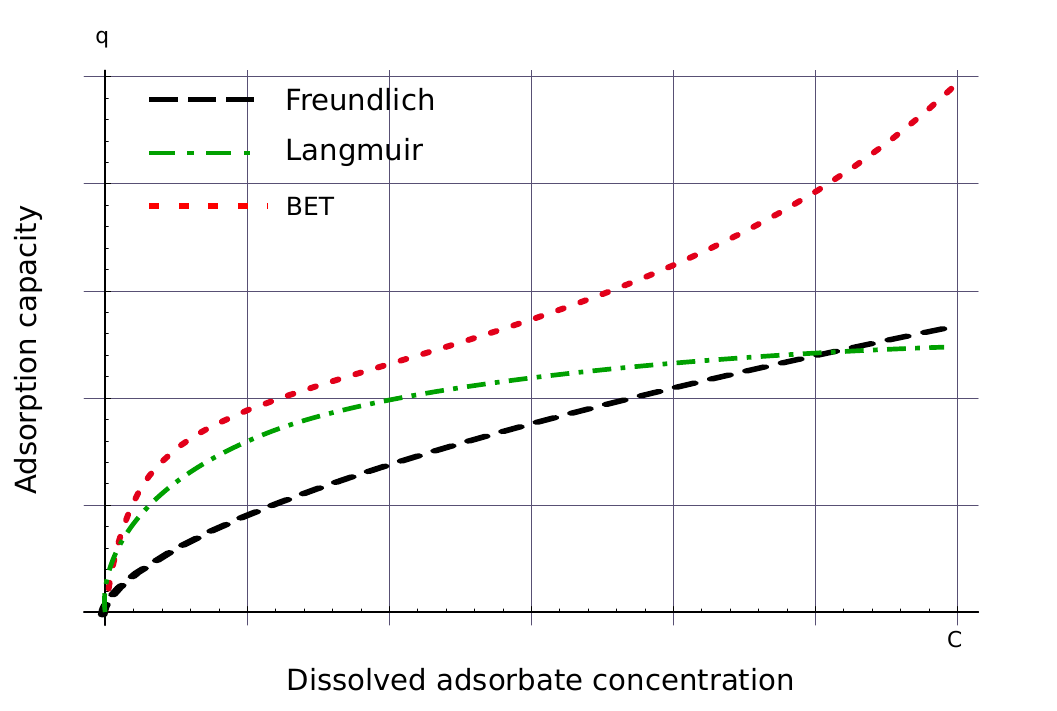}
\caption{Freundlich, Langmuir and BET adsorption isotherm}\label{fig:freundlich}
\end{figure}

Several years later, Langmuir proposed an adsorption isotherm based on the kinetic theory of gases. This theory allows the understanding of the monolayer surface adsorption on an ideal surface. The equation proposed by Langmuir is as follow:
\begin{equation}
q_e = N \times \frac{(K_L \times C_e)}{1 + (K_L \times C_e)}
\label{eq:langmuir}
\end{equation}
where $N$ represents the quantity adsorbed at saturation, i.e.~the maximum number of adsorption sites occupied per unit area, $C_e$ the concentration of adsorbate in equilibrium solution and $K_L$ the affinity constant of the adsorbate for the surface of the adsorbent or Langmuir constant. Most the modern work on adsorption is based on the works completed by Langmuir \cite{Langmuir1916, Langmuir1918, Swenson2019}. Langmuir was the first to deal with adsorption from a kinetic point of view. The main hypothesis of this theory was that adsorption takes place in the form of a monolayer of molecules adsorbed on the surface of the adsorbent. However in the case of the physisorption of molecules having a weak interaction with the adsorbent, it appeared obvious that it was necessary to take into account the formation of a multilayer of adsorbed molecules. To compensate this limitation, Stephen Brunauer, Paul Hugh Emmett and Edward Teller published for the first time in 1938 an article presenting an extension of Langmuir's theory to a multilayer adsorption, resulting in the BET equation which is used today everywhere to calculate the specific surface \cite{Brunauer1938}. The principle that was used was to apply the Langmuir method to each of the layers of adsorbed molecules. The relationship that governs this model is as follow:
\begin{equation}
q_e = \frac{q_{mBET}~C_{BET}~C_e}{(C_e - C_s) \left[1+(C_{BET}-1)\frac{C_e}{C_s}\right]}
\label{eq:BET}
\end{equation}
where $C_{BET}$ is related to the energy of interaction with the surface also known as the BET constant, $q_{mBET}$ maximum adsorption capacity of adsorbent corresponding to monolayer saturation and $C_s$ the adsorbate monolayer saturation concentration. The theories of Langmuir and Brunauer, Emmett and Teller (BET) are the most well-known theoretical treatments (Figure \ref{fig:freundlich}), however many studies have been carried out since then and have led to the introduction of numerous adsorption isotherm modeling which will not be developed here but which can be found elsewhere \cite{Do1998, Dabrowski2001, Lowell2004, Saadi2015, Karimi2019}.

At the same time, another important contribution by Brunauer, Deming, Deming and Teller \cite{Brunauer1940} dealt with the identification of five principal types of adsorption isotherms for gases and vapours. This identification is known as the BDDT classification and is recommended as the basis for a more complete classification introduced by IUPAC.

The different models of adsorption isotherms can be derived by assuming a thermodynamic equilibrium relationship between the quantity of molecules adsorbed by a unit mass of adsorbent and the quantity of adsorbate remaining in the bulk fluid phase at constant temperature and pH. These isotherms are a main source of information on adsorption and its mechanism, in addition to the calorimetric measurements of the adsorption heat. They give information on the distribution of the adsorbable solute between the liquid and solid phases at different equilibrium concentrations.

As mentioned earlier, adsorption is a consequence of surface energy. Initially, the Gibbs adsorption isotherm for multicomponent systems made it possible to link changes in concentration of a component in contact with a surface with changes in surface tension, which results in a corresponding change of surface energy \cite{Gibbs1874, Boutaric1940, Grumbach1912}. It represents an exact relationship between the adsorption and change in surface tension of a solvent due to presence of a solute. This equation was derived by J. Willard Gibbs (1878) \cite{Gibbs1878} and afterwards independently by J.J Thomson (1888) \cite{Thomson1888}. The Gibbs adsorption equation is application to binary system is the following:
\begin{equation}
d\gamma = \Gamma_1 d\mu_1 + \Gamma_2 d\mu_2
\label{eq:gibbs_1}
\end{equation}
where $\gamma$ is the surface tension, $\Gamma_i$ is the surface excess, i.e.~the concentration linked to the surface of a surfactant $i$ at the surface or at the interface ($mol/m^2$) \cite{Mitropoulos2008}, $\mu_i$ is the chemical potential of component $i$. Knowing that $d\mu = RT \ln c$, the Gibbs equation can be written in the form:
\begin{equation}
\Gamma = - \frac{1}{RT} \frac{d\gamma}{d c}
\label{eq:gibbs_2}
\end{equation}
where $c$ is the concentration of species i in the bulk phase.

\subsection{Kinetic adsorption}\label{kinetic-adsorption}

As just seen, the adsorption isotherms make it possible to determine the quantity adsorbed at the equilibrium for an adsorbate-adsorbent system. This isotherm is represented by the curve of the quantity adsorbed at equilibrium $(q_e)$ as a function of the equilibrium concentration of the solute $(C_e)$. To be adsorbed, an adsorbate molecule must find its way to the adsorbent particle by diffusing through the fluid film surrounding the particle, travelling by diffusion along the length of a pore until it finds a vacant adsorption site, and then adsorb onto the solid surface. As in any transport process, these mass-transfer steps are driven by a departure from equilibrium, with the equilibrium described by an isotherm equation as mentioned before.

Kinetic studies make it possible to determine the adsorption equilibrium time as a function of the adsorbent used under specific conditions. Analysis of kinetic data is necessary to understand the adsorption mechanism and predict diffusion, as well as the kinetic transport mechanism that controls the rate of adsorption. Understanding the adsorption dynamics and determining the system's adsorption rate are important in estimating the response time for the process.

\subsubsection{The different modes of transport}\label{the-different-modes-of-transport}

\begin{figure}[!ht]
\centering
\includegraphics[scale=0.5]{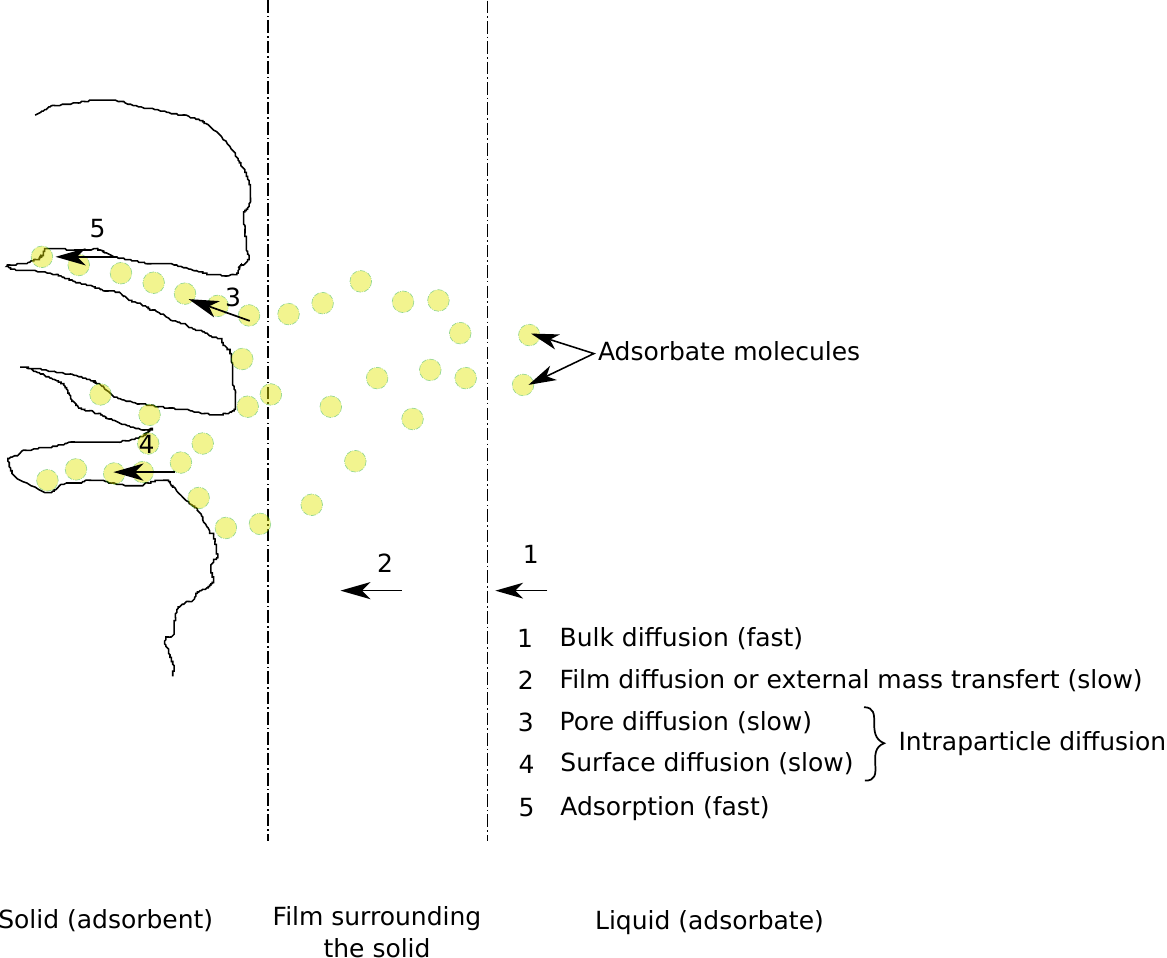}
\caption{The different modes of transport}\label{fig:transport}
\end{figure}

Understanding adsorption systems requires knowledge of equilibrium state and adsorption kinetics. The understanding of the kinetics is largely limited by the theoretical complexity of the adsorption mechanisms as we have seen previously. Many models of varying complexity have been developed to predict the adsorption rate of the adsorbate into the adsorbent \cite{Alberti2012, Plazinski2009}. As defined above, adsorption is a surface phenomenon in which adsorbent molecules (gaseous or liquid) weakly bind to a solid surface. In such circumstances, mass transfer effects are inevitable. The complete course of adsorption therefore includes a mass transfer and comprises four consecutive steps as shown in Figure \ref{fig:transport} \cite{Dabrowski2001, Zhang2018, Ho2000}.

\begin{itemize}
 
\item
  Step 1: Transport of the adsorbate from the bulk liquid phase to the hydrodynamic boundary layer located around the adsorbent.
\item
  Step 2: Film diffusion or External Mass Transfer Resistance (EMTR): transport through the boundary layer surrounding the surface of the particle. The external mass transfer is affected by the design of the characteristic of the material surface and also depends on the hydrodynamic conditions outside the particles. The driving force is the difference in concentration across the boundary layer that surrounds each particle, and the latter is affected by the hydrodynamic conditions outside the particles. Due to the complexities encountered with rigorous treatment of the hydrodynamics around particles, the mass transfer to and from the adsorbent is described in terms of the mass transfer coefficient $k_f$. The flux at the surface of the particles is:
  \begin{equation}
    N_i = k_f (c_i - c_s)
    \end{equation}
  where $c_i$ and $c_s$ are respectively the solute concentrations in the bulk fluid and at the particle surface. $k_f$ can be estimated from the correlations available in terms of the Sherwood number $sh = k_f~d_p/D_i$ and the Schmidt number $Sc = \nu/D_i$. An example of this type of correlation is that of \citet{Wakao1978}, valid for gas and liquid systems:
  \begin{equation}
    Sh = 2.0 + 1.1 Re^{0.6} Sc^{1/3}  \quad (3< Re < 10^4)
    \end{equation}
  Other correlations can be found in general references \cite{Perry1997, Ruthven1984}.
\item
  Step 3: Intraparticle diffusion or Internal Mass Transfer Resistance (IMTR): transport inside the adsorbent particle by diffusion in liquid contained within the pores (porous diffusion) and/or by diffusion in the adsorbed state along the internal surface (surface diffusion).

  \begin{itemize}
   
  \item
    In the case of pore diffusion, the pores are large enough for the adsorbent molecule to escape the force field of the adsorbent surface. Thus, this process is often referred to as macropore diffusion. The driving force of such a diffusion process can be approximated by the mole fraction gradient or, if the mole concentration is constant, by the concentration gradient of the diffusing species within the pores. Thus, the molecules propagate from the surface of the material towards their center through the pores. These transfers are generally described by effective diffusion coefficients which group together several migration phenomena \cite{Lowell2004}:

    \begin{itemize}
     
    \item
      molecular diffusion;
    \item
      Knudsen diffusion (involved when the mean free path of the molecule is larger than the pore size);
    \item
      Poiseuille flow in the case of large pores.
    \end{itemize}
  \item
    During surface diffusion, diffusion takes place in pores small enough so that the diffusing molecule never escapes the force field of the adsorbent surface. In this case, the transport can occur by an activated process involving jumps between adsorption sites \cite{Higashi1963, Okazaki1981, Yang1973}. The driving force of the process can thus be approximated by the concentration gradient of the species in its adsorbed state. Phenomenologically, the process is indistinguishable from that of a homogeneous diffusion which occurs inside a sorbent gel or in a pore-filling fluid immiscible with the external fluid.
    The phenomenological aspects of diffusional mass transfer in adsorption systems can be described according to Fick's law:
    \begin{equation}
      J_i = - D_i (c_i) \frac{\partial c_i}{\partial x}
      \label{eq:diffusion}
      \end{equation}
    This expression can be used to describe both porous and solid diffusion as long as the driving force is expressed in terms of the appropriate concentrations. Although the driving force should be more correctly expressed in terms of chemical potentials, equation \ref{eq:diffusion} provides a qualitatively and quantitatively correct representation of adsorption systems as long as diffusivity can be a function of adsorbate concentration. The diffusivity will only be constant for a thermodynamically ideal system, which is only an adequate approximation for a limited number of adsorption systems.
  \end{itemize}
\item
  Step 4: Adsorption onto Active Sites (AAS): attachment of the adsorbent to the internal surface of the sorbent at the final adsorption sites (adsorption to the adsorbent) by binding ions to the active sites.
\end{itemize}

The overall adsorption rate (as measured experimentally) is determined by the total resistance, which is the sum of the resistances of each of the steps in series. This rate can be controlled by any of these steps individually, as well as by combined effect of a several steps. Reducing the overall resistance value therefore increases the adsorption rate. One of the stages usually offers much greater resistance than the others, thus single-handedly limiting the speed of the process. Normally the first two steps are fast processes when compared to intraparticle diffusion (step 3) and the binding reaction at surface sites (step 4) as adequate mixing of the solution in contact with the solid results in removal of the boundary layer around the sorbate. If that condition is met, only the last 2 processes effectively control the kinetic mechanism of sorption \cite{Alberti2012, Unuabonah2019}. Overall, transport resistances depend on many factors, including the types of adsorbent and adsorption, their properties and operating conditions. The limited step may also change during the adsorption process.

In order to determine the contribution of each step, many kinetic models were used comparatively to predict the behavior of the experimental data. The most used are those which suppose that step 4 makes a significant contribution to the kinetics of a process. The crucial assumption behind this model is that the rate of transfer of solute molecules from a solution (located in the direct vicinity of the surface) to the adsorbed phase governs or is at least partially involved in the overall rate of the sorption process.

However, it would seem that in order to be able to correctly theoretically calculate the overall adsorption rate, many precise parameters need to be known. It has indeed been argued by \citet{Ho1999} that a knowledge of the rate law describing the adsorption system is necessary. The rate law is determined by experimentation and cannot be deduced by a simple examination of the overall chemical reaction equation. The three main requirements of a law on rates were highlighted by Spark \cite{Sparks1986}:

\begin{itemize}
 
\item
  knowledge of all the molecular details of the reaction, including energetics and stereochemistry,
\item
  interatomic distances and angles throughout the reaction,
\item
  the individual molecular steps involved in the mechanism.
\end{itemize}

\hypertarget{the-different-kinetic-models}{%
\subsubsection{The different kinetic models}\label{the-different-kinetic-models}}

In general, the mathematical models to describe the adsorption kinetics \cite{Belaid2011, Kajjumba2018, Kecili2018, Qiu2009, Wang2020} are separated into two groups:

\begin{itemize}
 
\item
  Adsorption reaction models,
\item
  Adsorption diffusion models.
\end{itemize}

\paragraph{Adsorption reaction models:}\label{adsorption-reaction-models}

The adsorption reaction models derived from the chemical reaction kinetics are based on the entire adsorption process without taking into account the steps mentioned in the adsorption diffusion model. Pseudo-first order (PFO) and pseudo-second order (PSO) models are the two most common empirical models used in liquid adsorption studies.

\begin{itemize}
 
\item
  The pseudo-first order model (PFO) was developed by \citet{Lagergren1898}. The main assumptions of this model kinetics are as follows:

  \begin{itemize}
  \item
    Adsorption occurs only at specific binding sites, located on the surface of the adsorbent.
  \item
    The adsorption energy does not depend on the formation of a layer on the adsorbent surface.
  \item
    A first order equation (i.e. equation \ref{eq:PFO}) governs the speed of the adsorption process.
  \item
    No interactions occur between the molecules adsorbed on the surface of the adsorbent.

    The equation of the pseudo-first order kinetic model is given by:
    \begin{equation}
    \frac{dq_t}{dt} = k_1~(q_e - q_t)
    \label{eq:PFO}
    \end{equation}
    where $q_e$ is the amount of compound adsorbed $(mg.g^{-1}$) to the adsorbent at equilibrium, $q_t$ is the amount of compound adsorbed $(mg.g^{-1}$) on the adsorbent at time t $(t$), and $k_1$ $(min^{-1}$) represents the rate constant for the first order adsorption process.
  \end{itemize}
\item
  The pseudo-second order kinetic model (PSO) has almost the same assumptions as the pseudo-first order model, the only difference is the model rate \cite{Ho1999b, Ho2006}. The PSO model assumes that the solute adsorption rate is proportional to the sites available on the adsorbent. The reaction rate depends on the amount of solute on the surface of the adsorbent; the driving force is proportional to the number of active sites available on the adsorbent. It has been demonstrated :
  \begin{equation}
    \frac{dq_t}{dt} = k_2~(q_e - q_t)^2
    \label{eq:PSO}
    \end{equation}
\end{itemize}

\paragraph{Adsorption diffusion models:}\label{adsorption-diffusion-models}

As previously mentioned, typical liquid/solid adsorption involves film diffusion, intraparticle diffusion, and mass action. For physical adsorption, mass action is a very fast process and may be negligible for the kinetic study. So, the adsorption kinetic process is always controlled by liquid film diffusion or intraparticle diffusion, i.e.~one of the processes should be the rate limiting step. Therefore, adsorption diffusion models are primarily constructed to describe the process of film diffusion and/or intraparticle diffusion.
They are developed on the basis of a three step process.

\begin{itemize}
\item
  Step 1 - Liquid film diffusion model: In this first step, it is assumed that diffusion occurs through the liquid film surrounding the adsorbent particles. The resistance to external mass transfer is associated with the existence of a laminar sublayer around the adsorbent particles (no slip condition at the solid boundary), which is called a \textit{film}. Mass transfer through an outer liquid film has been described as one-dimensional diffusion in a flat layer, in which the driving force is a concentration gradient between a region very close to the outer surface of the adsorbent and the solution in bulk. The models that describe these phenomena are as follows:

  \begin{itemize}
  \item
    Linear motive force rate law:

    In liquid/solid adsorption systems, the rate of solute accumulation in the solid phase is equal to that of solute transfer through the liquid film according to the law of mass equilibrium. The rate of solute accumulation is given by:
    \begin{equation}
      V_p \left(\frac{\partial \bar{q}}{\partial t}\right)  
    \end{equation}
    where $\bar{q}$ is the average concentration of solute in the solid and $V_p$ the volume of the particle. During this time, the rate of solute transfer through the liquid film is proportional to the area of the particle $A_s$ and the driving force of concentration $(C - C_i)$ ($mg.L^{-1}$). Therefore, it is equal to $k_f A_s (C - C_i)$ where $k_f$ is the film mass transfer coefficient ($h^{-1}$). We obtain the following relation for the external distribution:
    \begin{equation}
        V_p \left(\frac{\partial \bar{q}}{\partial t}\right) = k_f A_s (C - C_i)
    \end{equation}
    By taking into account the density of absorbent particles $\rho_p$ ($g.L^{-1}$), we obtain:
    \begin{equation}
        \left(\frac{\partial \bar{q}}{\partial t}\right) =\frac{k_f}{\rho_p} (C - C_i)
    \end{equation}
    This equation is called: \textit{linear driving force rate} is generally applied to describe the transfer of mass through the liquid film.

\item
  Film Diffusion Mass Transfer Rate Equation

  The film scattering mass transfer rate equation presented by \citet{Boyd1947} is:
  \begin{equation}
   \ln \left (1 - \frac{q_t}{q_e}\right) = - R^{'}t
   \end{equation}
  with $R^{'} = \frac{3D_e^{'}}{r_0 \Delta r_0 k^{'}}$ where $R^{'}$ ($min^{-1}$) is the diffusion constant of the liquid film, $D_e^{'}$ ($cm^2.min^{-1}$) is the effective diffusion coefficient of the liquid film, $r_0$ ($cm$) is the radius of the adsorbent beads, $\Delta r_o$ ($cm$) is the thickness of the liquid film, and $k^{'}$ is the adsorption equilibrium constant.
  \end{itemize}  
\item
  Step 2 - In the second step, the mechanism of intraparticle diffusion is assumed to be governed by the diffusion into the liquid contained in the pores and along the walls of the pores. The models that describe these phenomena are as follows:

  \begin{itemize}
  \item Homogeneous Solid Diffusion Model (HSDM):

    Intraparticle diffusion has been described by Fick's law \cite{Crank1956}, which is obtained from applying the mass conservation law by considering a spherical shell as the control volume. This typical intraparticle diffusion model is the \textit{Homogeneous Solid Diffusion Model (HSDM)}, which can describe mass transfer in an amorphous and homogeneous sphere \cite{Cooney1999}. The HSDM equation can be presented as:
    \begin{equation}
        \frac{\partial q}{\partial t} = \frac{D_s}{r^2} \frac{\partial}{\partial r}\left(r^2 \frac{\partial q}{\partial r}\right)
    \end{equation}
    where $D_s$ is the intraparticle diffusion coefficient, $r$ the radial position and $q$ the amount of solute adsorption in the solid varying with the radial position at time t.
    \citet{Crank1975} gave an exact solution to this equation for the case of the \textit{infinite bath} where the sphere is initially free of solute and the concentration of the solute at the surface remains constant. The strength of the outer film can be neglected depending on the constant surface concentration. This result is of the form:
    \begin{equation}
        \frac{\bar{q}}{q_{\infty}} = 1 - \frac{6}{\pi^2} \sum_{n=1}^{\infty} \frac{1}{n^2} \exp \left(\frac{- D_s n^2 \pi^2 t}{R^2}\right)
    \end{equation}
    where $q_{\infty}$ represents the average concentration in the solid at infinite time and R the total particle radius. It has been observed that the adsorption rate decreases with increasing size of the adsorbent particles and vice versa. Linearization gives:
    \begin{equation}
        \ln \left(1 - \frac{\overline{q}}{q_{\infty}}\right) = \frac{- D_s \pi^2}{R^2} t + \ln \frac{6}{\pi^2}
    \end{equation}
    However, the assumption of a constant surface concentration for HSDM is likely to be violated in the long term. Therefore, the equation discussed above is generally quite valid only within a short period of time \cite{Cooney1999}.
    
  \item Weber-Morris model:

    The intraparticle diffusion model is mainly used for the description of the adsorption process if the diffusion of the adsorbate to the porous adsorbent is the rate limiting step. Weber-Morris found that in many cases of adsorption, solute uptake varies almost proportionally with $t^{1/2}$ rather than with contact time $t$. They proposed this kinetic model \cite{Weber1963} and described it in the following formula:
    \begin{equation}
        q_t = k_p.\sqrt(t) + C
    \end{equation}
    where $C$ is the intersection and represents the thickness of the boundary layer. The higher the $C$ value, the greater the effect of the boundary layer. $q_t$ represents the quantity of compound adsorbed ($mg.g^{-1}$) on the adsorbent at time $t$ ($min$) and $k_p$ is the rate constant of the intraparticle diffusion ($mg.g^{-1}min^{-0.5}$). For the Weber-Morris model, it is essential that the $q_t \sim~ t^{1/2}$ graph passes through the origin if intraparticle diffusion is the only rate limiting step. However, this is not always the case and the adsorption kinetics can be controlled simultaneously by film diffusion and intraparticle diffusion.
    
  \item Dumwald-Wagne model:

    Dumwald-Wagner proposed another model which makes it possible to verify whether intrapaticle diffusion is the only step limiting the speed which has the following form:
    \begin {equation}
    F = \frac{q_t}{q_e} = 1 - \frac{6}{\pi^2} \sum_{n=1}^{\infty} \frac{1}{n^2} exp(-n^2 K t)
    \end{equation}
    where $k$ ($min^{-1}$) is the adsorption rate constant.
    \begin{equation}
    \log (1 - F^2) = - \frac{K}{2,303}t
    \end{equation}
    A plot of $\log(1-F^2) \sim t$ must be linear and the rate constant $K$ can be obtained from the slope.
    The Dumwald-Wagner model has been found to be adequate for modeling different types of adsorption systems, for example, adsorption of p-toluidine from aqueous solutions onto hypercrosslinked polymeric adsorbents \cite{Wang2004}.
  \end{itemize}
\item
  Step 3 - Adsorption on active sites (AAS): In the last step, adsorption and desorption take place between the adsorbate molecules and the active sites within the adsorbent. Models for AAS assume that adsorption at the active site is the slowest step and that the diffusion process is negligible. Two models can describe the AAS phenomenon.

  \begin{itemize}
  \item
    Langmuir kinetics model:
    The adsorption in adsorbent sites can be described by the Langmuir kinetic model such as:
    \begin{equation}
    \frac{dq_t}{dt} = k_a C_t (q_e - q_t) - k_d q_t
    \end{equation}
    where $k_a$ is the adsorption kinetic constant ($L.mg^{-1}.h^{-1}$) and $k_d$ is the desorption kinetic constant ($h^{-1}$). At adsorption equilibrium, this equation reduces to the Langmuir isothermal model \cite{Langmuir1918}. Efforts have been made to find the analytical solution to this equation. The resolution methods reported by previous studies have been shown to be complex, which can lead to difficulties in estimating the parameters.
  \item
    Phenomenological AAS model:

    The phenomenological AAS mode, developed on the basis of Langmuir's kinetic model \cite{Langmuir1918, Thomas1944}, can be described by \cite{Blanco2017, Sausen2018}:
    \begin{equation}
    \frac{dq_t}{dt} = k_a \left(C_0 - \frac{m q_t}{V}\right) (q_{max} - q_t) - \frac{k_a}{K_L} q_t
    \end{equation}
    where $C_0$ represents the initial concentration ($mg.L^{-1}$), $V$ the volume of the solution ($L$) and $K_L$ the Langmuir constant.
    The AAS model was adopted to study the adsorption mechanism \cite{Blanco2017, Guo2019}.
  \end{itemize}
\end{itemize}

\subsection{Thermodynamic aspects}\label{thermodynamic-aspects}

In evaluating the overall adsorption process, it is fundamental to study on a large scale:

\begin{itemize}
 
\item
  mass transfer phenomena,
\item
  thermodynamics,
\item
  equilibrium, in order to better understand the process.
\end{itemize}

In previous sections mass transfer was described via absorption kinetics and equilibrium via absorption isotherms.

The original concepts of thermodynamics assume that in an isolated system, where energy can neither be gained nor lost, the change in entropy $\Delta S^0$ is the driving force. Energy and entropy factors must be taken into account in order to determine which processes will occur spontaneously. The nature of the adsorption process can be predicted by calculating these thermodynamic parameters \cite{Gueu2007}. Gibbs' free energy $\Delta G^0$ is the fundamental criterion of spontaneity. The reactions (or sorption processes in this case) will spontaneously occur at a given temperature if $\Delta G^0$ is a negative amount. On the other hand, the enthalpy $\Delta H^0$ gives information on whether the adsorption process is endothermic or not. The adsorption process is endothermic when $\Delta H^0$ is a positive value.

In order to clarify information regarding energy changes associated with adsorption and the feasibility and characteristics of the process, the thermodynamics governing sorption processes were investigated. For this, the adsorption activation energy ($E_a$) was calculated using the Arrhenius expression by applying the parameters of phenomenological kinetic models (diffusion and adsorption). So we have:
\begin{equation}
k_i = A \exp\left(- \frac{E_a}{RT}\right)
\end{equation}
in which $k_i$ is one of the kinetic parameters of each kinetic model obtained at different temperature values, $A$ is the Arrhenius pre-exponential factor (the unit depends on the kinetic model used), $E_a$ the activation energy ($kJ.mol^{-1}$), $R$ is the universal gas constant ($J.mol^{-1}.K^{-1}$) and $T$ is the temperature ($K$).

The enthalpy of adsorption $\Delta H$ was determined using the equilibrium data. For this, Henry's constant $K_H$ was determined by:
\begin{equation}
K_H = \lim_{c_{eq}\rightarrow 0}\left(\frac{q_{eq}}{C_{eq}}\right)
\end{equation}
In the case of the Langmuir isothermal model, Henry's constant can be determined, by applying his definition of infinite dilution according to:
\begin{equation}
K_H = \lim_{c_{eq}\rightarrow 0} \left(\frac{q_{max} K_L C_{eq}}{1 + K_L C_{eq}} \frac{1}{C_{eq}}\right) = q_{max} K_L
\end{equation}
The enthalpy of adsorption $\Delta H$ was therefore determined by:
\begin{equation}
K_H = K_0 \exp \left(-\frac{\Delta H}{RT}\right)
\end{equation}
where $K_H$ is Henry's constant ($L.g^{-1}$), $K_0$ ($L.g^{-1}$) is Henry's pre-exponential factor and $\Delta H$ the enthalpy of adsorption ($kJ.mol^{-1}$).

The free energy of the sorption reaction, considering the sorption equilibrium constant, is given by the following equation:
\begin{equation}
\Delta G^0 = - R T \ln K_L
\end{equation}
where $\Delta G^0$ is the change of standard free energy ($J$); $R$ is the universal gas constant 8.314 $J.mol^{-1}.K^{-1}$ and $T$ is the absolute temperature ($K$).

In fact the Gibbs free energy change, $\Delta G^0$, can be represented as follows :
\begin{equation}
\Delta G^0 = \Delta H^0 - T \Delta S^0
\end{equation}
So the equilibrium adsorption distribution constant can be related to the changes in the standard entropy, $\Delta S^0$ and enthalpy, $\Delta H^0$ as:
\begin{equation}
\ln K_L = \frac{\Delta S^0}{R} - \frac{\Delta H^0}{RT}
\end{equation}

\subsection{Applications to polymer materials}\label{applications-to-polymers}

Adsorption models, derived from research on the phenomena of interaction between gases and mineral adsorbents initiated by Langmuir, were historically the first to be applied to polymer systems and remained the only theoretical tool until 1950 \cite{McLaren1951}.

The kinetics of sorption in an amorphous polymer is governed by:

\begin{itemize}
 
\item
  interactions between the polymer and the solvent,
\item
  the state of molecular mobility of the polymer,
\item
  the size of the diffusing molecules.
\end{itemize}

The transport of adsorbates therefore depends on their own capacity for movement and on the mobility of the chains of the polymer considered, but also on factors including the solubility and diffusivity of the penetrant in the polymer, the morphology, the fillers and the plasticization \cite{Karimi2011}. Apart from substances which react chemically with polymers (bases and strong acids for example), the molecules are liable to adsorb onto the polymers, then to penetrate into the polymers when their mass and their steric hindrance are not too great.

On the other hand, for ion exchange on synthetic polymers when convection in solution is low and/or degree of polymer crosslinking is low, it is likely that the main resistance to diffusion can be found in the liquid phase. Conversely, in the case of strong crosslinking, the resins and materials are more compact, so that the mass transfer inside the particles is more difficult \cite{Alberti2012}.

\subsubsection{Flory-Huggins model}\label{flory-huggins-model}

The sorption of molecules on the surface of the adsorbent is described as a phase change, while the adsorbent, which constitutes a proper phase of the system, is unaffected by this process. The phenomenon can be described using mechanistic methods (Langmuir, BET), or on thermodynamic bases. The mechanism which is at the base of the theories modeling the adsorption, if it adapts perfectly to rigid mineral structures, does not adapt well to the properties of the \textit{soft material} type manifested by polymers. A radically different approach was then taken by Flory and Huggins \cite{Flory1950}, treating the polymer/solvent mixture as one and as the same fluid phase (solution); an elaborate calculation of the entropic and enthalpic contributions of this type of mixture made it possible to arrive to the first model correctly accounting for certain atypical properties of polymers in solution (solubility, heat of mixing, osmotic pressure, sorption isotherms).

\begin{figure}[!ht]
\centering
\includegraphics[scale=0.4]{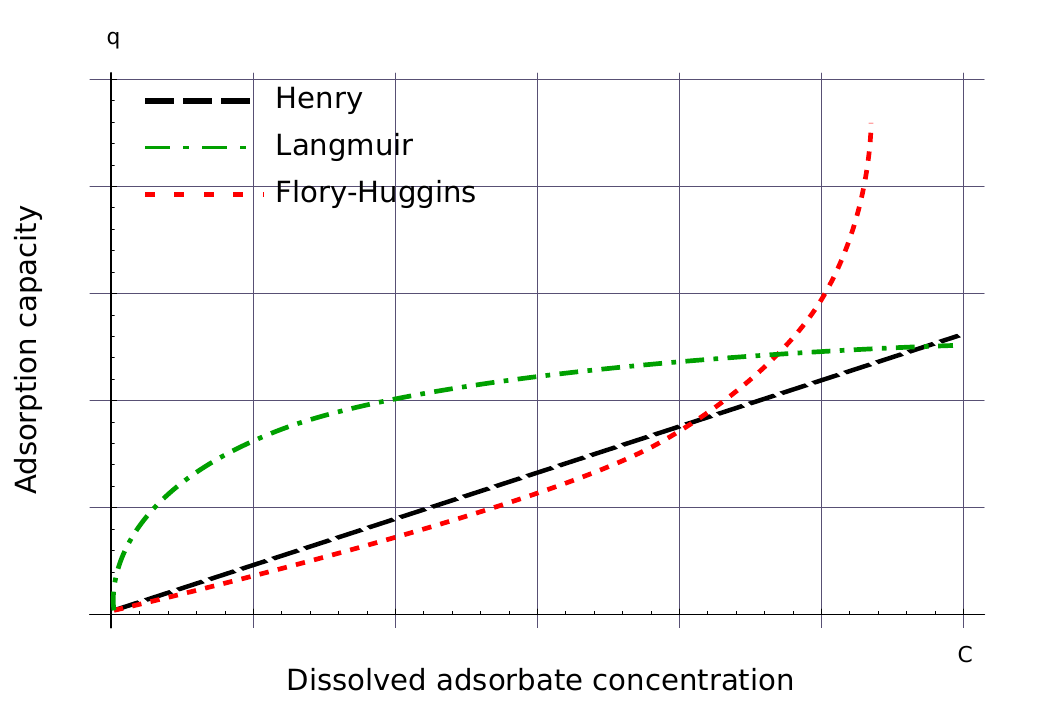}
\caption{Comparison between the adsorption isotherms of Henry, Langmuir and Flory-Huggins}\label{fig:flory}
\end{figure}

Very different types of isotherms can be obtained with polymer/solvent solutions depending on the nature of the species involved; the main types are shown in Figure \ref{fig:flory}. The current trend tends to treat differently the extreme case corresponding to the sorption of liquids or vapors in elastomeric materials, leading to exponential-type paces, for which the Flory-Huggins has enabled experimental results to be reported correctly in many cases.

The Flory-Huggins model is the most famous and the most used. It has been established for ideal systems, i.e.~for non-polar solvents which do not exhibit strong solvent-solvent and solvent-polymer interactions \cite{Flory1953, Huggins1942}, and therefore is not appropriate for systems using aqueous solvents. The Flory-Huggins isotherm describes the degree of surface coverage characteristics of the adsorbate on the adsorbent \cite{Ayawei2017}. The linear form of the Flory-Huggins equation is expressed as:
\begin{equation}
\ln \frac{\theta}{C_0} = \ln K_{FH} + n \ln (1 - \theta)
\end{equation}
where $\theta$ is the degree of surface coverage, $n$ is the number of adsorbates occupying the adsorption sites and $K_{FH}$ is the Flory-Huggins equilibrium constant ($L.mol^{-1}$).

This isothermal model can express the feasibility and spontaneity of an adsorption process. The equilibrium constant is used to calculate the Gibbs free energy of spontaneity as shown in the following expression:
\begin{equation}
\Delta G^0 = R T \ln K_{FH}
\end{equation}
where $\Delta G^0$ is the standard free energy change, $R$ is the universal gas constant 8.314 $J.mol^{-1}.K^{-1}$ and $T$ is the absolute temperature.
This sorption model describes well the sorption of a solvent into a polymer at a low glass transition temperature, but not the interactions of a dissolved solute with the solid polymer. In most cases, especially in the case of strong interactions between the polymer and the solvent, these simple models may not accurately describe the sorption of a solvent into a polymer. A combination of these models is then used to adjust the sorption isotherms and to interpret the phenomena involved in the sorption process.

\subsubsection{GAB (Guggenheim Anderson de-Boer) model}\label{gab-guggenheim-anderson-de-boer-model}

\begin{figure}[!ht]
\centering
\includegraphics[scale=0.4]{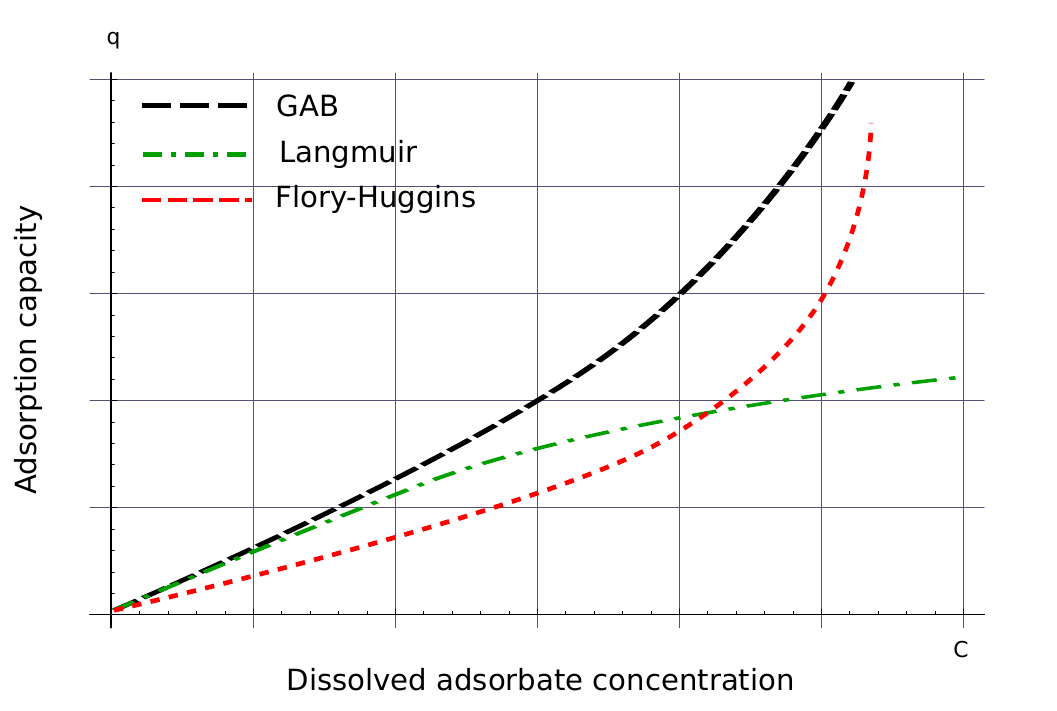}
\caption{Comparison between the adsorption isotherms of Langmuir, Flory-Huggins and GAB model}\label{fig:GAB}
\end{figure}

The GAB model can be described by the contribution of the Langmuir and Flory-Huggins sorption modes. The Guggenheim Anderson de-Boer (GAB) isotherm is a modification of the Langmuir and BET physical adsorption isotherms (Figure \ref{fig:GAB}). This isotherm necessarily includes an additional parameter, $K_G$, which is the criterion for the difference of the standard chemical potential between the molecules of the second and subsequent adsorption layers and those of molecules in liquid state. The GAB equation is determined as:
\begin{equation}
q_e = \frac{q_{mG}~ C_{GAB}~ K_G~ C_e}{(C_s - K_G C_e)\left[1 + (C_{GAB} - 1) K_G \frac{C_e}{C_s}\right]}
\end{equation}
where $C_{GAB}$ and $K_G$ are the GAB constants, which are related to the energies of interaction between the first and the further sorbed molecules at the individual sorption sites.

In this mode the solvent is absorbed in preferential sites and it is able to form solvent ``clusters'' around these sites, unlike a pure Langmuir mode in which a single solvent molecule is absorbed in an absorption site .

In order to estimate the size of molecular \textit{clusters}, several equations found in the literature have been proposed. These equations calculate an average size or amount of solvent molecules that are suspected of clumping, which is known as the \textit{Mean Cluster Size} or MCS \cite{Dolmaire2004, Jonquieres1998}.
\begin{equation}
MCS = -(1 - \phi) \left[\left(\frac{\phi}{\alpha_{gab} c_{gab}}\right)(-2 c_{gab} ka + 2ka + c_{gab} - 2) -1 \right]
\end{equation}
Where $\alpha_{GAB}$ is the single-layer solvent content, $C_{GAB}$ is the Guggenheim constant, $k$ is a factor correcting the properties of multilayer molecules relative to bulk liquid, and $a$ chemical activity of the solvent.

Another function of clustering is the Zimm-Lundberg relationship \cite{Zimm1956}. This equation was developed from the assumption that polar molecules tend to aggregate. The function is derived from statistical mechanics and indicates that the free energy of the polymer-solvent system depends on the first derivative of the chemical activity of the solvent relative to the volume fraction of the solvent. So Zimm and Lundberg developed an approach to estimate the probability of formation of clusters of penetrating in a polymer matrix, in the case of nonlinear sorption isotherms. The principle is based on the calculation of the probability P (i, j), for two molecules i and j, of occupying coordinate positions included in the space delimiting two elementary volumes $d(i)$ and $d(j)$ for a total volume system V \cite{Favre1994}. The Zimm-Lundberg function is given by:
\begin{equation}
\frac{G}{V} = -(1 - \phi) \left(\frac{\partial\left(\frac{a}{\phi}\right)}{\partial a}\right)_{P,T} - 1
\end{equation}
Where $G$ is the cluster integral, $V$ is the partial molecular volume of the solvent, $\phi$ is the volume fraction of the solvent absorbed by the polymer, and $a$ is the chemical activity of the solvent. This type of approach therefore has no predictive power, but offers a useful means of interpreting sorption data, based on geometric considerations. Its major advantage lies in the fact that it only requires sorption data (a: chemical activity of the solvent, the volume fraction of the solvent absorbed by the polymer), as well as an analytical function, in order to be implemented giving a correct description of the isotherm. From the Zimm-Lundberg clustering function, the mean cluster size (MCS) can be expressed as :
\begin{equation}
MCS  = 1 +\phi \frac{G}{V}
\end{equation}

\subsubsection{ENSIC (ENgaged Species Induced Clustering) clustering model}\label{ensic-engaged-species-induced-clustering-clustering-model}

The description of the sorption equilibria between adsorbates and polymers is of the utmost importance for studies dealing with mass transfer through dense polymer membranes, since transport mechanisms in such systems is generally assumed to follow a diffusion process in solution. Understanding upstream thermodynamic equilibrium is therefore an essential prerequisite to further separate the equilibrium (i.e.~solution) and kinetic (i.e.~diffusion) aspects in the resulting mass transfer. The inherent complexity in the equilibria of polymeric solvents (i.e.~a large difference in molar volumes) leads to highly non-ideal equilibrium behaviors, even for athermal solutions, as the pioneering work of Flory clearly demonstrates \cite{Flory1953}.

Given the difficulties encountered when taking into account the peculiarities of polymer solvents (non-regular enthalpy of mixing, elastic contribution of the network), other types of mixtures can hardly be described by purely thermodynamic models unless complicated equations are used. A simple mechanistic approach has been developed in order to circumvent these limitations; it is based on the assumption that the insertion of a solvent molecule into the polymeric solvent matrix will be governed by the intrinsic affinity of the solvent either for a polymer segment or for an already sorbed solvent molecule. The model allowing the description of the formation of clusters was named Clustering Induced by the ENgaged Species (ENSIC).

This model developed by E. Favre et al. \cite{Favre1993} is a mechanistic approach to the sorption process. It reports numerous sorption isotherms quantitatively and with great precision \cite{Favre1996, Jonquieres1998, Scott1949}. He considers that the sorption of a solvent molecule can only take place on two different sites of the swollen material: either on a \textit{solvent} site (ie. a site already occupied by a solvent molecule, or on a \textit{polymer} site. The swollen material is represented by a network of cells, of the same volume, containing a segment of the macromolecular chain or a solvent molecule. In this model, an infinitesimal increase in the activity of the supernatant results in the sorption of the solvent molecules on one or the other of the sites according to an elementary Henry's law with a probability $k_s$ or $k_p$ respectively representing the affinity of the solvent for itself or for the polymer. The mathematical development leads to the following expression:
\begin{equation}
dn_s= (k_p n_p + k_s n_s) dp
\label{eq:ENSIC}
\end{equation}
where $n_p$ and $n_s$ are respectively the polymer molecules and the number of solvent molecules in the polymer, and $k_p$ and $k_s$ are respectively the affinity parameters between a molecule newly absorbed with the polymer or with the solvent already in the polymer. By integrating equation \ref{eq:ENSIC}, the relation describing the supply of solvent according to its chemical activity in the preceding equation is obtained:
\begin{equation}
\phi = \frac{[exp(a(k_s - k_p))-1]}{(k_s - k_p)/k_p}
\end{equation}
thus, the mean cluster size (MCS) can be estimated from the ENSIC model :
\begin{equation}
MCS = \frac{ (1-\phi)(1-K \phi) \ln (1 + K \phi) }{ K \phi } + \phi
\end{equation}
with $K = \frac{k_s - k_p}{k_p}$

According to the values of the affinity constants, three cases appear:

\begin{itemize}
 
\item
  $k_s = k_p$ : The solvent molecule has no preference between the ``solvent'' sites or the ``polymer'' sites of the swollen material.
\item
  $k_s > k_p$ : The affinity of the solvent is stronger for a molecule of the same nature than for the polymer. There will be formation of aggregates and self-association of the solvent molecules.
\item
  $k_s < k_p$ : The isotherms obtained have a convex shape recalling the isotherms classically described by the Langmuir equation corresponding to a progressive saturation of the solvation sites.
\end{itemize}

This model has been shown to be effective in describing the sorption of solvents of increasing polarity on to polar polymers \cite{Jonquieres1998}.

ENSIC clustering model (ENgaged Species Induced Clustering) considers a probabilistic approach in which a solvent molecule is absorbed by a polymer-solvent system. The model is governed by the interactions between the newly absorbed molecule with the polymer and with the solvent already absorbed by the material. The ENSIC clustering function is defined by the increase in the number of molecules for an increase in the pressure of the solvent (or of the chemical activity) \cite{Favre1993, Favre1996, Haidong2006}.

\begin{sidewaystable*}
\caption{Adsorption kinetic models (\textit{$q_e$ :  quantity of adsorbate fixed per unit of adsorbent surface at the equilibrium; $C_e$: concentration of adsorbate in the equilibrium solution; $K_f$: Freundlich constant; $K_L$: Langmuir constant; $N$:  quantity adsorbed at saturation; $q_{mBET}$: maximum adsorption capacity of adsorbent corresponding to monolayer saturation;  $C_{BET}$: BET constant; $C_s$: adsorbate monolayer saturation concentration; $\theta$: degree of surface coverage; $K_{FH}$: Flory-Huggins equilibrium constant; $n$: number of adsorbates occupying the
adsorption sites); $C_{GAB}$ and $K_G$ GAB constants)}\label{tbl:table2}}
\begin{tabular}{p{2cm}|p{4cm}|p{7cm}|p{6cm}}
\toprule%
Date & Name & Significance & Equation\\
\midrule
1909 & Freundlich &  Empirical equation for representing the isothermal
variation of adsorption of a quantity of gas adsorbed by unit mass of solid adsorbent with pressure & $q_e = K_f~C_e^{1/n}$\\
\\
1918 & Langmuir & Adsorption isotherm based on the
kinetic theory of gases - interpreted the kinetics of surface reaction in terms of his monolayer equation & $q_e = n \times \frac{K_L \times C_e}{1 + (K_L \times C_e)}$\\
\\
1938 & BET & Take into account the formation of a multilayer of adsorbed molecules & $q_e = \frac{q_{mBET}~C_{BET}~C_e}{(C_e - C_s)\left[1 + (C_{BET} - 1) \frac{C_e}{C_s}\right]}$\\
\\

1940 & Flory-Huggings & Adaptation of Langmuir and BET to polymers by treating the polymer/solvent mixture as one and as the same fluid phase & $\ln \frac{\theta}{C_0} = \ln K_{FH} + n \ln (1 - \theta)$\\
\\
1945 & Guggenheim Anderson de-Boer (GAB) isotherm & Modification of the Langmuir and BET physical adsorption isotherms by including an additional parameter, $K_G$, which is the criterion for
the difference of the standard chemical potential between the molecules of the second and subsequent adsorption layers and those of molecules in liquid state & $q_e = \frac{q_{mG}~ C_{GAB}~ K_G~ C_e}{(C_s - K_G C_e)\left[1 + (C_{GAB} - 1) K_G \frac{C_e}{C_s}\right]}$\\
\\
2014 & ENSIC (ENgaged Species Induced Clustering) model & Probabilistic approach in which a solvent molecule is absorbed by a polymer-solvent system & \\
\end{tabular}
\end{sidewaystable*}

\section{Absorption - Permeation}\label{absorption---permeation}

\subsection{Absorption definition}\label{absorption}

Absorption is a physical and chemical phenomenon or process in which atoms, molecules or ions enter into another, different gas, liquid or solid phase. If any specific absorption is a purely physical process, and isn't accompanied by any other physical or chemical processes, it will generally follows Nernst's distribution law (1891), or partition law, which governs the distribution of a solute between two non-miscible solvents.

A \textit{partition coefficient (P)} or \textit{distribution coefficient (D)} is the ratio of concentrations of a compound in a mixture of two immiscible solvents at equilibrium. This ratio is therefore a comparison of the solubilities of the solute in these two liquids.

\begin{itemize}

\item
  The \textit{partition coefficient P} generally refers to the concentration ratio of the non-ionized species of the compound. Partition coefficients are useful for estimating the distribution of drugs in the body. $\log P$, also called $\log kow$, is a measure of differential solubility of chemical compounds in two solvents. Experimentally, the most used system is octanol/water system, but other systems have also been used for partition coefficient determination. When one of the solvents is water and the other is a non-polar solvent, then the $\log P$ value is a measure of lipophilicity or hydrophobicity. For example, in a biphasic system of n-octanol (hereafter simply \textit{octanol}) and water:
  \begin{equation}
  \log P_{oct/wat} = \log \left(\frac{[solute]_{octanol}^{un-ionized}}{[solute]_{water}^{un-ionized}}\right)
  \end{equation}
\item
  The \textit{distribution coefficient D}, $\log D$, is the ratio of the sum of the concentrations of all forms of the compound (ionized plus un-ionized) in each of the two phases, one essentially always aqueous; as such, it depends on the pH of the aqueous phase, and $\log D = \log P$ for non-ionizable compounds at any pH, it can be expressed as:
    \begin{equation}
  \log D_{oct/wat} = \log \left(\frac{[solute]_{octanol}^{ionized} + [solute]_{octanol}^{un-ionized}}{[solute]_{water}^{ionized} + [solute]_{water}^{un-ionized}}\right)
  \end{equation}
\end{itemize}

\textit{Chemical absorption} or \textit{reactive absorption} is a chemical reaction between the absorbed and the absorbing substances, leading to the covalent bonding of the reacting moieties. Sometimes it combines with physical absorption. This type of absorption depends upon the stoichiometry of the reaction and the concentration of its reactants.

\subsection{Solution-diffusion mechanism}\label{solution-diffusion-model}

For polymeric materials, the transport of liquids (including gases) through the material can be described by the \textbf{solution-diffusion mechanism} . This process was first proposed by Sir Thomas Graham in 1866 \cite{Graham1866}. He proposed that a gas that penetrates through a non-porous polymer takes place in three stages:

\begin{enumerate}
\def\labelenumi{\arabic{enumi}.}
 
\item
  Sorption (S): or adsorption/absorption/condensation of the permeating species on the surface of the polymer (at the limit of the upstream surface).
\item
  Diffusion (D): the penetrating species diffuses through the membrane along chemical or pressure gradients.
\item
  Desorption: or evaporation from the polymer on the opposite side (or the downstream limit) of the membrane.
\end{enumerate}

Steps 1 and 3 are very fast compared to step 2, so the diffusion through the polymer is the step limiting the rate of mass transport through a membrane.

\subsubsection{Principle of diffusion}\label{principle-of-diffusion}

Diffusion is the process by which a small molecule is transferred through the system due to random molecular movements. It is therefore a kinetic term which reflects the mobility of the penetrant in the polymer phase. From a general point of view, diffusion is a phenomenon which tends to homogenize the concentrations of molecules in a matrix. It is in fact a phenomenon of random transport governed most often by Fick's laws and which has the overall consequence of a displacement of the constituents from the most concentrated areas to the less concentrated areas. Diffusion can be considered as the process by which a molecule will pass through a polymer phase by a succession of random movements.

\paragraph{Fick's Laws:}\label{ficks-laws}

The first mathematical treatment of diffusion was established by Fick \cite{Fick1855} who developed a one-dimensional diffusion law. This equation is also known as Fick's first law (Eq \ref{eq:fick1}). In the case of diffusion without convection and a unit area:
\begin{equation}
J = -D \frac{\partial C}{\partial x}
\label{eq:fick1}
\end{equation}
where $J$ is the flow of solvent in the polymer, $D$ is the diffusion coefficient, which is constant in the Fickian regime, $C$ is the concentration of the solvent in the polymer, $x$ the distance and $\frac{\partial C}{\partial x}$ the concentration gradient along the diffusion axis. This first law can only be applied directly to diffusion in steady state, whereas the concentration does not vary with time. This equation \ref{eq:fick1} is the starting point for many diffusion models in polymer systems.

In an unstable state where the penetrant accumulates in some element of the system, Fick's second law describes the diffusion process as given by Eq \ref{eq:fick2}:
\begin{equation}
\frac{\partial c}{\partial t} = \frac{\partial }{\partial x}\left(D \frac{\partial c}{\partial x}\right)
\label{eq:fick2}
\end{equation}

\paragraph{Mechanism of diffusion:}\label{mechanism-of-diffusion}

To understand diffusion mechanisms at a microscopic level, it is important to study the polymer-solute interactions. The structure of the polymer is an important parameter to take into account since the transport phenomena in glassy polymers are totally different from those in a rubbery polymer. Diffusion in a polymer matrix can be classified into three categories which depend on the relative mobilities of the penetrant and the polymer. There are several diffusion mechanisms depending on the kinetics of gas molecules and the relaxation modes of the macromolecular chains \cite{Crank1975, Rogers1985, Aminabhavi1988}.

\begin{itemize}
 
\item
  \textbf{Case I: Fickian diffusion}. In this case, which is treated by the laws presented above, the fickian diffusion is characterized by a diffusion rate of the solvent, $R_{diff}$, slower than the relaxation rate of the polymer $R_{relax}$ ($R_{diff} \ll R_{relax}$). The boundary conditions are independent of the time and of the swelling kinetics of the polymer. This case is encountered for most systems when the temperature is above the glassy temperature $T_g$ of the polymer. The timescale of the diffusion process is much larger than the timescale of the change in the internal structure of the polymer, that is, the diffusion rate is much smaller than the rate of relaxation process. In this type of diffusion, the polymer is considered to reach (and stay in) its equilibrium state so quickly that the temporal evolution of its internal structure is not taken into account.
\item
  \textbf{Case II: Non-Fickian diffusion}. This diffusion refers to a process where the diffusion is much faster than the relaxation rate ($R_{diff} \gg R_{relax}$) of the polymer chains. That means that the change in the internal structure of the polymer, which can for example induce internal stresses, plays an important role in the diffusion process \cite{Thomas1978}. This is particularly the case with organic vapors in polymers heated up to $T_g=$+15°C.
\item
  \textbf{Case III: Abnormal diffusion}. Gas diffusion and polymer relaxation modes have a similar rate ($R_{diff} \sim R_{relax}$). Diffusion and relaxation rates are comparable. It can be considered as a combination of cases I and II. The temporal evolution of the polymer structure must be taken into account when modeling type I or II transport processes. It is therefore important to couple the rheology of the polymer with the diffusion process.
\end{itemize}

The diffusion phenomenon is a complex mechanism and it is not uncommon that in a Fickien process, the diffusion speed depends on the concentration of diffusing molecules in the polymer. This is particularly the case if the molecules influence the properties of the polymer due to strong interactions. This is then referred to as plasticization. In this case, the diffusion rate varies depending on the concentration of the absorbed molecules. Several models have been developed to take these concentration effects into account. In 1946, Barrer proposed the following linear law \cite{Barrer1946}:
\begin{equation}
D = D_{C=0}(1 + aC)
\end{equation}
with $D_{C = 0}$ the diffusion rate at infinite dilution; and $a$ a coefficient associated with the diffusion dependence on $C$. Later, Aitken and Barrer used this law to study the diffusion of alkanes in rubber and compare their results to those of Prager and Long who studied the sorption of the same alkanes in polyisobutylene. These two authors proposed an exponential law to explain their results:
\begin{equation}
D = D_{C = 0} \exp ^ {\gamma ~ C}
\end{equation}
where $\gamma$ is called the plasticization parameter. This model is still widely used because of its simplicity and relevance \cite{Clement2004}.

It is important to point out that when a polymer is highly plasticized by a penetrant, the diffusion and solubility coefficients can become a function of the concentration and of the time, therefore, the problem to be solved is then non-fickian. To get an idea of the transport mechanism involved, it is common practice to adjust the sorption results \cite{Crank1968, Aminabhavi1988} by a law of the type:
\begin{equation}
\frac{M_t} {M_{\infty}} = k t^n
\end{equation}
where $M_t$ and $M_{\infty}$ respectively represent the mass absorption of the absorbed molecules at time $t$ and at long times, i.e.~when equilibrium is reached, $k$ is a constant which depends on the diffusion coefficient and the thickness of the film. The value of $n$ indicates the type of transport mechanism. A value of $n$ = 0.5 indicates a Fickian diffusion while $n$ = 1 concerns non-fickian diffusion II. The intermediate values of $n$ suggest a combination of these two mechanisms. 
The exponent $n$ can take several values:

\begin{itemize}
\item
  $n>$ 1: Supercase II
\item
  $n =$ 1: Case II
\item
  $1> n> 1/2$: Anomalous
\item
  $1/2 > n$: Pseuso-Fickian
\end{itemize}

On the other hand, the differences between the experimental curves of sorption and desorption are caused by different kinetics, linked to different diffusion coefficients function of the concentration \cite{Crank1975}.
\subsubsection{The fundamental parameters of permeation}\label{the-fundamental-parameters}

The rate of transport through a membrane is called permeability. As we mentioned earlier, this multi-step process is called the Solution-diffusion mechanism. The first step concerns the phenomenon of sorption. This involves dissolving the penetrant onto the membrane surface from the gas phase on the high pressure supply side. Once on the membrane surface, the penetrant diffuses into and through the membrane. This diffusion step is the speed control step of the process, because the penetrant has to wait for the polymer chains to make enough space for it to \textit{jump} in. The penetrant performs a series of these \textit{jumps} until it reaches the low pressure face of the membrane, where it is desorbed in the permeate stream.

There are three basic parameters used to describe the ability of a membrane to transport gas:
\begin{itemize}
    \item Permeability ($P$),
    \item Solubility ($S$),
    \item Diffusion ($D$).
\end{itemize}

These three parameters form the fundamental equation modelling the performance of a polymer membrane. The permeability of a polymer membrane to gases can be defined as the property to allow itself to be penetrated and then crossed by molecules. In a Fickien mechanism and in STP condition (Standard condition of Temperature and Pressure), the permeability coefficient $P$ is defined as being the product of the diffusion $D$ and the solubility $S$ coefficients:
\begin{equation}
P = D \times S
\end{equation}
Its dimension is as follows:\\
\begin{center}
    $P = \frac{(\text{volume of gas}) \times (\text{thickness of the membrane})} {(\text{surface of the membrane}) \times (\text{time}) \times (\text{pressure})}$
\end{center}

\paragraph{Coefficient of permeability (P) or permeability}\label{coefficient-of-permeability-p-or-permeability}

The permeation of a gas through a polymer is described by a diffusion model, using the laws of Henry and Fick to obtain the expression which relates the rate of permeation to the surface and to the thickness of the film. We use Fick's first law (\ref{eq:fick1}) and we obtain:
\begin{equation}
J = -D \frac{\Delta C} {L}
\label{eq:fick1b}
\end{equation}
where $L$ represents the thickness of the membrane.

When the diffusion mechanism is in its stationary state, the equilibrium of the gas of concentration $C$ at the surface and partial pressure of gas $p$ obeys to Henry's law. When the permeant is a gas, it is more convenient to measure the vapor pressure $p$, so that $\Delta C$ can be replaced by $S\Delta p$, where $S$ is the solubility coefficient which reflects the amount of permeant in the polymer and $\Delta p$ is the pressure difference across the film. The equation \ref{eq:fick1} becomes:
\begin{equation}
J = -D \left(\frac{S \Delta p} {L}\right)
\end{equation}

The product $DS$ is indicated as permeability coefficient (or constant) or permeation coefficient or simply as permeability $P$.

Permeability is the flow of a gas through the membrane after it has been normalized for thickness and pressure gradient across the membrane. It can be thought of as the flow of a gas through the membrane. The coefficient of permeability ($P$) is a measure of the speed at which a penetrant can pass through a membrane. It is often referred to simply as permeability. Permeability can be calculated easily for an individual component using equation \ref{eq:permeability}.
\begin{equation}
P = \frac{J L} {p_1 - p_2}
\label{eq:permeability}
\end{equation}

The ratio $P/L$ is called permeance.

The temperature will also have a direct effect on $P$ (increasing T will decrease P). The experimental data at different temperatures will follow the Arrhenius relationship as follows:
\begin{equation}
P = P_0 \exp(-\Delta E_P/RT)
\end{equation}
with $\Delta E_p$ the activation energy for permeation.

\paragraph{Diffusion coefficient (D)}\label{diffusion-coefficient-d}

The second parameter is the diffusion coefficient ($D$), and it describes the ease with which a penetrant can migrate through a membrane. The diffusion coefficient of a penetrant within a polymer is a kinetic parameter, and it is a function of the properties of the penetrant and of the polymer that makes up the membrane. The properties of the penetrant affect the diffusion coefficient in several ways. Larger sized particles have more difficulty to move than smaller sized particles, and as such, the larger the size of a particle, the smaller its diffusion coefficient will be. Published studies show that there is an inversely proportional trend between $D$ and the Van der Waals volume of a penetrant \cite{Chern1985, Park1986}.

The diffusion coefficient also depends on the chemical structure of the polymer. For a penetrant to diffuse through a membrane, its movement must coincide with the inter-segmental movements of the polymer chains. These inter-segmental movements must result in a space large enough for a penetrant to enter. As such, the chemical structure of the polymer plays an important role in the value of $D$. The incorporation of flexible bonds, such as $-O-$, allows for easy intra-segmental movement. With a polymeric structure strongly influencing the diffusivity of a component, it follows that there will be drastic differences in the diffusion values for a gas between glassy and rubbery membranes. Rubbery polymers are known to have diffusion coefficients which are much higher than glassy polymers for the same gas (For example, natural rubber has a $D_{O_2}$ of $\rm 158 \times 10^{-8}~cm^2/s$, while glassy polycarbonate
has a $D_{O_2}$ value of $\rm 5.1 \times 10^{-8}~cm^2/s$.) \cite{Crank1968}.

The temperature of the system will also influence the value of the diffusion coefficient as it will increase the segmental movements of the polymer chains and lead to the development of more spaces large enough for a gas molecule to enter. In addition, it will communicate more kinetic energy to the penetrant, thus allowing it to diffuse more quickly. The relation between the temperature $T$ and $D$ is generally modelled by an Arrhenius type relation and can be expressed as follows:
\begin{equation}
D = D_0 \exp(-\Delta E_D/RT)
\end{equation}
where $E_D$ is the activation energy of the diffusion, $D_0$ is a constant and $R$ is the ideal gas constant.

The amount of free-volume trapped in a polymer will affect the diffusivity of a penetrant and this amount is often referred to as \textit{Fractional Free-Volume} ($FFV$). The $FFV$ is the amount of space that is not occupied by the polymer chain. As it consists of empty space, it evolves exponentially with the coefficient of diffusivity. The $FFV$ can be determined using the relationship below:
\begin{equation}
FFV = \frac{V - V_0} {V}
\label{FFV}
\end{equation}
where $FFV$ is the Fractionnal Free-Volume, $V$ is the specific volume of the polymer and $V_0$ is the volume that is occupied by the polymer, which is equal to 1.3 times the volume of Van der Waals of the polymer.
$D$ and $FFV$ can be linked by the expression of Doolittle \cite{Fujita1961}:
\begin{equation}
D = D_0 \exp \left(\frac{-B} {FFV}\right)
\label{eq:Doolittle}
\end{equation}
where $D$ is the diffusivity coefficient, $D_0$ and $B$ are characteristic constants of the polymer-penetrating system. $D_0$ is equal to ($RTA_f$) where $A_f$ depends on the size and shape of the gas molecule, and $B$ is a measure of the minimum local Free-volume needed to allow a diffusive jump. This relation shows that as the $FFV$ increases, the diffusion coefficient will increase exponentially with it.

\paragraph{Solubility parameter (S)}\label{solubility-parameter-s}

The last parameter is the solubility $S$, a thermodynamic term which makes it possible to calculate a concentration $C$ of gas solubilized in a polymer at equilibrium, knowing the pressure of this gas, according to:
\begin{equation}
C = S \times P
\end{equation}

Its dimension is:
\begin{equation}
S = \frac{(\text{volume ~ of ~ gas ~ solubilizes ~ under ~ the ~ standard temperature and pressure conditions ~ STP})} {(\text{volume ~ of ~ the ~ matrix}) \times (\text{pressure })}
\end{equation}

It describes the amount of penetrant that can be retained by the membrane at a time. A membrane has a maximum thermodynamic quantity of gas that it can hold at a given pressure. This quantity is called the solubility and is represented by $S$. The equation used to calculate the solubility coefficient is:
\begin{equation}
S = \frac{C}{p}
\label{eq:solubility}
\end{equation}
where $S$ is the solubility of the gas in the polymer at equilibrium, $C$ is the concentration of the component in the polymer and $p$ is the partial pressure of the gas.
The value of the solubility coefficient depends in part on the properties of the gas molecule. A gas molecule with a high critical temperature ($T_c$) will generally have a higher $S$ than a similar molecule with a lower $T_c$. The size of the gas molecule also influences $S$. As the size increases, generally the solubility will also increase. Solubility also tends to correlate well with the boiling point of gas molecules ($T_b$) and the Lennard-Jones force constant $\varepsilon / k_B$ which has the dimension of a temperature in Kelvin with $\varepsilon$ is the depth of the energy gap and $k_B$ the Boltzmann constant.

$S$ also depends on the interactions between the penetrant and the polymer. Polar gases can have favorable interactions with polymer functional groups. The temperature can increase or decrease the solubility coefficient depending on the interactions between the condensate and the polymer. The solubility in polymers can be expressed by:
\begin{equation}
S = S_0 \exp(-\Delta H_S / RT)
\end{equation}
where $S_0$ is a constant, $R$ is the ideal gas constant, $T$ is the temperature and $\Delta H_S$ is the partial molar enthalpy of sorption. $\Delta H_S$ can be calculated as follows by:
\begin{equation}
\Delta H_S = \Delta H_{cond} + \Delta H_{mix}
\end{equation}
where $\Delta H_{cond}$ is the enthalpy change of the penetrant from a gaseous penetrant to a condensed phase, and $\Delta H_{mix}$ is the enthalpy change associated with the creation of a gap in the polymer of sufficient size to accommodate the penetrant molecule\cite{Freeman1989}.

Solubility can also be affected by the amount of free-volume when working with glassy polymers. The free-volume in a glassy polymer results from the inability of the polymer chains to undergo conventional rearrangement fast enough to reach equilibrium once the temperature is below the polymer's $T_g$. This excess volume becomes a site of sorption for gaseous molecules. However, these sites can become saturated as the pressure increases, resulting in decreased concentration increases when the pressure is high. This reduces the value of the solubility parameter.

\hypertarget{parameters-defining-the-polymer-solvent-interactions}{%
\subsubsection{Parameters defining the polymer-solvent interactions}\label{parameters-defining-the-polymer-solvent-interactions}}

As seen in the previous sections, when a polymer is put into contact with a solvent at a given temperature, the solvent can diffuse through the polymer. The solvent will be absorbed at a certain rate or amount of time until the sorption equilibrium is reached, i.e.~no more solvent is absorbed by the polymer. The amount of solvent absorbed depends on the chemical activity of the solvent and the interactions between the solvent and the polymer according to the following equation:
\begin{equation}
C_{solv} = S_{solv} \times a_{solv}
\end{equation}
where $C_{solv}$ is the concentration of solvent in the polymer, $S_{solv}$ is the solubility parameter of the solvent and $a_{solv}$ is the chemical activity of the solvent. This relation corresponds to the Henry sorption law. The chemical activity $a_{solv}$ is defined by:
\begin{equation}
\ln a_{solv} = \frac{\mu_{solv} - \mu_{solv}^ 0} {RT}
\end{equation}
where $T$ is the temperature, $R$ is the ideal gas constant (8.314 $J/(mol.K)$), and $\mu_{solv}$ and $\mu^0_{solv}$ are the chemical potentials of the solvent in the polymer and that of the solvent in the reference state respectively, which is defined as the chemical potential of the solvent in the liquid state at the same temperature and pressure (1 atm).

From a simple thermodynamic point of view, the average energy per mole in a uniform system is the Gibbs partial molar function, $G_i$, or chemical potential $\mu_i$. It follows that the average energy per molecule is $\mu_i / N_A$ where $N_A$ is the Avogadro number. If this energy were dissipated by moving a molecule along the chemical potential gradient, the driving force, $X_i$, per species molecule, i, would be:
\begin{equation}
X_i = -(\frac{1}{N_A} \frac{d\mu_i}{dx})
\end{equation}

Assuming that there is a frictional resistance which depends on the speed of the fi molecules, and taking into account the concentration, the number of moles and using Fick's law, we obtain the diffusion coefficient:
\begin{equation}
D_i = (R T / N_A f_i) [1 + (c_i / \gamma_i) (d\gamma_i / dc_i)]
\end{equation}
where $\gamma_i$ is the activity coefficient of the considered species. In the non-ideal situation, the quantity $RT/N_Af$ is sometimes called the thermodynamic diffusion coefficient $D_T$ \cite{Park1986}.

\paragraph{Transition temperature}\label{transition-temperature}

The amorphous phases of polymers all have at least two types of phase transitions:

\begin{itemize}
 
\item
  One or more secondary transitions associated with the appearance of local movements;
\item
  and the main transition, or glass transition ($T_g$), which marks the passage from the glassy state to the rubbery state with the appearance of cooperative movements.
\end{itemize}

The glass transition being much more influential than the secondary transitions at the level of the mobility of the macromolecular chains, we will focus in this part only on this one. At a temperature below the $T_g$, the polymers are in the glassy state: there is no cooperative movement of the macromolecules, which tends to limit the diffusion of the molecules. At temperatures above $T_g$, the amorphous parts of the polymer are in their rubbery state. The polymer chains are much more mobile, the diffusion of penetrating molecules is then favored by the possible recombination of the free-volumes. It is also noted that when passing from the $T_g$, the specific volume of the polymers increases more rapidly, creating more free-volume. These phenomena can influence the solubility and diffusion of penetrating molecules. In terms of solubility, a change in sorption mode can be noted. At temperatures below $T_g$, the presence of elementary voids makes it frequent to observe a double-mode type sorption (see paragraph \ref{DMS}). Whereas at temperatures above $T_g$, the disappearance of these sorption sites generally results in a Henry-type sorption mechanism. Regarding the diffusion of penetrating molecules in an amorphous polymer, we note that it often increases more rapidly with the temperature above $T_g$ than below. Thus, the classical pattern of the variation of the diffusion coefficient $D$ of a molecule in a polymer as a function of temperature resembles that representing the variation of the specific volume of a polymer as a function of temperature. This results in a change in slope at the $T_g$. This phenomenon, which results in a higher permeability in the rubbery domain (when $T> T_g$) than in the vitreous domain, is nevertheless linked to the size of the diffuser relative to the size of the voids available. Thus, it is common not to observe a break in the slope of the curve representing the permeability as a function of temperature at the passage of the glass transition, for small molecules such as helium or hydrogen. On the other hand, this break in slope is often clearly observed for gases with a larger kinetic diameter.

\begin{table}[!ht]
\caption{Diffusions of case I (Fick) or case II according to the initial state of the material and the temperature $T$}\label{tbl:diffu}%
\begin{tabular}{@{}lllll@{}}
\toprule
& $c=0$ & & $c=\infty$ \\
\midrule
$T > T_g$ & rubbery & $\xrightarrow[\text{D big}]{Fick}$ & rubbery \\
$T \le T_g$ & glassy & $\xrightarrow[]{\text{case II}}$ & rubbery \\
$T < T_g$ & glassy & $\xrightarrow[\text{D small}]{Fick}$ & glassy \\
\end{tabular}
\end{table}

In view of the previous notions, here is a summary table of the diffusions of cases I and II according to the glass transition temperature (see Table \ref{tbl:diffu}).

\subsubsection{Diffusion models}\label{diffusion-model}

\paragraph{Fickian and case II diffusion regimes}\label{fickian-and-case-ii-diffusion-regimes}

Concerning Fick's diffusion, the flow of solvent in the polymer is governed by the concentration gradient of this solvent as defined by the 1st law of Fick (Eq \ref{eq:fick1}). From the equivalence of the 1st and 2nd laws of Fick (Eq \ref{eq:fick2}), the profile of the solvent concentration in the polymer can be determined by:
\begin{equation}
c (x) = \frac{M}{\sqrt{4~D~t}}\exp \left(\frac{x^2}{4Dt}\right)
\end{equation}

\begin{figure}[!ht]
\centering
\includegraphics[scale=0.4]{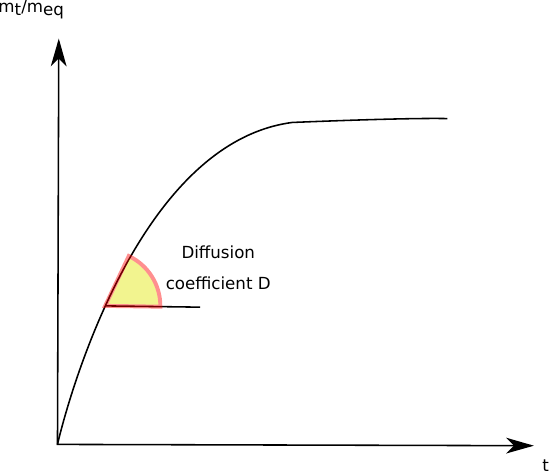}
\caption{Graphic diagram of the sorption curve when the solvent diffuses according to the fickian regime (case I)}\label{fig:fick1regime}
\end{figure}

In fickian diffusions (case I), the solvent uptake varies linearly with the square root of time, and if the solvent uptake at a given instant $m_t$ is normalized by the sorption of the solvent at equilibrium $m_{eq}$, the diffusion coefficient $D$ of the absorbed solvent can be obtained from the slope of the graph obtained (Figure \ref{fig:fick1regime}).

The diffusion coefficient $D$ is considered to be constant over the entire diffusion of the solvent. It can then be calculated from the following relation:
\begin{equation}
    \frac{m_t}{m_{eq}} = 1 - \frac{8}{\pi^2}\exp \left(-\frac{\pi^2 D t} {L^2}\right)
\end{equation}
where $m_{eq}$ and $m_t$ respectively represent the quantity of the penetrant at equilibrium and at a given instant, $t$ is the diffusion time and $L$ is the thickness of the polymer film.

\begin{figure}[!ht]
\centering
\includegraphics[scale=0.4]{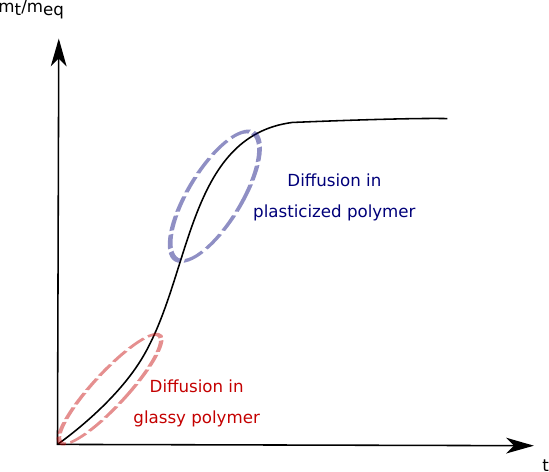}
\caption{Graphic diagram of the sorption curve when the solvent diffuses according to the non-fickian regime (case II)}\label{fig:fick2regime}
\end{figure}

In the case of a Case II non-fickian diffusion, if the solvent supply is plotted against the square root of time, a change in the diffusion regime is observed when the solvent begins to locally plasticize the polymer (see the sorption curve is schematized in Figure \ref{fig:fick2regime}).

Berens et.al. \cite{Berens1978} consider that a diffusion of case II can be described by a Fickian regime plus a relaxation process caused by the presence of the solvent in the polymer. In this model, the diffusion and plasticization processes start at the same time. However, it has been suggested that the lamination process would start later after the start of the diffusion \cite{Ramani2001}. A modified heuristic diffusion-relaxation model taking into account a relaxation time has been proposed in the literature \cite{McDowell1999}:
\begin{equation}
\frac{m_t} {m_{eq}} = \phi_F \left(1 - \sum_{n = 0}^{\infty} \frac{8} {(2n +1)^2 \pi^2} \exp \left[\frac{-(2n +1)^2 \pi^2 d t} {l^2}\right]\right) + \phi_R (1 - \exp \left[-k_r (t -t_d)\right])
\end{equation}
where $\phi_F = m_{eq-F}/m_{eq}$ and $\phi_R = m_{eq-R}/m_{eq}$ knowing that ($\phi_F + \phi_R = 1$) are the fractions volumes of solvent contributing to the Fickian and relaxation processes respectively, $D$ is the diffusion coefficient, $k_R$ is the relaxation constant and $t_D$ is the time at which the polymer chain relaxation begins. It should be noted that if this relation makes it possible to adjust the diffusion regimes of case II, this model does not consider a distribution of the relaxation times of the polymer chains but rather considers that the relaxation of the polymer occurs suddenly at a single instant. $t_D$.

\subparagraph{\texorpdfstring{Glassy polymer $(T <T_g$):}{Glassy polymer (T \textless T\_g):}}\label{glassy-polymer-t-t_g}

The transport mechanisms in polymers at the molecular level are not fully understood when $T <T_g$. All the models proposed in the literature are phenomenological and contain one or more adjustable parameters which must be determined experimentally and are only suitable for a limited number of systems \cite{Stern1994}.

A polymer in a glassy state has a specific volume $V_s$ greater than the specific equilibrium volume $V$. This difference, due to the non-equilibrium nature of the glassy state, is at the origin of the non-linearity of the sorption isotherms (which depends on the solubility coefficient and on the pressure) \cite{Barrer1958}. On the other hand, when a polymer is exposed to a vapor, the gas molecules dissolved in the matrix can modify the microstructure of the polymer. This process can lead to a decrease in $T$ and an increase in the specific volume $V_g$, i.e.~a plasticization effect \cite{Zhang1998}. In some extreme cases, for high pressure, the opposite effect can be observed: the matrix $T$ will rise, unlike $V$, due to compression of the polymer, which will reduce its segmental movements. The phenomena are much more complex in glassy polymers. To describe the solution and the diffusion of molecules in glassy polymers, we will detail the two-mode (dua) sorption model, the most commonly used \cite{Barrer1958}, then the ``time-lag'' model.

\underline{Dual Mode Sorption Model (DMS)\label{DMS}}:

This model, initially proposed by Barrer et al. \cite{Barrer1958} as a combination of Henry and Langmuir modes to explain the concentration dependence of the solubility coefficient found for glassy polymers, was then extended by Koros and Paul \cite{Koros1976} for the diffusion coefficient. This model postulates that there are two distinct populations of diffusing molecules (with a local equilibrium between them):

\begin{itemize}
 
\item
  molecules dissolved in the polymer by an ordinary dissolution process of concentration $C_D$, and
\item
  a second population corresponding to molecules trapped by adsorption on specific sites (microvides or holes) with a concentration $C_H$.
\end{itemize}

The total concentration of penetrant $(C$) in a glassy amorphous polymer can be expressed as follows:
\begin{equation}
C = C_D + C_H
\end{equation}

\begin{figure}[!ht]
\centering
\includegraphics[scale=0.4]{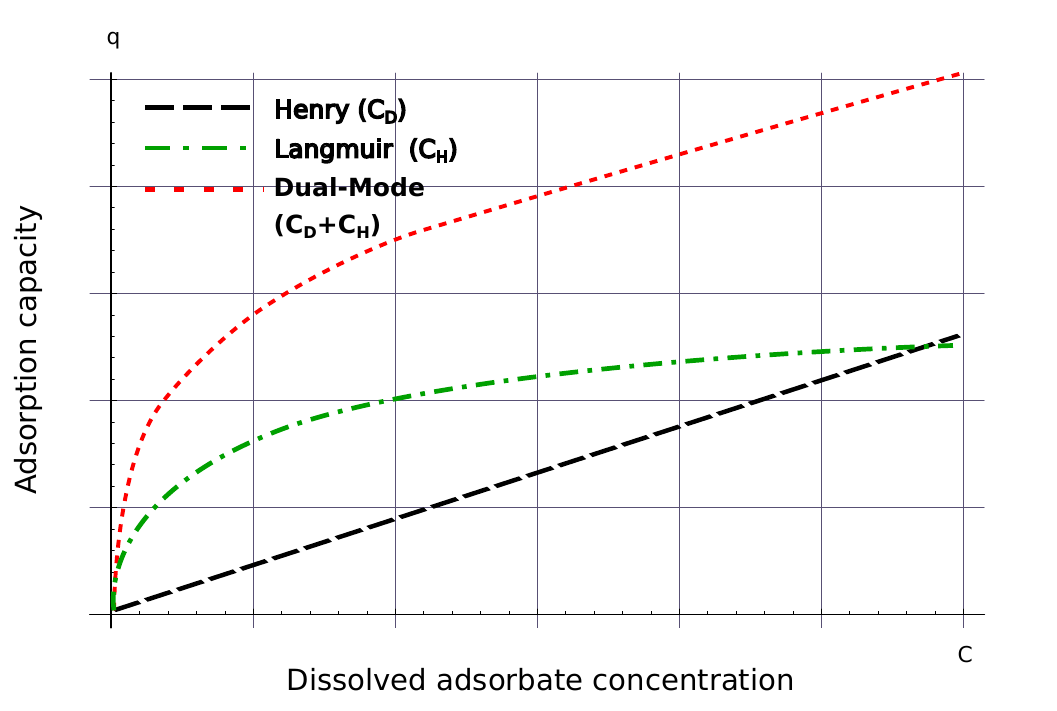}
\caption{Dual-mode sorption}\label{fig:dualmode}
\end{figure}

In this model, the equilibrium concentration of a gas in a glassy polymer behaves very differently from that in a rubbery polymer. In rubbery polymers, the concentration is directly proportional to the applied pressure. In glassy membranes, on the other hand, the equilibrium concentration is generally non-linear, and begins to approach an asymptotic value as pressure increases. This is due to the fact that the holes are saturated and can no longer accommodate gas molecules (See Figure \ref{fig:dualmode}).

Generally, it is believed that the holes are generated by the slow relaxation processes associated with the glass transition or by the unreleased volume frozen during the glass transition \cite{Stastna1995}. The relative proportions of the penetrant in each state depend on the total concentration. Furthermore, it is assumed that at the equilibrium pressure $p$, the concentration of molecules dissolved in the polymer by an ordinary dissolution mechanism, $C_D$, obeys Henry's law $C~=~k_D~p$, and that the concentration of molecules sorbed in a number of pre-existing microcavities is given by the Langmuir equation:
\begin{equation}
C_H = \frac{C_H^{'} b p} {1 + bp}
\end{equation}
where $C_H^{'}$ is called the \textit{Langmuir saturation} constant, is directly related to the overall volume of specific sites, i.e.~a measure of the \textit{hole} concentration, and b is a constant, characterizing the affinity for these sites.

In this model, the solubility coefficient has the expression:
\begin{equation}
S = \frac{C} {p} = k_D + \frac{C_H^{'} B} {1 + b p}
\end{equation}

This model can be expressed in a different way, giving a physical meaning to the parameter $F$. The population of diffusing molecules is divided into two categories:

\begin{itemize}
 
\item
  all the species dissolved by ordinary dissolution plus a fraction $F$ of the molecules trapped in the microcavities are totally mobile and have a diffusion coefficient $D_D$. It represents a concentration $C_D + F C_H$;
\item
  a fraction $(1 - F)$ of this last species is totally immobilized representing a concentration $(1 - F) C_H$.
\end{itemize}

Therefore, the coefficient of permeability is expressed as follows:
\begin{equation}
P = K_D D_D \left(1 + \frac{K F} {1 + b p}\right)
\label{eq:dualmode}
\end{equation}

The dual-mode sorption has satisfactorily represented the solubility and transport of gases in glassy polymers, polar or not, and has therefore been widely applied to the study of structure/permeability relationships of polymers \cite{Aminabhavi1988} but it presents some major drawbacks. The main limitation of phenomenological models is that they are not predictive because the model parameters are not directly related to the chemical structure of polymers \cite{Stern1994}.

\underline{The Time Lag Method}:
\label{Time-lag}

The concept of delayed diffusion was developed in the early twenties by Daynes \cite{Daynes1920} and was further extended for a variety of geometric shapes of polymer films \cite{Barrer1940}. A digital solution of the timed diffusion integrating the double sorption model was tried by Toi et al. \cite{Toi1983} for gas permeation in a polymer film obeying a nonlinear isotherm. This study established a new experimental tool that can be used to examine how the mechanism of the double sorption model applies to timed diffusion in order to reveal the detailed mechanism of diffusion in glassy polymers.

When a gas or vapor is placed on one side of a plastic film, the molecules of the gas will tend to dissolve on the surface of the film and diffuse under a concentration gradient through the film reaching the other side (or the low pressure side) of the film. After a short period of time, a steady state will be reached in which gas will diffuse through the film at a constant rate, provided that a constant pressure difference is maintained across the film. No permeation will occur when the concentration gradient is zero and extremely high rates of permeation when the gradient is very large.

\begin{figure}[!ht]
\centering
\includegraphics[scale=0.4]{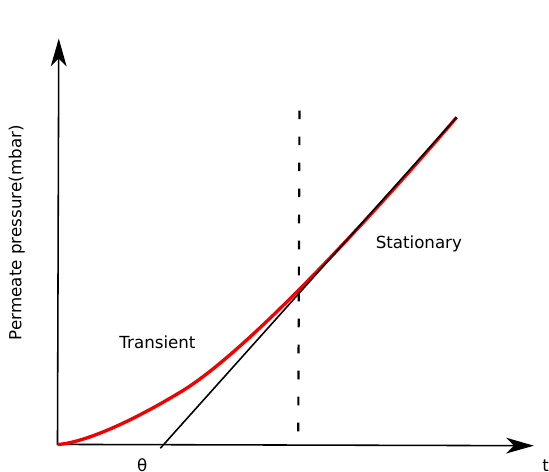}
\caption{Time lag before reaching equilibrium state}\label{fig:timelag}
\end{figure}

On the graph representing $J$ the flow of solvent in the polymer as a function of time $t$ for a material, we note that a short period of time must elapse before a linear relation is reached (Figure \ref{fig:timelag}).

The first part of the curve is nonlinear and shows that the evolution is non-stationary and that in this case Fick's second law applies. Once this step is over, the relationship becomes linear and it can be deduced that the steady state of diffusion has been reached. The equation that has found considerable use in evaluating $D$, $P$, and $S$ is the Barrer's equation \cite{Barrer1939}, often referred to as the \textit{time-lag} equation or method that highlights evidence of a time lag which represents the time it takes for a gas or vapor to penetrate through a membrane. It can be defined as:
\begin{equation}
D = \frac{L^2}{6\tau}
\end{equation}
where $L$ is the thickness of the film and $\tau$ is called the \textit{time offset} which can be found by extrapolating the linear part of the curve to the time axis (black tangent on Figure \ref{fig:timelag}).

In this method, it is assumed that the penetrant molecules sorbed by the Langmuir isothermal mode are completely immobilized and do not contribute to the diffusive flux. Under these conditions, the permeability is independent of the upstream pressure. Petropoulos \cite{Petropoulos1970} proposed the partial immobilization of gas molecules sorbed by the Langmuir mode. Paul and Roros \cite{Paul1976} formulated the effect of partial immobilization of sorption on permeability and time lag. They assigned different diffusion coefficients to the gas sorbed by each of the two sorption mechanisms. The expression describing the partial immobilization model for permeability and diffusion retardation depends on the upstream gas pressure. Partial immobilization has been tested for several polymer penetration systems \cite{Paul1976}. All of these studies confirmed that the diffusivity due to the Langmuir mode species is much lower than the diffusivity due to the Henry's law species. For the total immobilization model, the permeation has the form:
\begin{equation}
P = K_D D
\end{equation}
and taking into account the \textit{time-lag} we find the formula of dual mode (See equation \ref{eq:dualmode}).

\subparagraph{\texorpdfstring{Rubbery polymer ($T> T_g$):}{Rubbery polymer (T\textgreater{} T\_g):}}\label{rubbery-polymer-t-t_g}

To describe the diffusion of small molecules in polymers above their glass transition temperature, a number of molecular models and theories derived from free-volume theory have been proposed \cite{Crank1968, Comyn1985}. The different models developed to describe the diffusion of small gas molecules in polymers are based on the analysis of the relative mobility of diffusing molecules and polymer chains taking into account the relevant inter-molecular forces. We have two categories:

\begin{itemize}
 
\item
  molecular models analyze specific movements of penetrants and chains as well as relevant intermolecular forces, and,
\item
  free-volume models attempt to elucidate the relationship between the diffusion coefficient and the free-volume of the system, regardless of microscopic description.
\end{itemize}

\underline{Molecular model}:

Molecular models generally assume that fluctuating microcavities or \textit{holes} exist in the polymer matrix and that at equilibrium a defined size distribution of these holes is established on a time averaged basis. A hole of sufficient size may contain a dissolved penetrant molecule, which can \textit{jump} into a nearby hole once it acquires sufficient energy. Diffusive motion occurs only when the holes that have become vacant in this way are occupied by other penetrant molecules. A net diffusive flux occurs in a preferred direction in response to a driving force, otherwise the molecules will diffuse in random directions since their motion has a Brownian nature. The magnitude of the flow depends on the concentration of holes that are large enough to accommodate a penetrant molecule. Molecular models include these features largely to describe the Arrhenius behavior of the scattering coefficients observed experimentally. Barrer (1937) was the first to show that the diffusion of molecules in polymers was a thermally activated process. Then a number of formulations are based on energy considerations. Here are a few examples of the main models:

\begin{itemize}
 
\item
  Meares model: The Meares model \cite{Meares1954} is of only historical interest because it was the first molecular model of diffusion in polymers. Meares found that the activation energy for diffusion correlates linearly with the square of the penetrant diameter, but not with the cubic diameter. Therefore, he deduced that the elemental diffusion step is not governed by the energy required to create a hole that can accommodate a penetrant molecule, but by the energy required to separate the polymer chains so that a cylindrical vacuum is produced which allows the penetrant molecule to \textit{jump} from one equilibrium position to another. These first molecular models only allow the prediction of the activation energy of diffusion and not of the diffusion coefficients.
\item
  Brandt's model: Brandt (1959) \cite{Brandt1959} subsequently suggested defining energy from considerations of the molecular structure of the polymer. In this theory, a molecular model is formulated where the activation energy is broken down into two terms: $E_D = E_i + E_b$. The first term $E_i$ characterizes the inter-molecular energy required to overcome the forces of attraction between the chains and create a \textit{hole} in the polymer structure of the penetrant. The second term represents the intra-molecular energy $E_b$ used to bend the neighboring chains of the penetrant. These two terms mainly depend on the diameter of the penetrant $\sigma_p$ ($E_i \approx \sigma_p$; $E_b \approx \sigma_p^2$), on the length of chain involved in the diffusion, on the length of a jump elementary, etc \dots 
\item
  Model of DiBenedetto and Paul \cite{DiBenedetto1964}: They developed two different approaches, one based on statistical mechanics (theory of fluctuations) applied to diffusion in glassy polymers, and the other on molecular theory of transport in rubbery areas. In the latter approach, the activation energy of the diffusion is equal to the difference in potential energy between the \textit{normal} dissolved state and the \textit{activated} state in which the cylindrical cavity allowing the movement of the penetrant is present. This variation in energy interaction between macromolecules is defined by the Lennard-Jones potential 6-12.
\item
  Pace and Datyner model \cite{Pace1979, Pace1980}: They came up with a diffusion theory, which incorporated both DiBenedetto and Paul's model and Brandt's theory. They suggested that the transport process could be due to two separate mechanisms: diffusion along the direction of the chain and jumps perpendicular to the direction of the main chain. It is these last jumps through the chains that control the time scale of the transport phenomenon and define the activation energy of the diffusion. A good agreement between the predictions of their model and the experimental results, for amorphous or semi-crystalline polymers, is obtained to express the dependence of the activation energy on the diameter of the penetrant \cite{Pace1979a}.
\end{itemize}

The recent development of numerical microstructure simulation techniques makes it possible to model transport phenomena in polymers, in particular by applying Monte-Carlo simulation techniques, molecular dynamics (MD) and Brownian motions. In recent studies, MD simulations have made it possible to estimate fairly precisely the diffusion coefficients in rubbery polymers, for example for $-CH-$ in PE \cite{Pant1993}, for $He$ and $-CH-$ in PDMS \cite{Sok1992}. The transport of diffusing molecules in glassy polymers is much more complex because these polymers are not in a state of true thermodynamic equilibrium given the scale of the diffusion measurements. Moreover, the diffusion coefficients in such matrices are smaller, by several orders of magnitude, which requires much longer computation times \cite{Gusev1994}.

An important and useful concept, which appeared in the early sixties, for understanding the permeation mechanisms in amorphous polymers, is fractional free-volume, the definition of which was given earlier (Equation \ref{FFV}). Cohen and Turnbull (1959) \cite{Cohen1959} considered the diffusion of hard spheres in a liquid as the result of a redistribution of the free-volume inside the liquid without associated energy variation. Moreover, in 1953, Bueche gave an analysis of segmental mobility in polymers which was based on the theory of free-volume fluctuations \cite{Bueche1953, Slater1940}.

\underline{Free-volume theory}:

The concept of free-volume in polymer science is well known. The free-volume was defined as the volume not occupied by matter and refers to the empty space between molecules (cf. Eq. \ref{FFV}). More generally, the free-volume can be specified as the volume of a given system at the study temperature minus the volume of the same system at 0K. Thus, the rearrangement of the free-volume creates holes through which the scattering particles can pass. The free-volume is provided by all the species present in the system, solvent, solute(s) and polymer. Free-volume theories are based on the assumption that free-volume is the main factor controlling the rate of diffusion of molecules.
In the amorphous domains of the polymer, there are certain regions of low electron density called \textit{free-volume holes or cavities}. These cavities provide pathways for the diffusing species. The amount of free-volume in a polymer is called the \textit{fractional free-volume (FFV)} or \textit{free-volume content} which describes the kinetics of diffusion. The existence of \textit{free-volume holes} in polymers has been proposed to explain molecular motion and the physical behavior of glassy and liquid states. Since then, this theory has been widely used in polymer science due to its simplicity in understanding many properties of polymers at the molecular level.
A theoretical description of the diffusion of small molecules in the polymer falls into several categories. Accordingly, a number of models have been proposed for the transport of gases in polymers. Models of gas transport in \textit{rubbery} polymers are mainly based on free-volume concepts. In free-volume approaches, the concentration dependence of the diffusivity of the penetrant is described by considering the average spaces between the chains. Such models generally relate the mutual diffusion coefficients for a gas / polymer system to the free-volume of the system.
This type of approach does not offer a precise description of the phenomenon but links, thanks to statistical mechanical considerations, the diffusion coefficient to the free-volume of the system. These models developed for rubbery polymers were, in some cases, applied to glassy polymers \cite{Stern1990}.

\begin{itemize}
 
\item
  Fujita model
\end{itemize}

One of the most promising and oldest free-volume models, developed by Fujita \cite{Fujita1961} in the early 1960s, has enjoyed a very long popularity. This approach used the William-Landel-Ferry (WLF) modification of the Doolittle equation \cite{Williams1955}. The simplest theory of ``free-volume'' is that of Fujita \cite{Fujita1961}, which owes its origin to the theory of free-volume of Cohen and Turnbull \cite{Cohen1959}. Fujita extended the theory of free-volume diffusion to take into account the dependence of the diffusion coefficient ($D$) on the concentration of diffusing molecules in polymer films \cite{Fujita1961}. Fujita \cite{Fujita1960, Fujita1961} suggested that molecular transport was due to free-volume redistribution and not thermal activation. The basic idea of this theory is that a diffusing molecule can only move from one position to another when, in its vicinity, the local free-volume exceeds a certain critical value. The dependence of $D$ on parameters such as the concentration, shape and size of the penetrant, the temperature and the glass transition temperature of the polymer can be explained by the theory of free-volume. In the case of amorphous polymers, the coefficients $S$ and $D$ can be linked to the free-volume fraction \cite{Cohen1959, Peterlin1975} by:
\begin{equation}
f = \frac{V_f}{V_{tot}} = \frac{V_{tot} - V_{occ}}{V_{tot}}
\end{equation}
$V_{tot}$ is the total volume considered to be the sum of the occupied volume (Van der Waals), $V_{occ}$, and the free occurrence volume, $V_f$. An expression similar to that of Doolittle \cite{Doolittle1951, Doolittle1952} for viscosity gives the expression of the previously mentioned diffusion coefficient (Eq \ref{eq:Doolittle}) .

This model has been successfully applied to a number of vapor-rubbery organic polymer systems in which $D$ is highly concentration dependent. However, Fujita et al. \cite{Fujita1960} have found it unsuitable for the diffusion of small molecules like water in rubbery polymers, which is largely independent of concentration. A correction was made to this model to decrease its limitation. Indeed, it is important to underline that the free-volume fraction depends on three thermodynamic variables \cite{Frisch1970, Stern1972, Kulkarni1983}: the temperature $T$; the hydrostatic pressure $p$ applied to the system which is in fact the pressure of the penetrant; and the penetrant concentration which can be expressed as a volume fraction:
\begin{equation}
f(T, p, C) = f_{ref}(T_{ref}, p_{ref}, 0) + \alpha (T- T_{ref}) - \beta (p - p_{ref}) + \gamma C
\end{equation}

The first term of this equation represents the fractional free-volume of the system in a reference state, which is the pure polymer in $T$ and in $p$.
The second term characterizes the increase in $f$ due to thermal expansion with $\alpha$, the coefficient of thermal expansion of the free-volume.
The third term shows how the free-volume decreases during hydrostatic compression, $\beta$ being the compressibility ($\beta = \chi_l - \chi_g$, the coefficients $\chi$ are the compressibilities of the liquid and glassy states) .
The last term is a measure of the penetrant's effectiveness in increasing free-volume, $\gamma$ being a concentration coefficient.
These three parameters ($\alpha, \beta, \gamma$) are characteristics of the considered system.
With such modifications taking into account the effect of gas pressure, Stern et al. \cite{Stern1983, Stern1986} have shown that Fujita's model can be applied to small, nonpolar molecules in polymers.

\begin{itemize}
 
\item
  Model of Vrentas and Duda
\end{itemize}

A major contribution to the development of free-volume theory was made by Vrentas and Duda and his colleagues reviewed and improved the free-volume model over the years \cite{Vrentas1977, Vrentas1977a, Vrentas1985, Vrentas1985a}. They extended the free-volume theory to a wide range of temperatures and polymer concentrations \cite{Vrentas1993}. The contributions in free-volume of the solvent and of the polymer are taken into account using the thermodynamic theory of Flory as well as the theory of entanglement of Bueche (1962) (idea of the coefficient of friction) \cite{Bueche1962}. This model is relevant at temperatures above and below the glass transition temperature. A polymer/solvent mutual diffusion coefficient is calculated based on the solvent concentration, but it requires values of at least ten parameters.

Therefore, Fujita's free-volume model \cite{Fujita1961} has emerged as a special case of the new model of Vrentas and Duda \cite{Vrentas1993}. Along with the many improvements, Vrentas and Duda's free-volume theory takes into account several physical parameters such as temperature, activation energy, polymer concentration, solvent size and diffusor molecular weight. Ganesh et al. \cite{Ganesh1992} extended this free-volume model to the gas permeability of polymers. Empirical correlations between gas permeability or diffusion coefficients and free-volume of polymers, or penetrating/polymer systems, have also been proposed \cite{Lee1980}. Free-volume or fractional free-volume has been defined and described by group contribution methods and also by d-spacing obtained by wide-angle X-ray diffraction spectra of polymers, taken as the mean inter-segment distance \cite{Stern1994}.

A recent theory based on free-volume and dependent on molecular shape and size was carried out by Mauritz et al \cite{Mauritz1990}, a theory applicable for large molecules in the rubbery state. In the rubbery state, the polymers can be thought of as being in a dynamic fluid-like state and the penetrant molecule can be imagined as ``swimming'' in this liquid-like medium \cite{Coughlin1991}. In addition, this theory dwells on the concept of the combination of free-volume scattering and energy activated scattering that was addressed earlier by Meares' model \cite{Meares1954}. However, this theory is not suitable for glassy polymers because it was specifically developed for rubbery polymers. Although diffusion in the case of glassy polymers is somewhat complex, it depends to a large extent on the extent of the non-equilibrium free-volume available in the polymer matrix \cite{Bartos1996, Muruganandam1987}.

Previous studies have resulted in a better understanding of the morphology and structure of polymers \cite{Korsmeyer1981}, transport phenomena \cite{Hariharan1993} and, more recently, the controlled release of drugs from polymeric carriers \cite{Peppas1992}. Additionally, these studies have led to theoretical descriptions of the diffusion of solvents and / or solutes in polymer solutions, gels and even solids \cite{Fujita1961, Fricke1924}. These physical models are based on different physical concepts (obstruction effects, hydrodynamic interactions and free-volume theory) and their applicability varies \cite{Masaro1999}.

\subsection{Liquid permeation - pervaporation}\label{liquid-permeation---pervaporation}

\begin{figure}[!ht]
\centering
\includegraphics[scale=1]{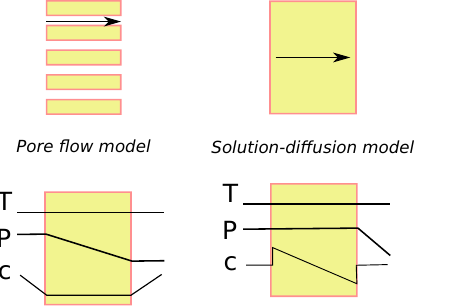}
\caption{Illustration of the two mass transfer mechanisms, where T represent the temperature, P the pressure and C the concentration}\label{fig:masstransfertpermeation}
\end{figure}

We have already discussed the phenomenon of permeability in general terms (cf \ref{diffusion-model}).  It should be remembered that permeability, as reported in the literature, is defined as the quantification of the transmission of permeate, gas or vapor, through a resistant material. Generally, we talk about permeation when an organic gas or vapor comes into contact with a polymer. More specifically, permeability is a flow of vapor through a matrix. This flow starts from an area where the vapor has a certain concentration to another area where the concentration level is lower. Permeability consists in measuring over time the quantity of a species which passes right through a polymer film subjected to a difference in activity of the species considered.
These two diffusion laws were used to describe the permeability of polymer membranes according to two mass transfer mechanisms: the \textbf{flow system in porous media} and the \textbf{solution - diffusion model} (Figure \ref{fig:masstransfertpermeation}).

\begin{itemize}
 
\item
  In the case of the \textbf{flow model in porous media} (or \textit{pore flow model}), the membrane is considered as being heterogeneous. The penetrating molecules form a continuous medium in the membrane and the permeability depends on the viscosity of the diffusing phase. Thus, different mechanisms are observed depending on the nature of the diffusing phase (gas or liquid). The driving force behind permeability is the pressure gradient and the transport mechanism is convection. The pressure decreases linearly between the upstream and downstream side of the membrane. The concentration in the membrane is constant, lower than the concentrations on each side of the membrane.
\item
  The \textbf{solution - diffusion model} considers the membrane as dense and homogeneous. The penetrating molecules are dispersed in the polymer matrix, they are solubilized like a gas in a liquid. The driving force behind the transport of diffusing molecules is the difference in concentrations. The viscosity as well as the state of the diffusing phase therefore has no influence on the diffusion rate. This model has been validated by Wijman et al. \cite{Wijmans1995}. The pressure applied to the membrane is felt identically over the entire membrane. The concentration of penetrating molecules decreases linearly between the upstream side where it is maximum and the downstream side where it is close to zero.
\end{itemize}

In both cases, the temperature $T$ is considered constant and the chemical potential $\mu$ decreases linearly. These two mechanisms with different concepts do not apply to the same systems. Thus, the flow model in porous media can be applied in the case of porous membranes. In the case of dense polymer membranes, the solution-diffusion model is appropriate.

As we saw earlier, many researchers contributed to the development of this model. In 1920, Daynes wrote an article in which he first described "time lag" (cf \ref{Time-lag}) also known as time delay \cite{Daynes1920}. He used Fick's second law to develop this concept, which is still used today to determine the coefficient of gas diffusion in polymers. It was in 1939 that Barrer adapted this dynamic method to measure diffusion constants \cite{Barrer1939}. In 1975, Crank published "The Mathematics of Diffusion" \cite{Crank1975}. This book details the mathematical laws used to describe diffusion. Fick's laws are developed there in particular for several geometries and specific cases: plane membranes, cylinders, spheres, constant diffusion, diffusion dependent on concentration, etc$\dots$ Let us end this non-exhaustive history with the chapter written by Rogers in 1985 \cite{Rogers1985} which summarizes the definitions and basic equations used to describe the permeability of polymers and describes the main influencing parameters.

The general mechanism of the liquid permeation process is similar to gas permeation, but there are significant differences in the state of the permeable membrane for liquid permeation and in the speed of material transport between the liquid charge phase and the membrane.

In gas permeation, the characteristics of the polymer film are not significantly changed by the permeation gas because it has a very low degree of solubility in the film; therefore, it is possible to calculate the composition of the permeate from the permeation rates of the pure gases and the composition of the feed mixtures. When a stable flow state is established, the amount of gas entering the unit surface of the polymer film in unit time satisfies a form of Fick's first law of diffusion (Eq \ref{eq:permeability}).

When it comes to the permeation of organic vapors \cite{Kokes1953}, or in the case of liquid permeation process, it is not possible to calculate the composition of the permeate from a knowledge of the composition of the charge and permeation rates of the pure components. The swollen membrane (i.e.~the membrane containing the dissolved permeability components) has a significantly different permeability from that of the original membrane; likewise, the state of the membrane upon penetration of individual compounds differs from that upon penetration of a mixture of different compounds. The most basic form of Fick's first law should be used to describe the process of steady-state liquid permeation:
\begin{equation}
J = \frac{D K (C_2 - C_1)}{L}
\label{eq:permeabiliteliquide}
\end{equation}
where $J$ is the amount of liquid that permeates the unitary film surface in units of time, $K$ the partition coefficient and $(C_2 - C_1)$ is the concentration differential between the two sides of the film in some coherent units which express the number of molecules per unit of volume.

Although the mechanism of gas permeation has been well established, it does not adequately describe the permeation of liquids as there are important differences between these two processes. The rate of gas permeation is much lower than that of liquids, we have a linear relationship between rate and film thickness and selectivity independent of film thickness.

Another important difference is the condition of the film under permeation conditions. Upon permeation of liquid, the permeation liquid dissolves to an appreciable extent in the polymer film; therefore, the permeable film under operating conditions is a swollen \textit{solution} of polymer and penetrating organic compounds. This is entirely different from the state of the \textit{dry} film that exists in gas permeation. Therefore, the composition of the permeate permeate cannot be calculated with any certainty from the composition of the feed mixture and the known permeation rates for the pure components. Further, liquid permeation differs from gas permeation in that the permeation rate is independent of the pressure difference across the film under liquid permeation conditions due to the huge concentration gradient that inherently exists in a liquid-film-vapor system. However, the permeation of liquid and gas follows Fick's first law in that the rate of stability is inversely proportional to the thickness of the film. There is also an exponential dependence of the concentration of diffusivity \cite{McCall1957}. Liquid permeation, similar to gas permeation, is apparently a special case of ordinary diffusion and can be explained by a classical diffusion model \cite{Long1965}.

\begin{figure}[!ht]
\centering
\includegraphics[scale=0.75]{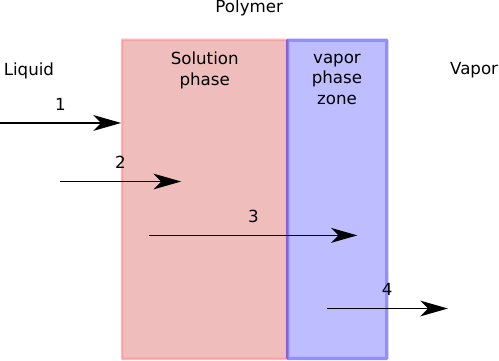}
\caption{Different step of liquid permeation through polymeric membrane}\label{fig:liquidpermeation}
\end{figure}

Component transport across the membrane is described by the solution desorption model that results from these serial processes (Figure \ref{fig:liquidpermeation})

\begin{enumerate}
\def\labelenumi{\arabic{enumi}.}
 
\item
  diffusion of the component through the liquid boundary layer towards the surface of the membrane,
\item
  sorption/diffusion in the membrane,
\item
  transport through the membrane and
\item
  diffusion through the vapor phase boundary layer in most of the permeance.
\end{enumerate}

A mechanism seems to have been demonstrated in the context of liquid permeation highlighting the existence of a \textit{solution phase}, which comprises the major part of the film, and of a \textit{vapor phase zone} in which vaporization of the permeation material occurs. This latter area occupies a relatively minor part of the film structure. There is a very rapid movement of liquid in the \textit{solution phase} and between the liquid charge phase and the \textit{solution phase}. Some selectivity may occur in the movement through the interface between the liquid charge phase and the \textit{solution phase} Most of the selectivity occurs at the interface between the \textit{solution phase} and the \textit{vapor phase}. Then the material diffuses through the \textit{vapor phase}, and this diffusion is considered the speed control step in the process.

The so-called \textit{solution phase} can be visualized as a highly swollen state of the polymer in which there is a high concentration of permeation liquid in the polymer structure. The \textit{vapor phase} can be thought of as a region in which the permeation molecules are much more dispersed and the structure of the polymer corresponds more closely to that of the dry polymer film. Two completely different phases such as these can exist inside the film because the opposing surfaces of the film are subjected to such different conditions. The existence of a \textit{solution phase} as described herein is demonstrated by the swollen state of a plastic film submerged in a liquid which is almost a solvent for the film. It makes sense to assume that there is a \textit{vapor phase zone} because vaporization must occur in an area close to the surface of the film permeate.
There must be very rapid movement of liquid in the \textit{solution phase} and between the liquid feed phase and the \textit{solution phase} to maintain selectivity in steady state operation. The very rapid movement of the liquid in \textit{solution phase} is supported by observations of sorption and desorption of organic vapors by polymer films \cite{Kokes1953}.

Selectivity must occur at one or both of the interfaces between the liquid feed phase, the solution phase and the vapor phase, since selectivity is not a function of film thickness; therefore, the selectivity function must occur at points or interfaces in the film.

The transport of gas, vapor or liquid through a dense non-porous membrane is described as \cite{Wijmans1995}:

Permeability (P) = solubility (S) × diffusivity (D).

The solubility of gases is described by Henry's law. For ideal systems, the solubility is independent of the concentration and therefore the sorption isotherm is linear and the concentration inside the polymer is proportional to the applied pressure. For liquids, because they are not ideal, Henry's Law does not apply. In addition, the affinity between the liquid and the polymer is much greater; therefore, crosslinking is sometimes necessary to prevent dissolution of the polymer. High solubility implies high permeability because it influences diffusivity by making polymer chains more flexible. The sorption isotherm is nonlinear and the behavior is described by the free-volume models and the thermodynamics of Flory Huggins. Diffusion coefficients increase with temperature, while sorption is generally exothermic \cite{Wijmans1995, Feng1996, Hillaire1999}. For the effect of temperature on the permeability of organic compounds through a membrane, different results have been reported in the literature, in some cases the permeability decreases with temperature \cite{Feng1996, Hillaire1999}, in others the permeability increases with temperature, and in some cases the permeability is not affected by the temperature in the range studied \cite{Baudot1997, Wijmans1996}.

Diffusivity depends on the geometry of the penetrant (as the molecular size increases, the diffusion coefficient decreases) and on the concentration. The size dependence is determined by the Stokes-Einstein equation, which is the relationship between the frictional resistance $f$ and the radius of the diffusing component $r$:
\begin{equation}
f = 6 \pi \eta r
\end{equation}
where $\eta$ represents the viscosity of the fluid.
And the diffusion coefficient is inversely proportional to the frictional resistance as:
\begin{equation}
D = \frac{K T}{f}
\end{equation}

Large molecules with the ability to swell the polymer can have a large diffusion coefficient. Diffusivity is determined by the permeation method or the time shift method. The quantity of penetrant ($Q_l$) crossing the membrane at time $t$ is given by:
\begin{equation}
Q_l = \frac{D c_i}{l} \left[t - \frac{l^2}{2d}\right]
\end{equation}

Instead of $Q_l$, the pressure ($p$) can also be monitored and the graph of $p$ versus time ($t$) will give the time shift as the intercept and the permeability as the slope from the part steady state of this permeation experiment. Since $P = S \times D$, therefore $S$ can be found once $D$ and $P$ are known.

Alternatively, maintaining the polymer in a closed volume and applying pressure to the chamber can determine diffusivity. Due to sorption, the pressure decreases over time and equilibrium is reached.

\section{Leaching mechanisms}\label{migration-mechanism}

\subsection{Introduction}\label{introduction}

The manufacture of "plastic" polymer materials requires the use of many compounds, including additives, which are added to confer certain properties to the host material or influence the manufacturability of the polymer, such as antioxydants, plasticizers, UV-protectors, catalysts, etc.... These additives, as well as other chemical compounds such as synthesis residues but also breakdown compounds and non intentionally added substances (NIAS \cite{Tumu2024}) tend to diffuse out (leach out) of the polymer and end up in the product, and can lead to environmental and health issues \cite{Muncke2021}. 

Among the main compounds that can leach out we can mention:

\begin{itemize}
 
\item
  The constituents of the polymers:

  \begin{itemize}
   
  \item
    residual monomers such as styrene, terephthalic acid $\dots$;
  \item
    prepolymers such as mono or dihydroxyethyl terephthalates;
  \item
    oligomers that result from incomplete polymerization, such as low molecular weight polystyrene.
  \end{itemize}
\item
  Degradation products of synthetic polymers: polymers can degrade over time or during their processing. For example, the photooxidation of polyolefins, like the hydrolysis of polyesters, breaks up the carbon chains into smaller molecules that are more easily transferred, but rarely characterized.
\item
  The adjuvants of synthetic or natural polymers:

  \begin{itemize}
   
  \item
    the agents necessary for the polymerization, such as surfactants, catalysts $\dots$;
  \item
    agents necessary for implementation or use, such as lubricants, antistats, dyes, $\dots$;
  \item
    mechanical property modifiers, such as fillers or plasticizers;
  \item
    stabilizing agents such as antioxidants and photoprotective agents.
  \end{itemize}
\end{itemize}

Theses compounds, when not chemically attached to polymer chains, can under certain conditions leave the polymer by migration, evaporation or extraction with liquids. This loss can be problematic during its use because it leads to:

\begin{enumerate}
\def\labelenumi{\arabic{enumi}.}
 
\item
  possible undesirable changes in material properties (e.g.~weaker mechanical properties)
\item
  contamination of the surrounding environment.
\end{enumerate}

For the sake of simplicity, plasticizers will be taken as model compounds for the rest of this article.

\subsection{Migration and diffusion of plasticizers}\label{migration-and-diffusion-of-the-plasticizer}

The loss of plasticizer will increase stiffness and often strength but will decrease flexibility, stretchability and toughness of the polymer. For polymers, which are brittle in the unplasticized state, the loss of plasticizer can lead to failure in its use, while the loss of flexibility limits the field of application of plasticized polymers. A detailed understanding of the mechanisms and kinetics of plasticizer loss is essential to assess the short and long term performance of plasticized polymers. For example, with a known correlation between mechanical properties and plasticizer concentration, the plasticizer loss kinetics can be used to predict the change in mechanical properties associated with a particular use condition, and thus predict the lifetime of plasticizers products. Understanding the plasticizer loss mechanisms and their kinetics is also useful for developing new methods to eliminate/limit plasticizer migration \cite{Arvanitoyannis2004,Ekelund2007,Chiellini2013}.
\subsubsection{Diffusion rate of plasticizers}\label{diffusion-rate-of-the-plasticizer}

The diffusion of additives in polymeric materials is too diverse to be described by a simple relation applicable to all cases. Several types of distribution may apply:

\begin{itemize}
 
\item
  Fickian and non-fickian diffusion can be used to describe certain processes during the incorporation, fusion and gelation of the plasticizer
\item
  Mutual (or cooperative) diffusion describes the removal of the plasticizer from the material with simultaneous replacement by solvent extraction, but it is also descriptive for any exchange process (e.g.~movement of additives in the composition, movement of polymer chains, etc $\dots$)
\item
  Limited-supply, non-Fickian diffusion process, may apply to  applied stress reactions craze formation.
\end{itemize}

The differences in these descriptions are in the availability of the penetrant, the concentration gradient, the reaction of the surrounding molecules (relaxation rate) and the physical conditions. These differences make the description of the process quite complex.

An element that must also be taken into account is the size and shape of the molecules. Plasticizer molecules are far from spherical but have very irregular shapes and must be described using a complex algorithm which considers the volume of the plasticizer in a three-dimensional space. The plasticizer volume data comes from conformational analysis and is used to calculate the plasticizer diffusion coefficients based on the experimentally obtained values of the plasticizer efficiency parameter $k$ \cite{Coughlin1990}. This parameter is obtained from the glass transition depression caused by the addition of a certain amount of plasticizer:
\begin{equation}
T_g = T_{g2} - k w_1
\label{eq:param_plastifiant}
\end{equation}
where $T_g$ glass transition temperature of the system containing $\omega_1$ plasticizer weight fraction, $T_{g2}$ pure polymer glass transition temperature and $k$ the plasticizer efficiency parameter.

\subsubsection{Plasticizer migration}\label{plasticizer-migration}

Migration is defined as the diffusion of an additive from a plastic to another material in contact \cite{Kovacic2002}. Several hypotheses have been formulated to carry out migration tests:

\begin{itemize}
 
\item
  the diffusion coefficient of the plasticizer is considered independent of the concentration
\item
  the surrounding environment is considered as having an infinite volume
\item
  boundary layer phenomena are ignored
\item
  migration may involve more than one component of the formulation
\item
  counter-percussion is a frequent phenomenon when the plasticizer is lost by extraction
\item
  the process involves diffusion to the surface and desorption from the surface (the two processes usually have different speeds and one of them can be the limiting step)
\item
  the diffusion process is well described by the second law of fickian diffusion
\item
  samples are considered to be very thin flat sheets with migration occurring mainly through the faces (negligible across the edges)
\end{itemize}

The migration of low molecular weight components is classicaly described as having three steps:

\begin{itemize}
 
\item
  diffusion of the plasticizer from the mass of the material to the surface
\item
  interface phenomena
\item
  sorption into the surrounding environment.
\end{itemize}

The sorption process having been presented previously, the diffusion process of the plasticizer will now be presented. Regarding the interface phenomena, they depend on the properties of the surrounding environment which can be a gas, liquid or penetrable or impenetrable solid.

The diffusion and accumulation of the plasticizer on the surface of the material is controlled by the compatibility between the plasticizer and the matrix, the surface energy of the liquid/solid interface and the volatility for a gas phase,solubility for a liquid phase) of the plasticizer. Volatility and migration are controlled by different properties of plasticizers, so they are not related. Plasticizer evaporation is a slow process due to their high boiling points. It can be calculated from the Hertz equation:
\begin{equation}
W = \frac{p}{(2 M k T)^{1/2}}
\label{eq:hertz}
\end{equation}
where $p$ is the partial pressure of the plasticizer at temperature $T$, $M$ the molecular weight of the plasticizer and $k$ the rate constant of transfer of low molecular weight substance from the material.

Plasticizer evaporation rates are 10 to 100 times lower than typical diffusion rates. If the plasticizer has good compatibility and good wetting characteristics for a particular solid surface, its surface spread will limit it's diffusion to the surface because diffusion is gradient-controlled.

Diffusion into a liquid can be described by the following equation:
\begin{equation}
\frac{M_t}{M_{\infty}}= 2\left(\frac{D t}{\pi l^2}\right)^{1/2}
\label{eq:liquide}
\end{equation}
where $M_t$ is the total amount of plasticizer desorbed at time $t$, $M_{\infty}$ the total amount of plasticizer desorbed after an infinite extraction time, $D$ the diffusion coefficient, $t$ the time and $l$ the half thickness of the sample. The graphical representation of this equation shows a linear relationship at the start of extraction, which is useful for obtaining the value of the diffusion coefficient. However, it should be noted that migration is independent of concentration and shows the influence of the medium in contact with the material containing the plasticizer on the rate of diffusion. It is very unlikely that a barrier will be present between the liquid and the plasticizer unless the two are immiscible or the character of the solid surface facilitates wetting by the surrounding liquid. With these exceptions, the surrounding liquids can be thought of as an infinite volume \cite{Papaspyrides1998}.

\subsection{Leaching model}\label{leaching-model}

As described previously, the loss of plasticizer involves the diffusion of the plasticizer in the mass of polymer towards the surface of the polymer and the subsequent transfer to an adjacent gas or liquid phase. 

Assuming that an additive is present in the polymer as a homogeneous solution, Angert et al. \cite{Angert1961} pointed out that the rate of additive loss is determined by two factors. Initially, the loss-rate is determined by the rate of transfert of the polymer surface material, which will act to create a concentration gradient at the surface. Subsequently, the depleted material on the surface must be replaced by diffusion from the mass so that the overall loss process depends on both mass transfer rate through the sample surface and diffusion rate in the sample.

The mechanism of additives loss from the matrix depends on which of these processes governs its loss. It can take place through three specifics processes \cite{Kovacic2002}:

\begin{itemize}
 
\item
  Evaporation: at which the additive enters the surrounding gas medium.
\item
  Extraction: During which the additive enters the surrounding liquid medium.
\item
  Migration: at which the additive, in direct contact with the surface of the other polymer material, migrates into this material.
\end{itemize}

An alternative process can occur when the additive is present in the polymer at a concentration greater than its saturation solubility. For the most common case where the melting point of the additive is lower than that of the polymer, it appears that supersaturated solutions are easily formed so that the additive can then precipitate either in the polymer or on the surface. This latter process is called \textit{Blooming}. Calvert et al. \cite{Calvert1979} attempted to establish a basic model which takes into account the involvement of solubility, volatility and mobility in both volatilization losses and blooming mechanisms. However, a complete model should also take into account the consumption of additives by the oxidation reactions and the variations in ultraviolet intensity within the depth of the sample during the loss of the stabilizer.

According to Calvert and Bilingham \cite{Calvert1979, Luston1993}, first the additive must be removed from the surface of the polymer in the surrounding environment (comparable to the emission of heat at the surface), then, in order to compensate for the resulting loss, the additive diffuses from the mass of polymer towards the interface (comparable to thermal conductivity). In fact, the mathematical model requires two parameters:

\begin{itemize}
\item
  a mass transfer constant characterizing the transfer across the boundary and
\item
  a constant characterizing the mass transfer within the bulk polymer.
\end{itemize}

In the case of the loss of additives, the mass parameter during migration can be described by Fick's law, and controlled by the diffusion coefficient $D$ of the molecule permeating through the polymer but the surface loss parameter is less obvious to define. 

When considering the interface between a polymer and air, a reasonable approximation might be to assume that the additive solution in the polymer behaves in a near-ideal manner, so that the vapour pressure above an unsaturated solution is the saturation value multiplied by the saturation solubility fraction. It can also be assumed that the volatilization rate of the additive is proportional to its vapour pressure under a given set of conditions, therefore that the rate of volatilization $V$ of the polymer additive would be related to the rate of volatilization $V_0$ pure additive per unit area by:
\begin{equation}
V=V_0 \frac{C_s}{S} = H C_s
\label{eq:volatilisation_rate}
\end{equation}
where $C_s$ is the additive concentration at the surface of the polymer and $S$ is the saturation solubility; the parameter $H$ is then the required mass transfer constant and can be evaluated by measurements of $V_0$ and $S$. The loss of additive in a flowing liquid is exactly the same, but it is much more difficult to assign values to $H$ since $V_0$ now becomes the rate of dissolution of the additive.

\subsubsection{Two modes of leaching}\label{two-modes-of-leaching}

The diffusion process can be described by Fick's second law (given here for a one-to-three-dimensional orthogonal system) \cite{Crank1975} as follows:
\begin{equation}
\frac{\partial C}{\partial t} = \sum_{i=1}^3 \frac{\partial }{\partial x_i} \left( D(C) \frac{\partial C}{\partial x_i}\right)
\label{eq:fick2_migration}
\end{equation}

where $t$, $x_i$, $C$ and $D(C)$ are respectively the diffusion time, the distance from the surface in the case of one to three dimensions, the concentration of the plasticizer and the diffusivity of the plasticizer. The diffusivity generally increases with the concentration of plasticizer, which can be explained by the increase in the free-volume and the mobility of the polymer molecules in the presence of plasticizers \cite{Huang2001, Storey1989}. This concentration-dependence of diffusivity can be described by the exponential function as follows:
\begin{equation}
D(C) = D_{C0}e^{\alpha C}
\label{eq:concentration-dependance}
\end{equation}
where $D_{C0}$ is the diffusivity of the plasticizer in a material without plasticizer, and $\alpha$ is the plasticizing power describes the degree of plasticization caused by the plasticizer. It is assumed here that all the aggregating effects of plasticizers, which can lead to a decrease in diffusivity with increasing plasticizer content, are weak or absent.

The diffusivity can also be expressed as a function of the volume fraction $v$ according to the semi-empirical exponential relation \cite{Philip1955}:
\begin{equation}
D(v) = D_{C0}e^{\alpha v}
\label{eq:concentration-dependance2}
\end{equation}

These equations are valid in both nonpolar and polar solute-polymer systems to describe the effects of exponential plasticization on the diffusivity of the solute \cite{Thornton2009}.
The average diffusivity over the entire concentration range ($D_{av}$) is obtained:
\begin{equation}
D_{av} = \frac{1}{C_{max}} \int_0^{C_{max}} D_{CO} e^{\alpha C} dC
\label{eq:diffusivite_moyenne}
\end{equation}

There are also other equations describing the concentration dependent diffusion of the solute. The concentration dependence on diffusivity can also be expressed as an increase in free-volume (fraction of free-volume: f) of the solute/polymer mixture relative to the pure polymer \cite{Koros2002, Sharma2017}. The equations \ref{eq:diffusivite1} and \ref{eq:diffusivite2} are examples of how f has been observed to relate to the diffusivity of the solute.
\begin{equation}
D = A e^{-B/f}
\label{eq:diffusivite1}
\end{equation}
where $A$ and $B$ are constants. We consider that A depends on the temperature and the size/shape of the solute and $B$ is related to the kinetic diameter of the solute \cite{Sharma2017, Thornton2009}.
\begin{equation}
D = \alpha_1 e^{\beta/f}
\label{eq:diffusivite2}
\end{equation}
where $\alpha_1$ (not to be confused with $\alpha$ of the equation \ref{eq:concentration-dependance}) is an empirical constant and $\beta$ depends linearly on the square of the kinetic diameter of the solute.

The boundary condition of evaporation in the x direction can be described according to \cite{Crank1975, Ekelund2008, Ekelund2010} by:
\begin{equation}
-D(C) \left(\frac{\partial C}{\partial x}\right) = F(C - C_e)
\label{eq:diffusivite3}
\end{equation}
which is equivalent to mass transfer through a given cross section of the sample to surface evaporation of the same mass from the same surface. The flux to the surface depends on the mobility of the plasticizer molecule ($D(C)$) and the concentration gradient just inside the surface. Evaporation takes place as long as the concentration of plasticizer just at the surface is greater than the concentration corresponding to the environment saturated with plasticizer ($C_e$) and as long as the evaporation coefficient ($F$) is not nul.

\begin{figure}[!ht]
\centering
\includegraphics[scale=0.4]{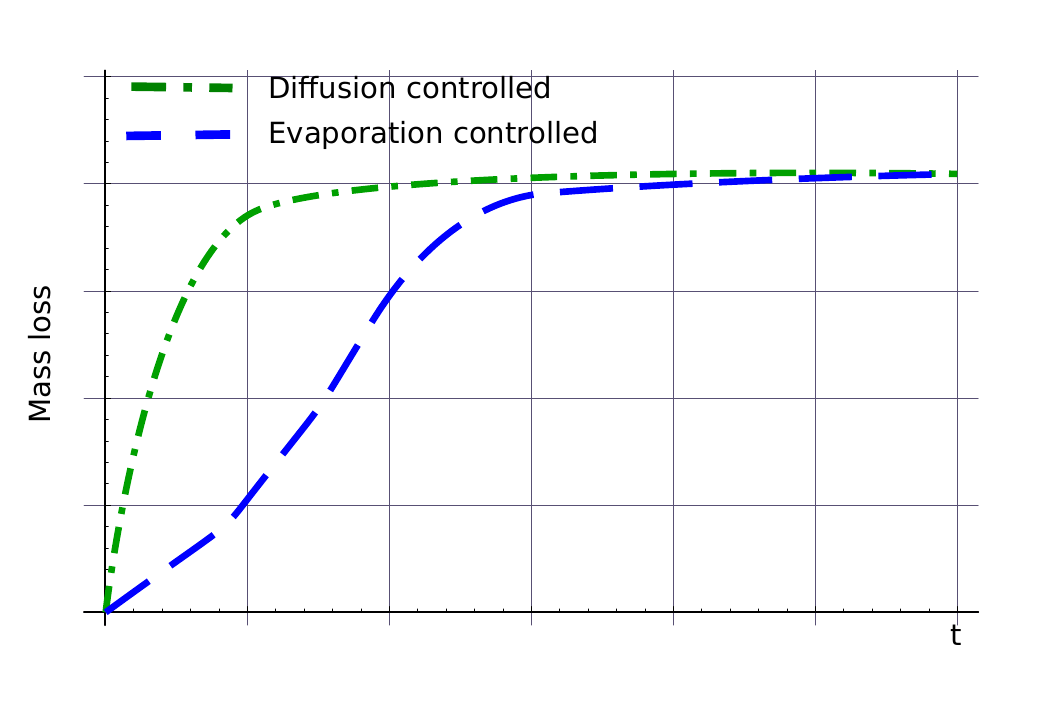}
\caption{Diffusion or evaporation of plasticizer}\label{fig:diffusionevaporation}
\end{figure}

Due to the coexistence of diffusion and evaporation, the overall rate of plasticizer loss is determined by the slower process, which means that the process is controlled by diffusion or evaporation. In the controlled diffusion case (Figure \ref{fig:diffusionevaporation}), evaporation is faster than the feed rate of the plasticizer to the surface, leading to a gradient in the concentration of the plasticizer in the region of surface of the sample and possibly also in the mass \cite{Smith2004}. When the system is controlled by evaporation, evaporation is slower than the rate of supply of plasticizer to the surface. In this case, depending on the system, a film of plasticizer may form on the surface.

\hypertarget{blooming}{%
\subsubsection{Blooming}\label{blooming}}

The model described above is valid if the additive is present below its saturation solubility; it can also be applied to an additive greater than its saturation value if it evaporates rapidly from the surface. However, a more likely mechanism of loss of supersaturated samples is the process of \textit{blooming} to produce crystals on the surface of the sample. Blooming has the effect of fixing the concentration at the surface of the sample and at a value $S$ equal to the saturation solubility. It is also the limiting concentration in the whole sample at equilibrium, and the concentration cannot fall below $S$ at any point if the crystals are removed from the surface to change the loss mechanism in case of evaporation. The surface concentration being fixed at saturation, the process of loss occurs essentially by diffusion through the concentration gradient in the sample. Only one parameter is required, $D$, and the mathematical model is the heat loss equivalent of a sample at constant surface temperature.

In case of plasticizer with a very high boiling point and if the material is used at low temperature, the plasticizer can accumulate on the surface and form a thin film on the surface. This can be reflected in two ways:

\begin{itemize}
 
\item
  Firstly, there is the case where the plasticizer loss rate is constant, at least over a certain period of time, which means that the loss is greatly limited by the evaporation process. The presence of the thin layer implies that the \textit{concentration} of plasticizer on the surface is constant over time. If the surface concentration decreases with time, the loss of plasticizer is no longer linear with time.
\item
  Second, the rate of evaporation of plasticizer from polymer surfaces is similar (of the same order of magnitude) to the rate of their pure liquid state, indicating that the constant concentration of plasticizer on the surface is close to that of pure liquid (100$\%$).
\end{itemize}

Surface blooming (due to crystallization) and bleeding are effects of the accumulation of migrated components on the surface \cite{Nouman2017}. Blooming occurs when the additive crystallizes on the surface. For example, in the case of plasticized PVC, the surface tends to become sticky over time, which is caused by a layer of plasticizer forming on the surface (bleeding).

The evaporation rate $v_0$ ($g.cm^{-1}.s^{-1}$) of a pure substance (plasticizer), from a band of width $l$ on which gas flows at the speed $u$ ($cm.s^{-1}$), can be calculated \cite{Treybal2004} by applying the mass transfer theory of evaporation from a stationary liquid in a turbulent gas:
\begin{equation}
v_0 = 0.33  \left(\frac{u^{\frac{1}{2}}}{l^{\frac{1}{2}}}\right )S_g D_g^{\frac{3}{4}} \left(\frac{\rho}{\mu}\right)^{1/6}
\label{eq:blooming}
\end{equation}
where $S_g$ ($g.cm^{-3}$) is the concentration of the substance in the gas phase and $D_g$ ($cm^2.s^{-1}$) is its diffusion coefficient in the gas phase. $\rho$ is the density ($g.cm^{-3}$), and $\mu$ is the viscosity ($g.cm^{-1}.s^{-1}$) of the gas in which the evaporation takes place. Bellobono et al. \cite{Bellobono1984} calculated the $v_0$ of several types of plasticizers using the equation \ref{eq:blooming} and the results obtained agree with the experimental data.

The formation of the plasticizer film results in a 100$\%$ plasticizer concentration on the surface, which hinders the diffusion process, which is gradient driven. A full understanding of the plasticizer film formation mechanisms and its effect on plasticizer diffusion and loss kinetics remains to be developed. The thin film of plasticizer can dramatically change the surface properties of polymers.

\subsection{Main factors influencing leaching}\label{main-factors}

Due to the formation of the plasticizer film, the kinetic loss of the plasticizer for a highly controlled process by evaporation is essentially independent of the polymer-plasticizer interactions, the concentration of the plasticizer and the gradient in the material and is essentially only influenced by temperature, characteristics of the plasticizer (size, shape, polarity, vapor pressure), the gas flow rate above the surface and the size of the gas volume surrounding the material \cite{Ekelund2010, Clausen2007}.

Evaporation is, in general, the process limiting the migration rate of the plasticizer at lower temperatures, while diffusion is the limiting process at higher temperatures. Therefore, the temperature dependence is stronger for the evaporation process than for the diffusion process. 

It is possible to define a ``transition'' temperature ($T_c$) at which there is a switch from evaporation being the rate-limiting factor to diffusion being the rate-limiting factor for plasticizer loss.

The diffusivity depends on the concentration of the plasticizer and decreases with the concentration (cf equation \ref{eq:fick2_migration}). In addition, a smaller concentration gradient at low plasticizer concentrations also contributes to the dominance of the diffusion-controlled mode. However, for the migration of plasticizer from samples with a high initial concentration of plasticizer, it is possible to observe both  migration modes as in the beginning the migration is first controlled by the diffusion process and secondly by evaporation when the plasticizer concentration decreases beyond a critical level.

According to the various experiments carried out, the most important factor to consider is the solubility of the additive in the polymer relative to the concentration at which it is used. If the additive is present below its saturation solubility, blooming cannot occur and the rate of loss in air is determined by the ratio $H/D$, i.e.~depending on whether volatilization or diffusion is dominant. Once the additive is soluble, the most important consideration is to reduce its volatility. The use of long flexible substituents is ideal because it increases solubility, reduces volatility and has little effect on the diffusion coefficient relative to the unsubstituted molecule. If such an additive is supersaturated in the polymer, the situation is totally different since blooming becomes the main wasting process and its rate is controlled by diffusion which is generally rapid. Supersaturation of additives with a low $H/D$ ratio is likely to be of little benefit since the additive will reflower relatively quickly to saturation.

For polymers in contact with liquids, a different situation arises since, whatever the solubility of the additive in the polymer, the rate of loss is determined by the rate of diffusion of the additive to the surface. Under these conditions, the most important thing is to reduce the diffusion coefficient, which is most easily achieved by increasing the size of the molecule or by devising methods of binding the additive to the polymer. Likewise, if the additive is supersaturated in a situation of contact with air, the most important factor is to lower the diffusion coefficient to avoid blooming.

\subsection{Extraction into a liquid}\label{migration-to-a-liquid}

In the case where the migration is controlled by diffusion, the following equation can be used:
\begin{equation}
M_t = 2 C_0 \rho (D_p t / \pi)
\label{eq:migration_liquide}
\end{equation}
where $M_t$ is the total polymer migrant over time $t$ (s); $C_0$ is the initial concentration of migrants in the polymer ($mg.g^{-1}$), $\rho$ is the density of the polymer ($g.cm^{-3}$); $D_p$ is the diffusion coefficient of the migrant in the polymer ($cm^2.s^{-1}$) and $t$ is the lifetime of the polymer in seconds.

An empirical approach was used whereby the known diffusion coefficients can be fitted to equation relative to Arrhenius' equation, as shown in the following equation:
\begin{equation}
D_p = 10^4 \exp[A^{-a}_p \times MW -b(1/t)]
\label{eq:migration_liquide2}
\end{equation}
where the coefficient $A_p$ (dimensionless constant) takes into account the effect of the polymer on diffusivity, $MW$ is the molecular weight of the additive/contaminant, $T$ is the temperature in K, and $a$ and $b$ are correlation constants for the effects of molecular weight and temperature on diffusion with values of 0.010 and 10450, respectively \cite{Begley1997,Baner1996}.

Another migration model has been described by the Crank equation and can be used to describe the migration of semi-infinite media according to the following equation (in the case of relatively short migration times):
\begin{equation}
m_t/A = 2 C^{'}_{p,0} (D_p t/ \pi)
\label{eq:migration_liquide3}
\end{equation}

\subsection{Prediction of plasticizer loss}\label{prediction-of-plasticizer-loss}

It appears that the only step which determines the possible speed of the plasticizer loss process is evaporation. In principle, when the whole process is kinetically controlled by evaporation, the concentration of plasticizer should remain homogeneously distributed throughout the thickness of the sample and a decrease in the rate of weight loss should lead to a decrease in the concentration of plasticizer in the surface layer. However, one might assume that an almost continuous plasticizer monolayer exists on the surface of the sample, making the rate of evaporation virtually insensitive to the bulk concentration of the plasticizer. In fact, when the latter decreases, the free-volume fraction and therefore the diffusivity of the plasticizer decreases and can reach a critical value beyond which the entire process becomes controlled by diffusion. A simple kinetic model therefore cannot describe the entire process of plasticizer loss.

The existence of a close correlation between the Arrhenius parameters $\ln r_0$ and $E_0$, often called compensation, allows the following relation to be expressed:
\begin{equation}
r = \exp a \exp-\left[\frac{e_0}{r}\left(\frac{1}{t} - \frac{1}{t_c}\right)\right]
\label{eq:arrhenius_evaporation}
\end{equation}
with $T_c = 1/bR$, $T_c$ is called \textit{compensation} or \textit{isokinetic} temperature. The existence of a compensation effect is generally interpreted in terms of the correlation between enthalpy and activation entropy. The studies carried out have shown that the processes of evaporation and diffusion of the plasticizer obey the Arrhenius law \cite{Audouin1992, Ekelund2010}. However, the activation energy of diffusion depends on the concentration, and it is also, as is the case with the energy of evaporation activation, dependent on the temperature. Therefore, when considering large concentrations of plasticizer and temperature intervals, this should not be ignored.

The rate of evaporation of a plasticizer from the polymer is similar (in an order of magnitude) to that of its liquid state when a plasticizer film forms on the surface. The evaporation rate should be able to be estimated by directly using the evaporation rate data for the neat plasticizer. In this approach, the temperatures for the accelerated tests can be above $T_c$, however, it is important that the high temperatures chosen for accelerated ageing do not cause any degradation of the plasticizer.

Regarding the extrapolation of diffusivity, it is more complex, since diffusivity depends on concentration and the desorption process at a certain temperature must be described by at least two factors, for example, $\alpha$ and $D_{C0}$ (equation \ref{eq:concentration-dependance}). In this case, the activation energies of $\alpha$ and $D_{C0}$, which are obtained by extrapolation, are necessary for the prediction. However, there is currently no study with data available on extrapolation of both $\alpha$ and $D_{C0}$.

\section{Conclusion} 

This document offers an in-depth analysis of sorption interactions ( adsorption, absorption, permeation and migration phenomena) between materials in general and more specifically polymers) and other chemical compounds present in the bulk or the liquid or gaseous environment in contact with it at the interface. Here are the key points to remember:
\begin{enumerate}
    \item \textbf{Sorption mechanisms}: Sorption can be physical (physisorption) or chemical (chemisorption), each type having different strengths and energies. Adsorption isotherms, such as the Freundlich, Langmuir and BET models, can be used to describe these processes.
    \item \textbf{Diffusion models}: The diffusion of molecules through polymers can follow different mechanisms, including Fickian and non-Fickian diffusion. The solution-diffusion model is particularly relevant for describing the transport of liquids and gases across polymer membranes.
    \item \textbf{Plasticizer migration}: Plasticizer migration is a complex phenomenon influenced by various factors, such as concentration, temperature and polymer-plasticizer interactions. Specific models can be used to predict plasticizer loss and its effects on material properties.
    \item \textbf{Polymer applications}: Flory-Huggins, GAB and ENSIC models are used to describe the sorption of molecules in polymers. These models take into account the specific interactions between polymers and solvents, as well as the thermodynamic aspects of sorption.
    \item \textbf{Thermodynamic aspects}: Thermodynamic parameters, such as Gibbs free energy, enthalpy and entropy, are essential for predicting the feasibility and nature of adsorption processes. 
\end{enumerate}

In summary, this paper provides a sound basis for understanding the complex mechanisms of container-content interactions in medical devices. It highlights the importance of these phenomena in ensuring the safety and efficacy of medical products.

\bibliographystyle{unsrtnat}
\bibliography{Sorption_mechanism_1}

\begin{thebibliography}{187}
\providecommand{\natexlab}[1]{#1}
\providecommand{\url}[1]{\texttt{#1}}
\expandafter\ifx\csname urlstyle\endcsname\relax
  \providecommand{\doi}[1]{doi: #1}\else
  \providecommand{\doi}{doi: \begingroup \urlstyle{rm}\Url}\fi

\bibitem[Von~Saussure(1814)]{Saussure1814}
Theodor Von~Saussure.
\newblock Beobachtungen \"uber die absorption der gasarten durch verschiedene k\"orper.
\newblock \emph{Annalen der Physik}, 47\penalty0 (6):\penalty0 113--183, 1814.
\newblock \doi{10.1002/andp.18140470602}.

\bibitem[Mitscherlich(1843)]{Mitscherlich1843}
E.~Mitscherlich.
\newblock Ueber die {G\"ahrung}.
\newblock \emph{Annalen der Physik}, 135\penalty0 (5):\penalty0 94--101, 1843.
\newblock ISSN 1521-3889.
\newblock \doi{10.1002/andp.18431350506}.

\bibitem[Kayser(1881{\natexlab{a}})]{Kayser1881}
Heinrich Kayser.
\newblock \"uber die {Verdichtung} von {Gasen} an {Oberfl\"achen} in ihrer {Abh\"angigkeit} von {Druck} und {Temperatur}.
\newblock \emph{Annalen der Physik}, 248\penalty0 (4):\penalty0 526--537, 1881{\natexlab{a}}.
\newblock \doi{10.1002/andp.18812480404}.

\bibitem[Langmuir(1916)]{Langmuir1916}
Irving Langmuir.
\newblock The constitution and fundamental properties of solids and liquids. {Part} 1. {Solids}.
\newblock \emph{Journal of the American Chemical Society}, 38\penalty0 (11):\penalty0 2221--2295, 1916.
\newblock \doi{10.1021/ja02268a002}.

\bibitem[Nollet(1743)]{Nollet1743}
Jean-Antoine Nollet.
\newblock \emph{Le\c{c}ons de physique exp\'erimentale. Par M. l'abb\'e Nollet, de l'Acad\'emie royale des sciences, \& de la Soci\'et\'e royale de Londres. Tome premier [-sixieme]}.
\newblock Frères Guérin, Paris, 1743.

\bibitem[Graham(1834)]{Graham1834}
Thomas Graham.
\newblock \emph{On the {Law} of {Diffusion} of {Gases}}.
\newblock Transactions of the {Royal Society} of {Edinburg}. The Society, Edinburg, 1834.

\bibitem[Fick(1855)]{Fick1855}
Adolf Fick.
\newblock Ueber diffusion.
\newblock \emph{Annalen der Physik}, 170\penalty0 (1):\penalty0 59--86, 1855.
\newblock ISSN 1521-3889.
\newblock \doi{10.1002/andp.18551700105}.

\bibitem[Darcy(1856)]{Darcy1856}
Henry Darcy.
\newblock \emph{Les fontaines publiques de la ville de Dijon : exposition et application des principes \`a suivre et des formules \`a employer dans les questions de distribution d'eau}.
\newblock Victor Dalmont, Paris, 1856.

\bibitem[West(1945)]{West1945}
James~R. West.
\newblock Some industrial aspects of adsorption.
\newblock \emph{Journal of Chemical Education}, 22\penalty0 (8):\penalty0 398, 1945.

\bibitem[Young and Crowell(1962)]{Young1962}
D.~M. Young and A.~D. Crowell.
\newblock \emph{Physical {Adsorption} of {Gases}}.
\newblock Butterworths, London, 1962.
\newblock ISBN 978-0-598-44626-8.

\bibitem[Israelachvili(2011)]{Israelachvili2011}
Jacob~N. Israelachvili.
\newblock \emph{Intermolecular and Surface Forces}.
\newblock Elsevier, Academic Press, Amsterdam, third edition edition, 2011.
\newblock ISBN 978-0-12-391927-4 978-0-12-375182-9.

\bibitem[Keesom(1915)]{Keesom1915}
WH~Keesom.
\newblock The second viral coefficient for rigid spherical molecules, whose mutual attraction is equivalent to that of a quadruplet placed at their centre.
\newblock \emph{Proc. R. Acad. Sci}, 18:\penalty0 636--646, 1915.

\bibitem[London(1930{\natexlab{a}})]{London1930}
F.~London.
\newblock Zur {Theorie} und {Systematik} der {Molekularkr\"afte}.
\newblock \emph{Zeitschrift f\"ur Physik}, 63\penalty0 (3-4):\penalty0 245--279, 1930{\natexlab{a}}.
\newblock \doi{10.1007/BF01421741}.

\bibitem[{Lennard-Jones}(1932)]{Lennard-Jones1932}
J.~E. {Lennard-Jones}.
\newblock Processes of adsorption and diffusion on solid surfaces.
\newblock \emph{Transactions of the Faraday Society}, 28:\penalty0 333--359, 1932.
\newblock ISSN 0014-7672.
\newblock \doi{10.1039/TF9322800333}.

\bibitem[Jones and Chapman(1924)]{Jones1924}
J.~E. Jones and Sydney Chapman.
\newblock On the determination of molecular fields. {II}. {From} the equation of state of a gas.
\newblock \emph{Proceedings of the Royal Society of London. Series A, Containing Papers of a Mathematical and Physical Character}, 106\penalty0 (738):\penalty0 463--477, 1924.
\newblock \doi{10.1098/rspa.1924.0082}.

\bibitem[{Le chatelier} and {Acad{\'e}mie des sciences (France)}(1884)]{Lechatelier1884}
Henri {Le chatelier} and {Acad{\'e}mie des sciences (France)}.
\newblock Sur un \'enonc\'e g\'en\'eral des lois des \'equilibres chimiques, 1884.

\bibitem[London(1930{\natexlab{b}})]{London1930a}
F.~London.
\newblock Ober einige {Eigenschaften} und {Anwendungen} der {Molekularkrafte}. {{Z}}. f.
\newblock \emph{phys. Chem. B}, 11:\penalty0 222--251, 1930{\natexlab{b}}.

\bibitem[De~Boer and Custers(1934)]{DeBoer1934}
JH~De~Boer and JFH Custers.
\newblock \"uber die natur der adsorptionskr\"afte.
\newblock \emph{Zeitschrift f\"ur Physikalische Chemie}, 25\penalty0 (1):\penalty0 225--237, 1934.
\newblock ISSN 2196-7156.

\bibitem[Lenel(1933)]{Lenel1933}
FV~Lenel.
\newblock Uber die {Adsorptions} warme von {Edelgasen} und {Kohlendioxyd} an {Ionenkristallen}.
\newblock \emph{Physik. Chem. B}, 23:\penalty0 379, 1933.

\bibitem[De~Boer(1956)]{DeBoer1956}
J.~H. De~Boer.
\newblock Adsorption phenomena.
\newblock In W.~G. Frankenburg, V.~I. Komarewsky, and E.~K. Rideal, editors, \emph{Advances in {{Catalysis}}}, volume~8, pages 17--161. Academic Press, London, 1956.
\newblock \doi{10.1016/S0360-0564(08)60538-6}.

\bibitem[Kayser(1881{\natexlab{b}})]{Kayser1881a}
Heinrich Kayser.
\newblock Ueber die verdichtung von gasen an oberfl\"achen in ihrer abh\"angigkeit von druck und temperatur.
\newblock \emph{Annalen der Physik}, 250\penalty0 (11):\penalty0 450--468, 1881{\natexlab{b}}.
\newblock ISSN 00033804, 15213889.
\newblock \doi{10.1002/andp.18812501105}.

\bibitem[Henneberg and Stohmann(1858)]{Henneberg1858}
W.~Henneberg and F.~Stohmann.
\newblock Ueber das {Verhalten} der {Ackerkrume} gegen {Ammoniak} und {Ammoniaksalze}.
\newblock \emph{Justus Liebigs Annalen der Chemie}, 107\penalty0 (2):\penalty0 152--174, 1858.
\newblock \doi{10.1002/jlac.18581070204}.

\bibitem[Van~Bemmelen and Ostwald(1910)]{Bemmelen1910}
J.M. Van~Bemmelen and W.~Ostwald.
\newblock \emph{Die Absorption: Gesammelte Abhandlungen \"Uber Kolloide Und Absorption}.
\newblock {T. Steinkopff}, Dresden, 1910.

\bibitem[Freundlich(1907)]{Freundlich1907}
Herbert Freundlich.
\newblock \"uber die {Adsorption} in {L\"osungen}.
\newblock \emph{Zeitschrift f\"ur Physikalische Chemie}, 57U\penalty0 (1), 1907.
\newblock \doi{10.1515/zpch-1907-5723}.

\bibitem[Henry(1803)]{Henry1803}
William Henry.
\newblock {III}. {Experiments} on the quantity of gases absorbed by water, at different temperatures, and under different pressures.
\newblock \emph{Philosophical Transactions of the Royal Society of London}, 93:\penalty0 29--274, 1803.
\newblock \doi{10.1098/rstl.1803.0004}.

\bibitem[Langmuir(1918)]{Langmuir1918}
Irving Langmuir.
\newblock The adsorption of gases on plane surfaces of glass, mica and platinum.
\newblock \emph{J Am Chem Soc}, 40\penalty0 (9):\penalty0 1361--1403, 1918.
\newblock ISSN 0002-7863.
\newblock \doi{10.1021/ja02242a004}.

\bibitem[Swenson and Stadie(2019)]{Swenson2019}
Hans Swenson and Nicholas~P. Stadie.
\newblock Langmuir's {Theory} of {Adsorption}: {A Centennial Review}.
\newblock \emph{Langmuir}, 35\penalty0 (16):\penalty0 5409--5426, 2019.
\newblock \doi{10.1021/acs.langmuir.9b00154}.

\bibitem[Brunauer et~al.(1938)Brunauer, Emmett, and Teller]{Brunauer1938}
Stephen Brunauer, P.~H. Emmett, and Edward Teller.
\newblock Adsorption of gases in multimolecular layers.
\newblock \emph{Journal of the American Chemical Society}, 60\penalty0 (2):\penalty0 309--319, 1938.
\newblock \doi{10.1021/ja01269a023}.

\bibitem[Do(1998)]{Do1998}
Duong~D Do.
\newblock \emph{Adsorption Analysis: {Equilibria} and {Kinetics}}.
\newblock Published by Imperial College Press and distributed By World Scientific Publishing Co., London, 1998.

\bibitem[D{\k a}browski(2001)]{Dabrowski2001}
A~D{\k a}browski.
\newblock Adsorption - from theory to practice.
\newblock \emph{Advances in Colloid and Interface Science}, 93\penalty0 (1):\penalty0 135--224, 2001.
\newblock ISSN 0001-8686.
\newblock \doi{10.1016/S0001-8686(00)00082-8}.

\bibitem[Lowell et~al.(2004)Lowell, Shields, Thomas, and Thommes]{Lowell2004}
S~Lowell, Joan~E Shields, Martin~A Thomas, and Matthias Thommes.
\newblock Adsorption mechanism.
\newblock In \emph{Characterization of Porous Solids and Powders: Surface Area, Pore Size and Density}, pages 15--57. Springer, Dordrecht, 2004.

\bibitem[Saadi et~al.(2015)Saadi, Saadi, Fazaeli, and Fard]{Saadi2015}
Reyhaneh Saadi, Zahra Saadi, Reza Fazaeli, and Narges~Elmi Fard.
\newblock Monolayer and multilayer adsorption isotherm models for sorption from aqueous media.
\newblock \emph{Korean Journal of Chemical Engineering}, 32\penalty0 (5):\penalty0 787--799, 2015.
\newblock ISSN 1975-7220.

\bibitem[Karimi et~al.(2019)Karimi, Yaraki, and Karri]{Karimi2019}
Samira Karimi, Mohammad~Tavakkoli Yaraki, and Rama~Rao Karri.
\newblock A comprehensive review of the adsorption mechanisms and factors influencing the adsorption process from the perspective of bioethanol dehydration.
\newblock \emph{Renewable and Sustainable Energy Reviews}, 107:\penalty0 535--553, 2019.
\newblock ISSN 1364-0321.
\newblock \doi{10.1016/j.rser.2019.03.025}.

\bibitem[Brunauer et~al.(1940)Brunauer, Deming, Deming, and Teller]{Brunauer1940}
Stephen Brunauer, Lola~S. Deming, W.~Edwards Deming, and Edward Teller.
\newblock On a theory of the van der waals adsorption of gases.
\newblock \emph{Journal of the American Chemical Society}, 62\penalty0 (7):\penalty0 1723--1732, 1940.
\newblock ISSN 0002-7863.
\newblock \doi{10.1021/ja01864a025}.

\bibitem[Gibbs and Tyndall(1874)]{Gibbs1874}
J.~Willard Gibbs and John Tyndall.
\newblock On the equilibrium of heterogeneous substances : First [-second] part.
\newblock \emph{Published by the Academy}, 1874.
\newblock \doi{10.5479/sil.421748.39088007099781}.

\bibitem[Boutaric(1940)]{Boutaric1940}
A~Boutaric.
\newblock Remarques sur la relation de {Gibbs} et les formules d'adsorption.
\newblock \emph{Journal de Physique et le Radium}, 1\penalty0 (3):\penalty0 99--102, 1940.

\bibitem[Grumbach(1912)]{Grumbach1912}
A.~Grumbach.
\newblock Sur la th\'eorie thermodynamique de l'adsorption - {Application} \`a l'\'electrisation de contact.
\newblock \emph{Journal de Physique Th\'eorique et Appliqu\'ee}, 2\penalty0 (1):\penalty0 283--297, 1912.
\newblock \doi{10.1051/jphystap:019120020028301}.

\bibitem[Gibbs(1878)]{Gibbs1878}
J.~Willard Gibbs.
\newblock \emph{On the equilibrium of heterogeneous substances}.
\newblock New Haven : Connecticut Academy of Arts and Sciences; distributed by Archon Books, Hamden, Conn., 1878.

\bibitem[Thomson(1888)]{Thomson1888}
Sir Thomson, Joseph~John).
\newblock \emph{Applications of Dynamics to Physics and Chemistry}.
\newblock Macmillan and Co, London, 1888.

\bibitem[Mitropoulos(2008)]{Mitropoulos2008}
A.~Ch. Mitropoulos.
\newblock What is a surface excess.
\newblock \emph{Journal of Engineering Science and Technology Review}, 1\penalty0 (1):\penalty0 1--3, 2008.
\newblock \doi{10.25103/jestr.011.01}.

\bibitem[Alberti et~al.(2012)Alberti, Amendola, Pesavento, and Biesuz]{Alberti2012}
Giancarla Alberti, Valeria Amendola, Maria Pesavento, and Raffaela Biesuz.
\newblock Beyond the synthesis of novel solid phases: Review on modelling of sorption phenomena.
\newblock \emph{Coordination Chemistry Reviews}, 256\penalty0 (1):\penalty0 28--45, 2012.
\newblock ISSN 0010-8545.
\newblock \doi{10.1016/j.ccr.2011.08.022}.

\bibitem[Plazinski et~al.(2009)Plazinski, Rudzinski, and Plazinska]{Plazinski2009}
Wojciech Plazinski, Wladyslaw Rudzinski, and Anita Plazinska.
\newblock Theoretical models of sorption kinetics including a surface reaction mechanism: {A} review.
\newblock \emph{Advances in Colloid and Interface Science}, 152\penalty0 (1):\penalty0 2--13, 2009.
\newblock ISSN 0001-8686.
\newblock \doi{10.1016/j.cis.2009.07.009}.

\bibitem[Zhang et~al.(2018)Zhang, Wang, Zhou, Zhou, Dai, Zhou, Chriestie, and Luo]{Zhang2018}
Haibo Zhang, Jiaqing Wang, Bianying Zhou, Yang Zhou, Zhenfei Dai, Qian Zhou, Peter Chriestie, and Yongming Luo.
\newblock Enhanced adsorption of oxytetracycline to weathered microplastic polystyrene: {Kinetics}, isotherms and influencing factors.
\newblock \emph{Environmental Pollution}, 243:\penalty0 1550--1557, 2018.
\newblock ISSN 02697491.
\newblock \doi{10.1016/j.envpol.2018.09.122}.

\bibitem[Ho et~al.(2000)Ho, Ng, and McKay]{Ho2000}
Y.~S. Ho, J.~C.Y. Ng, and G.~McKay.
\newblock Kinetics of pollutant sorption by biosorbents: Review.
\newblock \emph{Separation and Purification Methods}, 29\penalty0 (2):\penalty0 189--232, 2000.
\newblock \doi{10.1081/SPM-100100009}.

\bibitem[Wakao and Funazkri(1978)]{Wakao1978}
N.~Wakao and T.~Funazkri.
\newblock Effect of fluid dispersion coefficients on particle-to-fluid mass transfer coefficients in packed beds: {Correlation} of sherwood numbers.
\newblock \emph{Chemical Engineering Science}, 33\penalty0 (10):\penalty0 1375--1384, 1978.
\newblock ISSN 0009-2509.
\newblock \doi{10.1016/0009-2509(78)85120-3}.

\bibitem[Perry et~al.(1997)Perry, Green, and Maloney]{Perry1997}
Robert~H. Perry, Don~W. Green, and James~O. Maloney, editors.
\newblock \emph{Perry's Chemical Engineers' Handbook}.
\newblock McGraw-Hill, {New York}, 7th ed edition, 1997.
\newblock ISBN 978-0-07-049841-9.

\bibitem[Ruthven(1984)]{Ruthven1984}
D.~M Ruthven.
\newblock \emph{Principles of Adsorption and Adsorption Processes.}
\newblock {John Wiley and Sons, Inc.}, {New York (USA)}, 1984.
\newblock ISBN 978-0-471-86606-0.

\bibitem[Higashi et~al.(1963)Higashi, Ito, and Oishi]{Higashi1963}
K.~Higashi, H.~Ito, and J.~Oishi.
\newblock Surface diffusion phenomena in gaseous diffusion. i. surface diffusion of pure gas.
\newblock \emph{Nippon Genshiryoku Gakkaishi (Japan)}, 5, 1963.

\bibitem[Okazaki et~al.(1981)Okazaki, Tamon, and Toei]{Okazaki1981}
Morio Okazaki, Hajime Tamon, and Ryozo Toei.
\newblock Interpretation of surface flow phenomenon of adsorbed gases by hopping model.
\newblock \emph{AIChE Journal}, 27\penalty0 (2):\penalty0 262--270, 1981.
\newblock ISSN 1547-5905.
\newblock \doi{10.1002/aic.690270213}.

\bibitem[Yang et~al.(1973)Yang, Fenn, and Haller]{Yang1973}
R.~T. Yang, J.~B. Fenn, and G.~L. Haller.
\newblock Modification to the {Kigashi} model for surface diffusion.
\newblock \emph{AIChE J.; (United States)}, 19:5, 1973.
\newblock \doi{10.1002/aic.690190529}.

\bibitem[Unuabonah et~al.(2019)Unuabonah, Omorogie, and Oladoja]{Unuabonah2019}
Emmanuel~I. Unuabonah, Martins~O. Omorogie, and Nurudeen~A. Oladoja.
\newblock 5 - {Modeling} in {Adsorption}: {Fundamentals} and {Applications}.
\newblock In George~Z. Kyzas and Athanasios~C. Mitropoulos, editors, \emph{Composite {Nanoadsorbents}}, Micro and {Nano Technologies}, pages 85--118. Elsevier, 2019.
\newblock ISBN 978-0-12-814132-8.
\newblock \doi{10.1016/B978-0-12-814132-8.00005-8}.

\bibitem[Ho and McKay(1999{\natexlab{a}})]{Ho1999}
Yuh-Shan Ho and Gordon McKay.
\newblock Pseudo-second order model for sorption processes.
\newblock \emph{Process biochemistry}, 34\penalty0 (5):\penalty0 451--465, 1999{\natexlab{a}}.
\newblock \doi{10.1016/s0032-9592(98)00112-5}.

\bibitem[Sparks(1986)]{Sparks1986}
D.~L. Sparks.
\newblock Kinetics of reactions in pure and in mixed systems., 1986.

\bibitem[Belaid and Kacha(2011)]{Belaid2011}
Kumar~Djamel Belaid and Sma{\"i}l Kacha.
\newblock \'etude cin\'etique et thermodynamique de l'adsorption d'un colorant basique sur la sciure de bois.
\newblock \emph{Revue des sciences de l'eau / Journal of Water Science}, 24\penalty0 (2):\penalty0 131--144, 2011.
\newblock \doi{10.7202/1006107ar}.

\bibitem[Kajjumba et~al.(2018)Kajjumba, Emik, {\"O}ngen, {\"O}zcan, and Ayd{\i}n]{Kajjumba2018}
George~William Kajjumba, Serkan Emik, Atakan {\"O}ngen, H~Kurtulus {\"O}zcan, and Serdar Ayd{\i}n.
\newblock Modelling of adsorption kinetic processes\textemdash errors, theory and application.
\newblock In \emph{Advanced Sorption Process Applications}, page~19. IntechOpen, London, 2018.
\newblock ISBN 978-1-78984-818-2.

\bibitem[Kecili and Hussain(2018)]{Kecili2018}
Rustem Kecili and Chaudhery~Mustansar Hussain.
\newblock Chapter 4 - {Mechanism} of {Adsorption} on {Nanomaterials}.
\newblock In Chaudhery~Mustansar Hussain, editor, \emph{Nanomaterials in {Chromatography}}, pages 89--115. Elsevier, Cambridge, 2018.
\newblock ISBN 978-0-12-812792-6.

\bibitem[Qiu et~al.(2009)Qiu, Lv, Pan, Zhang, Zhang, and Zhang]{Qiu2009}
Hui Qiu, Lu~Lv, Bing-cai Pan, Qing-jian Zhang, Wei-ming Zhang, and Quan-xing Zhang.
\newblock Critical review in adsorption kinetic models.
\newblock \emph{Journal of Zhejiang University-SCIENCE A}, 10\penalty0 (5):\penalty0 716--724, 2009.
\newblock ISSN 1862-1775.
\newblock \doi{10.1631/jzus.A0820524}.

\bibitem[Wang and Guo(2020)]{Wang2020}
Jianlong Wang and Xuan Guo.
\newblock Adsorption kinetic models: {Physical} meanings, applications, and solving methods.
\newblock \emph{Journal of Hazardous Materials}, 390:\penalty0 122156, 2020.
\newblock ISSN 0304-3894.
\newblock \doi{10.1016/j.jhazmat.2020.122156}.

\bibitem[Lagergren(1898)]{Lagergren1898}
S.~Lagergren.
\newblock Zur theorie der sogenannten adsorption geloster stoffe.
\newblock \emph{Kungliga Svenska Vetenskapsakademiens. Handlingar}, 24:\penalty0 1--39, 1898.

\bibitem[Ho and McKay(1999{\natexlab{b}})]{Ho1999b}
Y.S Ho and G~McKay.
\newblock A kinetic study of dye sorption by biosorbent waste product pith.
\newblock \emph{Resources, Conservation and Recycling}, 25\penalty0 (3-4):\penalty0 171--193, 1999{\natexlab{b}}.
\newblock ISSN 09213449.
\newblock \doi{10.1016/S0921-3449(98)00053-6}.

\bibitem[Ho(2006)]{Ho2006}
Yuh-Shan Ho.
\newblock Review of second-order models for adsorption systems.
\newblock \emph{Journal of Hazardous Materials}, 136\penalty0 (3):\penalty0 681--689, 2006.
\newblock ISSN 0304-3894.
\newblock \doi{10.1016/j.jhazmat.2005.12.043}.

\bibitem[Boyd et~al.(1947)Boyd, Adamson, and Myers]{Boyd1947}
G.~E. Boyd, A.~W. Adamson, and L.~S. Myers.
\newblock The exchange adsorption of ions from aqueous solutions by organic zeolites. {II}. {Kinetics1}.
\newblock \emph{Journal of the American Chemical Society}, 69\penalty0 (11):\penalty0 2836--2848, 1947.
\newblock ISSN 0002-7863.
\newblock \doi{10.1021/ja01203a066}.

\bibitem[Crank(1956)]{Crank1956}
J.~Crank.
\newblock \emph{The Mathematics Of Diffusion}.
\newblock Clarendon Press, 1956.
\newblock \doi{10.1137/1.9781611971972}.

\bibitem[Cooney(1999)]{Cooney1999}
David~O. Cooney.
\newblock \emph{Adsorption Design for Wastewater Treatment}.
\newblock Lewis Publishers, Boca Raton, Fl, 1999.
\newblock ISBN 978-1-56670-333-8.

\bibitem[Crank(1975)]{Crank1975}
J~Crank.
\newblock The mathematics of diffusion 2nd {Edition}.
\newblock \emph{Oxford Science Publications,}, page~32, 1975.

\bibitem[Weber and Morris(1963)]{Weber1963}
Walter~J. Weber and J.~Carrell Morris.
\newblock Kinetics of {Adsorptio} on {Carbon} from {Solution}.
\newblock \emph{Journal of the Sanitary Engineering Division}, 89\penalty0 (2):\penalty0 31--60, 1963.

\bibitem[Wang et~al.(2004)Wang, Chen, Zhai, Chen, and Zhang]{Wang2004}
Hai-ling Wang, Jin-long Chen, Zhi-cai Zhai, Yi-liang Chen, and Quan-xing Zhang.
\newblock {Study on thermodynamics and kinetics of adsorption of p-toluidine from aqueous solution by hypercrosslinked polymeric adsorbents}.
\newblock \emph{Environmental Chemistry}, pages 188--192, 2004.

\bibitem[Thomas(1944)]{Thomas1944}
Henry~C. Thomas.
\newblock Heterogeneous {Ion Exchange} in a {Flowing System}.
\newblock \emph{Journal of the American Chemical Society}, 66\penalty0 (10):\penalty0 1664--1666, 1944.
\newblock ISSN 0002-7863.
\newblock \doi{10.1021/ja01238a017}.

\bibitem[Blanco et~al.(2017)Blanco, Scheufele, M{\'o}denes, {Espinoza-Qui{\~n}ones}, Marin, Kroumov, and Borba]{Blanco2017}
Silvia Priscila Dias~Monte Blanco, Fabiano~Bisinella Scheufele, Aparecido~Nivaldo M{\'o}denes, Fernando~R. {Espinoza-Qui{\~n}ones}, Pricila Marin, Alexander~Dimitrov Kroumov, and Carlos~Eduardo Borba.
\newblock Kinetic, equilibrium and thermodynamic phenomenological modeling of reactive dye adsorption onto polymeric adsorbent.
\newblock \emph{Chemical Engineering Journal}, 307:\penalty0 466--475, 2017.
\newblock ISSN 1385-8947.
\newblock \doi{10.1016/j.cej.2016.08.104}.

\bibitem[Sausen et~al.(2018)Sausen, Scheufele, Alves, Vieira, {Da Silva}, Borba, and Borba]{Sausen2018}
Mateus~Gustavo Sausen, Fabiano~Bisinella Scheufele, Helton~Jos{\'e} Alves, Melissa Gurgel~Adeodato Vieira, Meuris Gurgel~Carlos {Da Silva}, Fernando~Henrique Borba, and Carlos~Eduardo Borba.
\newblock Efficiency maximization of fixed-bed adsorption by applying hybrid statistical-phenomenological modeling.
\newblock \emph{Separation and Purification Technology}, 207:\penalty0 477--488, 2018.
\newblock ISSN 1383-5866.
\newblock \doi{10.1016/j.seppur.2018.07.002}.

\bibitem[Guo and Wang(2019)]{Guo2019}
Xuan Guo and Jianlong Wang.
\newblock Sorption of antibiotics onto aged microplastics in freshwater and seawater.
\newblock \emph{Marine Pollution Bulletin}, 149:\penalty0 110511, 2019.
\newblock ISSN 0025-326X.
\newblock \doi{10.1016/j.marpolbul.2019.110511}.

\bibitem[Gueu et~al.(2007)Gueu, Yao, Adouby, and Ado]{Gueu2007}
S.~Gueu, B.~Yao, K.~Adouby, and G.~Ado.
\newblock Kinetics and thermodynamics study of lead adsorption on to activated carbons from coconut and seed hull of the palm tree.
\newblock \emph{International Journal of Environmental Science \& Technology}, 4\penalty0 (1):\penalty0 11--17, 2007.
\newblock ISSN 1735-2630.
\newblock \doi{10.1007/BF03325956}.

\bibitem[McLaren and Rowen(1951)]{McLaren1951}
A.~D. McLaren and John~W. Rowen.
\newblock Sorption of water vapor by proteins and polymers: {{A}} review.
\newblock \emph{Journal of Polymer Science}, 7\penalty0 (2-3):\penalty0 289--324, 1951.
\newblock ISSN 1542-6238.
\newblock \doi{10.1002/pol.1951.120070214}.

\bibitem[Karimi(2011)]{Karimi2011}
Mohammad Karimi.
\newblock Diffusion in polymer solids and solutions.
\newblock In Jozef Marko, editor, \emph{Mass Transfer in Chemical Engineering Processes}, page Ch. 2. IntechOpen, London, 2011.
\newblock ISBN 978-953-307-619-5.
\newblock \doi{10.5772/23436}.

\bibitem[Flory(1950)]{Flory1950}
Paul~J. Flory.
\newblock Statistical mechanics of swelling of network structures.
\newblock \emph{The Journal of Chemical Physics}, 18\penalty0 (1):\penalty0 108--111, 1950.
\newblock \doi{10.1063/1.1747424}.

\bibitem[Flory(1953)]{Flory1953}
P.J. Flory.
\newblock \emph{Principles of {Polymer Chemistry}}.
\newblock Baker Lectures 1948. Cornell University Press, Ithaca, 1953.
\newblock ISBN 978-0-8014-0134-3.

\bibitem[Huggins(1942)]{Huggins1942}
Maurice~L. Huggins.
\newblock Some properties of solutions of long-chain compounds.
\newblock \emph{The Journal of Physical Chemistry}, 46\penalty0 (1):\penalty0 151--158, 1942.
\newblock ISSN 0092-7325.
\newblock \doi{10.1021/j150415a018}.

\bibitem[Ayawei et~al.(2017)Ayawei, Ebelegi, and Wankasi]{Ayawei2017}
Nimibofa Ayawei, Augustus~Newton Ebelegi, and Donbebe Wankasi.
\newblock Modelling and interpretation of adsorption isotherms, 2017.

\bibitem[Dolmaire et~al.(2004)Dolmaire, Espuche, M{\'e}chin, and Pascault]{Dolmaire2004}
N.~Dolmaire, E.~Espuche, F.~M{\'e}chin, and J.-P. Pascault.
\newblock Water transport properties of thermoplastic polyurethane films: {Thermoplastic Polyurethane Films}.
\newblock \emph{Journal of Polymer Science Part B: Polymer Physics}, 42\penalty0 (3):\penalty0 473--492, 2004.
\newblock ISSN 08876266.
\newblock \doi{10.1002/polb.10716}.

\bibitem[Jonqui{\`e}res et~al.(1998)Jonqui{\`e}res, Perrin, Durand, Arnold, and Lochon]{Jonquieres1998}
Anne Jonqui{\`e}res, Laurent Perrin, Alain Durand, St{\'e}phanie Arnold, and Pierre Lochon.
\newblock Modelling of vapour sorption in polar materials: {Comparison} of {Flory}\textendash{Huggins} and related models with the {ENSIC} mechanistic approach.
\newblock \emph{Journal of Membrane Science}, 147\penalty0 (1):\penalty0 59--71, 1998.
\newblock ISSN 0376-7388.
\newblock \doi{10.1016/S0376-7388(98)00111-2}.

\bibitem[Zimm and Lundberg(1956)]{Zimm1956}
Bruno~H. Zimm and John~L. Lundberg.
\newblock Sorption of {Vapors} by {High Polymers}.
\newblock \emph{The Journal of Physical Chemistry}, 60\penalty0 (4):\penalty0 425--428, 1956.
\newblock \doi{10.1021/j150538a010}.

\bibitem[Favre(1994)]{Favre1994}
{\'E}ric Favre.
\newblock \emph{Sorption, Diffusion and Pervaporation of Water and Alcanols through Dense Polydimethylsiloxane Membranes: A Transport Analysis}.
\newblock Theses, Institut National Polytechnique de Lorraine, 1994.

\bibitem[Favre et~al.(1993)Favre, Clement, Nguyen, Schaetzel, and Nee]{Favre1993}
Eric Favre, Robert Clement, Quang~Trong Nguyen, Pierre Schaetzel, and Jean Nee.
\newblock Sorption of organic solvents into dense silicone membranes.
\newblock \emph{J. Chem. Soc.{,} Faraday Trans.}, 89:\penalty0 7, 1993.

\bibitem[Favre et~al.(1996)Favre, Nguyen, Cl{\'e}ment, and N{\'e}el]{Favre1996}
E.~Favre, Q.T. Nguyen, R.~Cl{\'e}ment, and J.~N{\'e}el.
\newblock The engaged species induced clustering ({ENSIC}) model: A unified mechanistic approach of sorption phenomena in polymers.
\newblock \emph{Journal of Membrane Science}, 117\penalty0 (1):\penalty0 227--236, 1996.
\newblock ISSN 0376-7388.
\newblock \doi{10.1016/0376-7388(96)00060-9}.

\bibitem[Scott(1949)]{Scott1949}
Robert~L. Scott.
\newblock The {Thermodynamics} of {High Polymer Solutions}. {IV}. {Phase Equilibria} in the {Ternary System}: {Polymer}\textemdash{Liquid} 1\textemdash{Liquid} 2.
\newblock \emph{Journal of Chemical Physics}, 17:\penalty0 268--279, 1949.
\newblock \doi{10.1063/1.1747238}.

\bibitem[Haidong(2006)]{Haidong2006}
Feng Haidong.
\newblock Prediction of infinite dilution solvent activity coefficient in rubbery polymer solutions using {Engaged Species Induced Clustering} ({ENSIC}) model.
\newblock \emph{Journal of Polymer Science Part B: Polymer Physics}, 44\penalty0 (12):\penalty0 1668--1675, 2006.
\newblock ISSN 1099-0488.
\newblock \doi{10.1002/polb.20811}.

\bibitem[Graham(1866)]{Graham1866}
Thomas Graham.
\newblock {LV}. {On} the absorption and dialytic separation of gases by colloid septa.
\newblock \emph{The London, Edinburgh, and Dublin Philosophical Magazine and Journal of Science}, 32\penalty0 (218):\penalty0 401--420, 1866.
\newblock ISSN 1941-5982.
\newblock \doi{10.1080/14786446608644207}.

\bibitem[Rogers(1985)]{Rogers1985}
C.~E. Rogers.
\newblock Permeation of {Gases} and {Vapours} in {Polymers}.
\newblock In J.~Comyn, editor, \emph{Polymer {Permeability}}, pages 11--73. Springer Netherlands, Dordrecht, 1985.
\newblock ISBN 978-94-009-4858-7.
\newblock \doi{10.1007/978-94-009-4858-7_2}.

\bibitem[Aminabhavi et~al.(1988)Aminabhavi, Aithal, and Shukla]{Aminabhavi1988}
Tejraj~M. Aminabhavi, U.~Shanthamurthy Aithal, and Shyam~S. Shukla.
\newblock An overview of the theoretical models used to predict transport of small molecules through polymer membranes.
\newblock \emph{Journal of Macromolecular Science, Part C}, 28\penalty0 (3-4):\penalty0 421--474, 1988.
\newblock ISSN 1532-1797.
\newblock \doi{10.1080/15583728808085382}.

\bibitem[Thomas and Windle(1978)]{Thomas1978}
Noreen Thomas and A.~H. Windle.
\newblock Transport of methanol in poly(methyl methacrylate).
\newblock \emph{Polymer}, 19\penalty0 (3):\penalty0 255--265, 1978.
\newblock ISSN 0032-3861.
\newblock \doi{10.1016/0032-3861(78)90218-5}.

\bibitem[Barrer(1946)]{Barrer1946}
R.~M. Barrer.
\newblock Measurement of diffusion and thermal conductivity "constants" in non-homogeneous media, and in media where these "constants" depend respectively on concentration or temperature.
\newblock \emph{Proceedings of the Physical Society}, 58\penalty0 (3):\penalty0 321--331, 1946.
\newblock ISSN 0959-5309.
\newblock \doi{10.1088/0959-5309/58/3/313}.

\bibitem[Cl{\'e}ment et~al.(2004)Cl{\'e}ment, Jonqui{\`e}res, Sarti, Sposata, Custal~Teixidor, and Lochon]{Clement2004}
Robert Cl{\'e}ment, Anne Jonqui{\`e}res, Ilaria Sarti, Maria~Federica Sposata, Maria~Angels Custal~Teixidor, and Pierre Lochon.
\newblock Original structure\textendash property relationships derived from a new modeling of diffusion of pure solvents through polymer membranes.
\newblock \emph{Journal of Membrane Science}, 232\penalty0 (1):\penalty0 141--152, 2004.
\newblock ISSN 0376-7388.
\newblock \doi{10.1016/j.memsci.2003.12.009}.

\bibitem[Crank and Park(1968)]{Crank1968}
John Crank and Geoffrey~Sheard Park.
\newblock \emph{Diffusion in Polymers}.
\newblock Academic Press, London; New York, 1968.
\newblock ISBN 978-0-12-197050-5.

\bibitem[Chern et~al.(1985)Chern, Koros, Hopfenberg, and Stannett]{Chern1985}
R.~T. Chern, W.~J. Koros, H.~B. Hopfenberg, and V.~T. Stannett.
\newblock Material selection for membrane-based gas separations.
\newblock In Douglas~R. Lloyd, editor, \emph{Materials Science of Synthetic Membranes}, volume 269, pages 25--46. American Chemical Society, Washington, D.C., 1985.
\newblock ISBN 978-0-8412-0887-2 978-0-8412-1098-1.
\newblock \doi{10.1021/bk-1985-0269.ch002}.

\bibitem[Park(1986)]{Park1986}
Geoffrey~S. Park.
\newblock Transport {Principles---Solution}, {Diffusion} and {Permeation} in {Polymer Membranes}.
\newblock In P.~M. Bungay, H.~K. Lonsdale, and M.~N. {de Pinho}, editors, \emph{Synthetic {Membranes}: {Science}, {Engineering} and {Applications}}, pages 57--107. Springer Netherlands, Dordrecht, 1986.
\newblock ISBN 978-94-009-4712-2.

\bibitem[Fujita(1961)]{Fujita1961}
Hiroshi Fujita.
\newblock Diffusion in polymer-diluent systems.
\newblock In \emph{Fortschritte Der Hochpolymeren-Forschung}, Advances in {Polymer Science}, pages 1--47, Berlin, Heidelberg, 1961. Springer.
\newblock ISBN 978-3-540-37048-2.
\newblock \doi{10.1007/BFb0050514}.

\bibitem[Freeman(1989)]{Freeman1989}
Benny~D. Freeman.
\newblock Mutual diffusion in polymeric systems.
\newblock In Geoffrey Allen and John~C. Bevington, editors, \emph{Comprehensive Polymer Science and Supplements}, pages 167--198. Pergamon, Amsterdam, 1989.
\newblock ISBN 978-0-08-096701-1.
\newblock \doi{https://doi.org/10.1016/B978-0-08-096701-1.00227-5}.
\newblock URL \url{https://www.sciencedirect.com/science/article/pii/B9780080967011002275}.

\bibitem[Berens and Hopfenberg(1978)]{Berens1978}
A.~R Berens and H.~B Hopfenberg.
\newblock Diffusion and relaxation in glassy polymer powders: 2. {Separation} of diffusion and relaxation parameters.
\newblock \emph{Polymer}, 19\penalty0 (5):\penalty0 489--496, 1978.
\newblock ISSN 0032-3861.
\newblock \doi{10.1016/0032-3861(78)90269-0}.

\bibitem[Ramani and Ranganathaiah(2001)]{Ramani2001}
R~Ramani and C~Ranganathaiah.
\newblock Free-volume microprobe study of iodine diffusion in polymers.
\newblock \emph{Polymer International}, 50\penalty0 (2):\penalty0 237--248, 2001.
\newblock ISSN 0959-8103, 1097-0126.
\newblock \doi{10.1002/1097-0126(200102)50:2<237::AID-PI613>3.0.CO;2-Q}.

\bibitem[McDowell et~al.(1999)McDowell, Freeman, and McNeely]{McDowell1999}
C.~C McDowell, B.~D Freeman, and G.~W McNeely.
\newblock Acetone sorption and uptake kinetic in poly(ethylene terephthalate).
\newblock \emph{Polymer}, 40\penalty0 (12):\penalty0 3487--3499, 1999.
\newblock ISSN 0032-3861.
\newblock \doi{10.1016/S0032-3861(98)00403-0}.

\bibitem[Stern(1994)]{Stern1994}
Alexander~S. Stern.
\newblock Polymers for gas separations: The next decade.
\newblock \emph{Journal of Membrane Science}, 94\penalty0 (1):\penalty0 1--65, 1994.
\newblock ISSN 03767388.
\newblock \doi{10.1016/0376-7388(94)00141-3}.

\bibitem[Barrer et~al.(1958)Barrer, Barrie, and Slater]{Barrer1958}
R.~M. Barrer, J.~A. Barrie, and J.~Slater.
\newblock Sorption and diffusion in ethyl cellulose. {Part III}. {Comparison} between ethyl cellulose and rubber.
\newblock \emph{Journal of Polymer Science}, 27\penalty0 (115):\penalty0 177--197, 1958.
\newblock ISSN 1542-6238.
\newblock \doi{10.1002/pol.1958.1202711515}.

\bibitem[Zhang and Handa(1998)]{Zhang1998}
Zhiyi Zhang and Y.~Paul Handa.
\newblock An in situ study of plasticization of polymers by high-pressure gases.
\newblock \emph{Journal of Polymer Science Part B: Polymer Physics}, 36\penalty0 (6):\penalty0 977--982, 1998.
\newblock ISSN 1099-0488.
\newblock \doi{10.1002/(SICI)1099-0488(19980430)36:6<977::AID-POLB5>3.0.CO;2-D}.

\bibitem[Koros et~al.(1976)Koros, Paul, and Rocha]{Koros1976}
W.~J. Koros, D.~R. Paul, and A.~A. Rocha.
\newblock Carbon dioxide sorption and transport in polycarbonate.
\newblock \emph{Journal of Polymer Science: Polymer Physics Edition}, 14\penalty0 (4):\penalty0 687--702, 1976.
\newblock ISSN 1542-9385.
\newblock \doi{10.1002/pol.1976.180140410}.

\bibitem[Stastna(1995)]{Stastna1995}
Jiri Stastna.
\newblock \emph{Transport {Properties} in {Polymers}}.
\newblock CRC Press, Boca Raton, Fl, 1995.
\newblock ISBN 978-1-56676-282-3.

\bibitem[Daynes and Smith(1920)]{Daynes1920}
H.~A. Daynes and Samuel Walter~Johnson Smith.
\newblock The process of diffusion through a rubber membrane.
\newblock \emph{Proceedings of the Royal Society of London. Series A, Containing Papers of a Mathematical and Physical Character}, 97\penalty0 (685):\penalty0 286--307, 1920.
\newblock \doi{10.1098/rspa.1920.0034}.

\bibitem[Barrer(1940)]{Barrer1940}
R.~M. Barrer.
\newblock Stationary and non-stationary states of flow of hydrogen in palladium and iron.
\newblock \emph{Transactions of the Faraday Society}, 36\penalty0 (0):\penalty0 1235--1248, 1940.
\newblock ISSN 0014-7672.
\newblock \doi{10.1039/TF9403601235}.

\bibitem[Toi et~al.(1983)Toi, Maeda, and Tokuda]{Toi1983}
Keio Toi, Yasushi Maeda, and Taneki Tokuda.
\newblock Numerical solution of the time-lag diffusion incorporating the dual sorption model.
\newblock \emph{Journal of Applied Polymer Science}, 28\penalty0 (11):\penalty0 3589--3592, 1983.
\newblock ISSN 1097-4628.
\newblock \doi{10.1002/app.1983.070281123}.

\bibitem[Barrer and Rideal(1939)]{Barrer1939}
R.~M. Barrer and Eric~K. Rideal.
\newblock Permeation, diffusion and solution of gases in organic polymers.
\newblock \emph{Transactions of the Faraday Society}, 35\penalty0 (0):\penalty0 628--643, 1939.
\newblock ISSN 0014-7672.
\newblock \doi{10.1039/TF9393500628}.

\bibitem[Petropoulos(1970)]{Petropoulos1970}
J.~H. Petropoulos.
\newblock Quantitative analysis of gaseous diffusion in glassy polymers.
\newblock \emph{Journal of Polymer Science Part A-2: Polymer Physics}, 8\penalty0 (10):\penalty0 1797--1801, 1970.
\newblock ISSN 1542-9377.
\newblock \doi{10.1002/pol.1970.160081014}.

\bibitem[Paul and Koros(1976)]{Paul1976}
D.~R. Paul and W.~J. Koros.
\newblock Effect of partially immobilizing sorption on permeability and the diffusion time lag.
\newblock \emph{Journal of Polymer Science: Polymer Physics Edition}, 14\penalty0 (4):\penalty0 675--685, 1976.
\newblock ISSN 1542-9385.
\newblock \doi{10.1002/pol.1976.180140409}.

\bibitem[Comyn(1985)]{Comyn1985}
J.~Comyn.
\newblock Introduction to polymer permeability and the mathematics of diffusion.
\newblock In J.~Comyn, editor, \emph{Polymer {Permeability}}, pages 1--10. Springer Netherlands, Dordrecht, 1985.
\newblock ISBN 978-94-009-4858-7.
\newblock \doi{10.1007/978-94-009-4858-7_1}.

\bibitem[Meares(1954)]{Meares1954}
Patrick Meares.
\newblock The {Diffusion} of {Gases Through Polyvinyl Acetate} {\textsuperscript{1}}.
\newblock \emph{Journal of the American Chemical Society}, 76\penalty0 (13):\penalty0 3415--3422, 1954.
\newblock ISSN 0002-7863, 1520-5126.
\newblock \doi{10.1021/ja01642a015}.

\bibitem[Brandt(1959)]{Brandt1959}
W.~Wilfried Brandt.
\newblock Model calculation of the temperature dependence of small molecule diffusion in high polymers.
\newblock \emph{The Journal of Physical Chemistry}, 63\penalty0 (7):\penalty0 1080--1085, 1959.
\newblock ISSN 0022-3654.
\newblock \doi{10.1021/j150577a012}.

\bibitem[DiBenedetto and Paul(1964)]{DiBenedetto1964}
A.~T. DiBenedetto and D.~R. Paul.
\newblock An interpretation of gaseous diffusion through polymers using fluctuation theory.
\newblock \emph{Journal of Polymer Science Part A: General Papers}, 2\penalty0 (2):\penalty0 1001--1015, 1964.
\newblock ISSN 1542-6246.
\newblock \doi{10.1002/pol.1964.100020234}.

\bibitem[Pace and Datyner(1979{\natexlab{a}})]{Pace1979}
R.~J. Pace and A.~Datyner.
\newblock Statistical mechanical model of diffusion of complex penetrants in polymers. {I}. {Theory}.
\newblock \emph{Journal of Polymer Science: Polymer Physics Edition}, 17\penalty0 (10):\penalty0 1675--1692, 1979{\natexlab{a}}.
\newblock ISSN 00981273, 15429385.
\newblock \doi{10.1002/pol.1979.180171005}.

\bibitem[Pace and Datyner(1980)]{Pace1980}
R.~J. Pace and A.~Datyner.
\newblock Statistical mechanical model of sorption and diffusion of simple penetrants in polymers.
\newblock \emph{Polymer Engineering and Science}, 20\penalty0 (1):\penalty0 51--58, 1980.
\newblock ISSN 0032-3888, 1548-2634.
\newblock \doi{10.1002/pen.760200109}.

\bibitem[Pace and Datyner(1979{\natexlab{b}})]{Pace1979a}
R.~J. Pace and A.~Datyner.
\newblock Statistical mechanical model for diffusion of simple penetrants in polymers. {II}. {Applications}\textemdash nonvinyl polymers.
\newblock \emph{Journal of Polymer Science: Polymer Physics Edition}, 17\penalty0 (3):\penalty0 453--464, 1979{\natexlab{b}}.
\newblock ISSN 00981273, 15429385.
\newblock \doi{10.1002/pol.1979.180170310}.

\bibitem[Pant and Boyd(1993)]{Pant1993}
P.~V.~Krishna Pant and Richard~H. Boyd.
\newblock Molecular-dynamics simulation of diffusion of small penetrants in polymers.
\newblock \emph{Macromolecules}, 26\penalty0 (4):\penalty0 679--686, 1993.
\newblock ISSN 0024-9297.
\newblock \doi{10.1021/ma00056a019}.

\bibitem[Sok et~al.(1992)Sok, Berendsen, and {van Gunsteren}]{Sok1992}
R.~M. Sok, H.~J.~C. Berendsen, and W.~F. {van Gunsteren}.
\newblock Molecular dynamics simulation of the transport of small molecules across a polymer membrane.
\newblock \emph{The Journal of Chemical Physics}, 96\penalty0 (6):\penalty0 4699--4704, 1992.
\newblock ISSN 0021-9606.
\newblock \doi{10.1063/1.462806}.

\bibitem[Gusev et~al.(1994)Gusev, {M{\"u}ller-Plathe}, {van Gunsteren}, and Suter]{Gusev1994}
A.~A. Gusev, F.~{M{\"u}ller-Plathe}, W.~F. {van Gunsteren}, and U.~W. Suter.
\newblock Dynamics of small molecules in bulk polymers.
\newblock In Lucien Monnerie and U.~W. Suter, editors, \emph{Atomistic Modeling of Physical Properties}, Advances in {Polymer Science}, pages 207--247. Springer, Berlin, Heidelberg, 1994.
\newblock ISBN 978-3-540-48352-6.
\newblock \doi{10.1007/BFb0080200}.

\bibitem[Cohen and Turnbull(1959)]{Cohen1959}
Morrel~H. Cohen and David Turnbull.
\newblock Molecular transport in liquids and glasses.
\newblock \emph{The Journal of Chemical Physics}, 31\penalty0 (5):\penalty0 1164--1169, 1959.
\newblock ISSN 0021-9606.
\newblock \doi{10.1063/1.1730566}.

\bibitem[Bueche(1953)]{Bueche1953}
F.~Bueche.
\newblock Segmental mobility of polymers near their glass temperature.
\newblock \emph{The Journal of Chemical Physics}, 21\penalty0 (10):\penalty0 1850--1855, 1953.
\newblock ISSN 0021-9606, 1089-7690.
\newblock \doi{10.1063/1.1698677}.

\bibitem[Slater(1940)]{Slater1940}
C~Slater.
\newblock Introduction to chemical physics.
\newblock \emph{Journal of the Society of Chemical Industry}, 59\penalty0 (11):\penalty0 187--187, 1940.
\newblock ISSN 03684075, 19349971.
\newblock \doi{10.1002/jctb.5000591106}.

\bibitem[Stern and Trohalaki(1990)]{Stern1990}
S.~A. Stern and S.~Trohalaki.
\newblock Fundamentals of {Gas Diffusion} in {Rubbery} and {Glassy Polymers}.
\newblock In \emph{Barrier {Polymers} and {Structures}}, volume 423 of \emph{ACS Symposium Series}, pages 22--59. American Chemical Society, Washington, 1990.
\newblock ISBN 978-0-8412-1762-1.
\newblock \doi{10.1021/bk-1990-0423.ch002}.

\bibitem[Williams et~al.(1955)Williams, Landel, and Ferry]{Williams1955}
Malcolm~L. Williams, Robert~F. Landel, and John~D. Ferry.
\newblock The {Temperature Dependence} of {Relaxation Mechanisms} in {Amorphous Polymers} and {Other Glass-forming Liquids}.
\newblock \emph{Journal of the American Chemical Society}, 77\penalty0 (14):\penalty0 3701--3707, 1955.
\newblock ISSN 0002-7863.
\newblock \doi{10.1021/ja01619a008}.

\bibitem[Fujita et~al.(1960)Fujita, Kishimoto, and Matsumoto]{Fujita1960}
Hiroshi Fujita, Akira Kishimoto, and Kinya Matsumoto.
\newblock Concentration and temperature dependence of diffusion coefficients for systems polymethyl acrylate and n-alkyl acetates.
\newblock \emph{Transactions of the Faraday Society}, 56\penalty0 (0):\penalty0 424--437, 1960.
\newblock ISSN 0014-7672.
\newblock \doi{10.1039/TF9605600424}.

\bibitem[Peterlin(1975)]{Peterlin1975}
A.~Peterlin.
\newblock Dependence of diffusive transport on morphology of crystalline polymers.
\newblock \emph{Journal of Macromolecular Science, Part B}, 11\penalty0 (1):\penalty0 57--87, 1975.
\newblock ISSN 0022-2348.
\newblock \doi{10.1080/00222347508217855}.

\bibitem[Doolittle(1951)]{Doolittle1951}
Arthur~K. Doolittle.
\newblock Studies in {Newtonian Flow}. {II}. {The Dependence} of the {{Viscosity}} of liquids on free-space.
\newblock \emph{Journal of Applied Physics}, 22\penalty0 (12):\penalty0 1471--1475, 1951.
\newblock ISSN 0021-8979.
\newblock \doi{10.1063/1.1699894}.

\bibitem[Doolittle(1952)]{Doolittle1952}
Arthur~K. Doolittle.
\newblock Studies in {Newtonian Flow}. {III}. the dependence of the viscosity of liquids on molecular weight and free space (in homologous series).
\newblock \emph{Journal of Applied Physics}, 23\penalty0 (2):\penalty0 236--239, 1952.
\newblock ISSN 0021-8979.
\newblock \doi{10.1063/1.1702182}.

\bibitem[Frisch(1970)]{Frisch1970}
H.L. Frisch.
\newblock Pressure {Dependence} of {Diffusion} in {Polymers}.
\newblock \emph{Journal of Elastoplastics}, 2\penalty0 (2):\penalty0 130--132, 1970.
\newblock ISSN 0022-071X.
\newblock \doi{10.1177/009524437000200206}.

\bibitem[Stern et~al.(1972)Stern, Fang, and Frisch]{Stern1972}
S.~A. Stern, S.-M. Fang, and H.~L. Frisch.
\newblock Effect of pressure on gas permeability coefficients. {A} new application of ``free volume'' theory.
\newblock \emph{Journal of Polymer Science Part A-2: Polymer Physics}, 10\penalty0 (2):\penalty0 201--219, 1972.
\newblock ISSN 1542-9377.
\newblock \doi{10.1002/pol.1972.160100202}.

\bibitem[Kulkarni and Stern(1983)]{Kulkarni1983}
S.~S. Kulkarni and S.~A. Stern.
\newblock The diffusion of {CO2}, {CH4}, {C2H4}, and {C3H8} in polyethylene at elevated pressures.
\newblock \emph{Journal of Polymer Science: Polymer Physics Edition}, 21\penalty0 (3):\penalty0 441--465, 1983.
\newblock ISSN 1542-9385.
\newblock \doi{10.1002/pol.1983.180210310}.

\bibitem[Stern et~al.(1983)Stern, Kulkarni, and Frisch]{Stern1983}
S.~A. Stern, S.~S. Kulkarni, and H.~L. Frisch.
\newblock Tests of a ``free-volume'' model of gas permeation through polymer membranes. {I}. {Pure CO2}, {CH4}, {C2H4}, and {C3H8} in polyethylene.
\newblock \emph{Journal of Polymer Science: Polymer Physics Edition}, 21\penalty0 (3):\penalty0 467--481, 1983.
\newblock ISSN 1542-9385.
\newblock \doi{10.1002/pol.1983.180210311}.

\bibitem[Stern et~al.(1986)Stern, Sampat, and Kulkarni]{Stern1986}
S.~A. Stern, S.~R. Sampat, and S.~S. Kulkarni.
\newblock Tests of a ``free-volume'' model of gas permeation through polymer membranes. {II}. {Pure Ar}, {SF6}, {CF4}, and {C2H2F2} in polyethylene.
\newblock \emph{Journal of Polymer Science Part B: Polymer Physics}, 24\penalty0 (10):\penalty0 2149--2166, 1986.
\newblock ISSN 1099-0488.
\newblock \doi{10.1002/polb.1986.090241001}.

\bibitem[Vrentas and Duda(1977{\natexlab{a}})]{Vrentas1977}
J.~S. Vrentas and J.~L. Duda.
\newblock Diffusion in polymer\textendash solvent systems. {II}. {A} predictive theory for the dependence of diffusion coefficients on temperature, concentration, and molecular weight.
\newblock \emph{Journal of Polymer Science: Polymer Physics Edition}, 15\penalty0 (3):\penalty0 417--439, 1977{\natexlab{a}}.
\newblock ISSN 1542-9385.
\newblock \doi{10.1002/pol.1977.180150303}.

\bibitem[Vrentas and Duda(1977{\natexlab{b}})]{Vrentas1977a}
J.~S. Vrentas and J.~L. Duda.
\newblock Diffusion in polymer\textemdash solvent systems. {I}. {Reexaminatio} of the free-volume theory.
\newblock \emph{Journal of Polymer Science: Polymer Physics Edition}, 15\penalty0 (3):\penalty0 403--416, 1977{\natexlab{b}}.
\newblock ISSN 1542-9385.
\newblock \doi{10.1002/pol.1977.180150302}.

\bibitem[Vrentas et~al.(1985{\natexlab{a}})Vrentas, Duda, Ling, and Hou]{Vrentas1985}
J.~S. Vrentas, J.~L. Duda, H.-C. Ling, and A.-C. Hou.
\newblock Free-volume theories for self-diffusion in polymer\textendash solvent systems. {II}. {Predictive} capabilities.
\newblock \emph{Journal of Polymer Science: Polymer Physics Edition}, 23\penalty0 (2):\penalty0 289--304, 1985{\natexlab{a}}.
\newblock ISSN 1542-9385.
\newblock \doi{10.1002/pol.1985.180230205}.

\bibitem[Vrentas et~al.(1985{\natexlab{b}})Vrentas, Duda, and Ling]{Vrentas1985a}
J.~S. Vrentas, J.~L. Duda, and H.-C. Ling.
\newblock Free-volume theories for self-diffusion in polymer\textendash solvent systems. {I}. {Conceptual} differences in theories.
\newblock \emph{Journal of Polymer Science: Polymer Physics Edition}, 23\penalty0 (2):\penalty0 275--288, 1985{\natexlab{b}}.
\newblock ISSN 1542-9385.
\newblock \doi{10.1002/pol.1985.180230204}.

\bibitem[Vrentas et~al.(1993)Vrentas, Vrentas, and Duda]{Vrentas1993}
J.~S. Vrentas, C.~M. Vrentas, and J.~L. Duda.
\newblock Comparison of {Free-Volume Theories}.
\newblock \emph{Polymer Journal}, 25\penalty0 (1):\penalty0 99--101, 1993.
\newblock ISSN 1349-0540.
\newblock \doi{10.1295/polymj.25.99}.

\bibitem[Bueche(1962)]{Bueche1962}
Frederick~J Bueche.
\newblock \emph{Physical Properties of Polymers.}
\newblock {Interscience Publishers}, {New York}, 1962.
\newblock ISBN 978-0-470-11664-7.

\bibitem[Ganesh et~al.(1992)Ganesh, Nagarajan, and Duda]{Ganesh1992}
K.~Ganesh, R.~Nagarajan, and J.~Larry Duda.
\newblock Rate of gas transport in glassy polymers: A free volume based predictive model.
\newblock \emph{Industrial \& Engineering Chemistry Research}, 31\penalty0 (3):\penalty0 746--755, 1992.
\newblock ISSN 0888-5885, 1520-5045.
\newblock \doi{10.1021/ie00003a016}.

\bibitem[Lee(1980)]{Lee1980}
W.~M. Lee.
\newblock Selection of barrier materials from molecular structure.
\newblock \emph{Polymer Engineering \& Science}, 20\penalty0 (1):\penalty0 65--69, 1980.
\newblock ISSN 1548-2634.
\newblock \doi{10.1002/pen.760200111}.

\bibitem[Mauritz et~al.(1990)Mauritz, Storey, and George]{Mauritz1990}
Kenneth~A. Mauritz, Robson~F. Storey, and Scott~E. George.
\newblock A general free volume-based theory for the diffusion of large molecules in amorphous polymers above the glass temperature. {I}. {Application} to di-n-alkyl phthalates in {PVC}.
\newblock \emph{Macromolecules}, 23\penalty0 (2):\penalty0 441--450, 1990.
\newblock ISSN 0024-9297.
\newblock \doi{10.1021/ma00204a016}.

\bibitem[Coughlin et~al.(1991)Coughlin, Mauritz, and Storey]{Coughlin1991}
Christopher~S. Coughlin, Kenneth~A. Mauritz, and Robson~F. Storey.
\newblock A general free volume based theory for the diffusion of large molecules in amorphous polymers above {Tg}. 4. {Polymer-penetrant} interactions.
\newblock \emph{Macromolecules}, 24\penalty0 (7):\penalty0 1526--1534, 1991.
\newblock ISSN 0024-9297, 1520-5835.
\newblock \doi{10.1021/ma00007a014}.

\bibitem[Barto{\v s} et~al.(1996)Barto{\v s}, Kri{\v s}tiakov{\'a}, {\v S}au{\v s}a, and Kri{\v s}tiak]{Bartos1996}
J.~Barto{\v s}, K.~Kri{\v s}tiakov{\'a}, O.~{\v S}au{\v s}a, and J.~Kri{\v s}tiak.
\newblock Free volume microstructure of tetramethylpolycarbonate at low temperatures studied by positron annihilation lifetime spectroscopy: A comparison with polycarbonate.
\newblock \emph{Polymer}, 37\penalty0 (15):\penalty0 3397--3403, 1996.
\newblock ISSN 0032-3861.
\newblock \doi{10.1016/0032-3861(96)88487-4}.

\bibitem[Muruganandam et~al.(1987)Muruganandam, Koros, and Paul]{Muruganandam1987}
N.~Muruganandam, W.~J. Koros, and D.~R. Paul.
\newblock Gas sorption and transport in substituted polycarbonates.
\newblock \emph{Journal of Polymer Science Part B: Polymer Physics}, 25\penalty0 (9):\penalty0 1999--2026, 1987.
\newblock ISSN 1099-0488.
\newblock \doi{10.1002/polb.1987.090250917}.

\bibitem[Korsmeyer and Peppas(1981)]{Korsmeyer1981}
Richard~W. Korsmeyer and Nikolaos~A. Peppas.
\newblock Effect of the morphology of hydrophilic polymeric matrices on the diffusion and release of water soluble drugs.
\newblock \emph{Journal of Membrane Science}, 9\penalty0 (3):\penalty0 211--227, 1981.
\newblock ISSN 0376-7388.
\newblock \doi{10.1016/S0376-7388(00)80265-3}.

\bibitem[Hariharan and Peppas(1993)]{Hariharan1993}
Deepak Hariharan and Nikolaos~A. Peppas.
\newblock Modelling of water transport and solute release in physiologically sensitive gels.
\newblock \emph{Journal of Controlled Release}, 23:\penalty0 123--135, 1993.

\bibitem[Peppas and Scott(1992)]{Peppas1992}
Nikolaos~A. Peppas and Jill~E. Scott.
\newblock Controlled release from poly(vinyl alcohol) gels prepared by freezing-thawing processes.
\newblock \emph{Journal of Controlled Release}, 18\penalty0 (2):\penalty0 95--100, 1992.
\newblock ISSN 0168-3659.
\newblock \doi{10.1016/0168-3659(92)90178-T}.

\bibitem[Fricke(1924)]{Fricke1924}
Hugo Fricke.
\newblock A {Mathematical Treatment} of the {Electric Conductivity} and {Capacity} of {Disperse Systems I}. {The Electric Conductivity} of a {Suspension} of {Homogeneous Spheroids}.
\newblock \emph{Physical Review}, 24\penalty0 (5):\penalty0 575--587, 1924.
\newblock \doi{10.1103/PhysRev.24.575}.

\bibitem[Masaro and Zhu(1999)]{Masaro1999}
L~Masaro and X.~X Zhu.
\newblock Physical models of diffusion for polymer solutions, gels and solids.
\newblock \emph{Progress in Polymer Science}, 24\penalty0 (5):\penalty0 731--775, 1999.
\newblock ISSN 0079-6700.
\newblock \doi{10.1016/S0079-6700(99)00016-7}.

\bibitem[Wijmans and Baker(1995)]{Wijmans1995}
J.~G. Wijmans and R.~W. Baker.
\newblock The solution-diffusion model: {A} review.
\newblock \emph{Journal of Membrane Science}, 107\penalty0 (1):\penalty0 1--21, 1995.
\newblock ISSN 0376-7388.
\newblock \doi{10.1016/0376-7388(95)00102-I}.

\bibitem[Kokes and Long(1953)]{Kokes1953}
R.~J. Kokes and F.~A. Long.
\newblock Diffusion of {Organic Vapors} into {Polyvinyl Acetate1}.
\newblock \emph{Journal of the American Chemical Society}, 75\penalty0 (24):\penalty0 6142--6146, 1953.
\newblock ISSN 0002-7863.
\newblock \doi{10.1021/ja01120a011}.

\bibitem[McCall(1957)]{McCall1957}
D.~W. McCall.
\newblock Diffusion in ethylene polymers. {I}. {Desorption} kinetics for a thin slab.
\newblock \emph{Journal of Polymer Science}, 26\penalty0 (113):\penalty0 151--164, 1957.
\newblock ISSN 1542-6238.
\newblock \doi{10.1002/pol.1957.1202611303}.

\bibitem[Long(1965)]{Long1965}
R.~B. Long.
\newblock Liquid {Permeation} through {Plastic Films}.
\newblock \emph{Industrial \& Engineering Chemistry Fundamentals}, 4\penalty0 (4):\penalty0 445--451, 1965.
\newblock ISSN 0196-4313.
\newblock \doi{10.1021/i160016a015}.

\bibitem[Feng and Huang(1996)]{Feng1996}
Xianshe Feng and Robert Y.~M. Huang.
\newblock Estimation of activation energy for permeation in pervaporation processes.
\newblock \emph{Journal of Membrane Science}, 118\penalty0 (1):\penalty0 127--131, 1996.
\newblock ISSN 0376-7388.
\newblock \doi{10.1016/0376-7388(96)00096-8}.

\bibitem[Hillaire and Favre(1999)]{Hillaire1999}
Anne Hillaire and Eric Favre.
\newblock Isothermal and nonisothermal permeation of an organic vapor through a dense polymer membrane.
\newblock \emph{Industrial \& Engineering Chemistry Research}, 38\penalty0 (1):\penalty0 211--217, 1999.
\newblock ISSN 0888-5885.
\newblock \doi{10.1021/ie980491k}.

\bibitem[Baudot and Marin(1997)]{Baudot1997}
A.~Baudot and M.~Marin.
\newblock Pervaporation of aroma compounds: Comparison of membrane performances with vapour-liquid equilibria and engineering aspects of process improvement.
\newblock \emph{Food and Bioproducts Processing}, 75\penalty0 (2):\penalty0 117--142, 1997.
\newblock ISSN 0960-3085.
\newblock \doi{10.1205/096030897531432}.

\bibitem[Wijmans et~al.(1996)Wijmans, Athayde, Daniels, Ly, Kamaruddin, and Pinnau]{Wijmans1996}
J.~G. Wijmans, A.~L. Athayde, R.~Daniels, J.~H. Ly, H.~D. Kamaruddin, and I.~Pinnau.
\newblock The role of boundary layers in the removal of volatile organic compounds from water by pervaporation.
\newblock \emph{Journal of Membrane Science}, 109\penalty0 (1):\penalty0 135--146, 1996.
\newblock ISSN 0376-7388.
\newblock \doi{10.1016/0376-7388(95)00194-8}.

\bibitem[Tumu et~al.(2024)Tumu, Vorst, and Curtzwiler]{Tumu2024}
Khairun Tumu, Keith Vorst, and Greg Curtzwiler.
\newblock Understanding intentionally and non-intentionally added substances and associated threshold of toxicological concern in post-consumer polyolefin for use as food packaging materials.
\newblock \emph{Heliyon}, 10\penalty0 (1), January 2024.
\newblock ISSN 2405-8440.
\newblock \doi{10.1016/j.heliyon.2023.e23620}.
\newblock URL \url{https://www.cell.com/heliyon/abstract/S2405-8440(23)10828-0}.
\newblock Publisher: Elsevier.

\bibitem[Muncke(2021)]{Muncke2021}
Jane Muncke.
\newblock Tackling the toxics in plastics packaging.
\newblock \emph{PLOS Biology}, 19\penalty0 (3):\penalty0 e3000961, 2021.
\newblock ISSN 1545-7885.
\newblock \doi{10.1371/journal.pbio.3000961}.
\newblock URL \url{https://journals.plos.org/plosbiology/article?id=10.1371/journal.pbio.3000961}.
\newblock Publisher: Public Library of Science.

\bibitem[Arvanitoyannis and Bosnea(2004)]{Arvanitoyannis2004}
Ioannis~S. Arvanitoyannis and Loulouda Bosnea.
\newblock Migration of substances from food packaging materials to foods.
\newblock \emph{Critical Reviews in Food Science and Nutrition}, 44\penalty0 (2):\penalty0 63--76, 2004.
\newblock ISSN 1040-8398.
\newblock \doi{10.1080/10408690490424621}.

\bibitem[Ekelund et~al.(2007)Ekelund, Edin, and Gedde]{Ekelund2007}
M.~Ekelund, H.~Edin, and U.W. Gedde.
\newblock Long-term performance of poly(vinyl chloride) cables. {Part} 1: {Mechanical} and electrical performances.
\newblock \emph{Polymer Degradation and Stability}, 92\penalty0 (4):\penalty0 617--629, 2007.
\newblock ISSN 01413910.
\newblock \doi{10.1016/j.polymdegradstab.2007.01.005}.
\newblock URL \url{https://linkinghub.elsevier.com/retrieve/pii/S0141391007000110}.

\bibitem[Chiellini et~al.(2013)Chiellini, Ferri, Morelli, Dipaola, and Latini]{Chiellini2013}
Federica Chiellini, Marcella Ferri, Andrea Morelli, Lucia Dipaola, and Giuseppe Latini.
\newblock Perspectives on alternatives to phthalate plasticized poly(vinyl chloride) in medical devices applications.
\newblock \emph{Progress in Polymer Science}, 38\penalty0 (7):\penalty0 1067--1088, 2013.
\newblock ISSN 0079-6700.
\newblock \doi{https://doi.org/10.1016/j.progpolymsci.2013.03.001}.
\newblock URL \url{http://www.sciencedirect.com/science/article/pii/S007967001300018X}.
\newblock REVIEW.

\bibitem[Coughlin et~al.(1990)Coughlin, Mauritz, and Storey]{Coughlin1990}
Christopher~S. Coughlin, Kenneth~A. Mauritz, and Robson~F. Storey.
\newblock A general free volume based theory for the diffusion of large molecules in amorphous polymers above {Tg}. 3. {Theoretical} conformational analysis of molecular shape.
\newblock \emph{Macromolecules}, 23\penalty0 (12):\penalty0 3187--3192, 1990.
\newblock ISSN 0024-9297, 1520-5835.
\newblock \doi{10.1021/ma00214a026}.

\bibitem[Kova\v{c}i\'{c} and Mrkli\'{c}(2002)]{Kovacic2002}
Tonka Kova\v{c}i\'{c} and \v{Z}eljko Mrkli\'{c}.
\newblock The kinetic parameters for the evaporation of plasticizers from plasticized poly(vinyl chloride).
\newblock \emph{Thermochimica Acta}, 381\penalty0 (1):\penalty0 49--60, 2002.
\newblock ISSN 00406031.
\newblock \doi{10.1016/S0040-6031(01)00643-8}.
\newblock URL \url{https://linkinghub.elsevier.com/retrieve/pii/S0040603101006438}.

\bibitem[Papaspyrides and Tingas(1998)]{Papaspyrides1998}
C~D Papaspyrides and S.~G. Tingas.
\newblock Comparison of isopropanol and isooctane as food simulants in plasticizer migration tests.
\newblock \emph{Food additives and contaminants}, 15 6:\penalty0 681--9, 1998.
\newblock URL \url{https://api.semanticscholar.org/CorpusID:31262937}.

\bibitem[Angert et~al.(1961)Angert, Zenchenko, and Kuzminski{\u \i}]{Angert1961}
L.~G. Angert, A.~I. Zenchenko, and A.~S. Kuzminski{\u \i}.
\newblock Volatilization of {Phenyl-2-Naphthylamine} from rubber.
\newblock \emph{Rubber Chemistry and Technology}, 34\penalty0 (3):\penalty0 807--815, 1961.
\newblock ISSN 1943-4804, 0035-9475.
\newblock \doi{10.5254/1.3540251}.

\bibitem[Calvert and Billingham(1979)]{Calvert1979}
P.~D. Calvert and N.~C. Billingham.
\newblock Loss of additives from polymers: {A} theoretical model.
\newblock \emph{Journal of Applied Polymer Science}, 24\penalty0 (2):\penalty0 357--370, 1979.
\newblock ISSN 00218995, 10974628.
\newblock \doi{10.1002/app.1979.070240205}.

\bibitem[Lusto{\u n} et~al.(1993)Lusto{\u n}, Pastu{\u s}{\'a}kov{\'a}, and Va{\u s}{\u s}]{Luston1993}
J.~Lusto{\u n}, V.~Pastu{\u s}{\'a}kov{\'a}, and F.~Va{\u s}{\u s}.
\newblock Volatility of additives from polymers. {Concentration} dependence and crystallinity effects.
\newblock \emph{Journal of Applied Polymer Science}, 48\penalty0 (2):\penalty0 219--224, 1993.
\newblock \doi{10.1002/app.1993.070480205}.

\bibitem[Huang et~al.(2001)Huang, Liu, and Liu]{Huang2001}
Jan-Chan Huang, Helen Liu, and Yl~Liu.
\newblock Diffusion in polymers with concentration dependent diffusivity.
\newblock \emph{International Journal of Polymeric Materials}, 49\penalty0 (1):\penalty0 15--24, 2001.
\newblock ISSN 0091-4037, 1563-535X.
\newblock \doi{10.1080/00914030108035864}.

\bibitem[Storey et~al.(1989)Storey, Mauritz, and Cox]{Storey1989}
Robson~F. Storey, Kenneth~A. Mauritz, and B.~Dwain Cox.
\newblock Diffusion of various dialkyl phthalate plasticizers in {PVC}.
\newblock \emph{Macromolecules}, 22\penalty0 (1):\penalty0 289--294, 1989.
\newblock ISSN 0024-9297, 1520-5835.
\newblock \doi{10.1021/ma00191a053}.

\bibitem[Philip(1955)]{Philip1955}
J.~R. Philip.
\newblock Numerical solution of equations of the diffusion type with diffusivity concentration-dependent.
\newblock \emph{Transactions of the Faraday Society}, 51:\penalty0 885, 1955.
\newblock ISSN 0014-7672.
\newblock \doi{10.1039/tf9555100885}.

\bibitem[Thornton et~al.(2009)Thornton, Nairn, Hill, and Hill]{Thornton2009}
Aaron~W. Thornton, Kate~M. Nairn, Anita~J. Hill, and James~M. Hill.
\newblock New relation between diffusion and free volume: {I}. {Predicting} gas diffusion.
\newblock \emph{Journal of Membrane Science}, 338\penalty0 (1-2):\penalty0 29--37, 2009.
\newblock ISSN 03767388.
\newblock \doi{10.1016/j.memsci.2009.03.053}.

\bibitem[Koros et~al.(2002)Koros, Burgess, and Chen]{Koros2002}
William~J. Koros, Steven~K. Burgess, and Zhang Chen.
\newblock Polymer {Transport Properties}.
\newblock In {John Wiley \& Sons, Inc.}, editor, \emph{Encyclopedia of Polymer Science and Technology}, pages 1--96. Wiley, New Jersey, first edition, 2002.
\newblock ISBN 978-1-118-63389-2 978-0-471-44026-0.
\newblock \doi{10.1002/0471440264}.

\bibitem[Sharma et~al.(2017)Sharma, Tewari, and Arya]{Sharma2017}
Jyoti Sharma, Kshitij Tewari, and Raj~Kumar Arya.
\newblock Diffusion in polymeric systems\textendash{A} review on free volume theory.
\newblock \emph{Progress in Organic Coatings}, 111:\penalty0 83--92, 2017.
\newblock ISSN 03009440.
\newblock \doi{10.1016/j.porgcoat.2017.05.004}.

\bibitem[Ekelund et~al.(2008)Ekelund, Azhdar, Hedenqvist, and Gedde]{Ekelund2008}
M.~Ekelund, B.~Azhdar, M.S. Hedenqvist, and U.W. Gedde.
\newblock Long-term performance of poly(vinyl chloride) cables, {Part} 2: {Migration} of plasticizer.
\newblock \emph{Polymer Degradation and Stability}, 93\penalty0 (9):\penalty0 1704--1710, 2008.
\newblock ISSN 01413910.
\newblock \doi{10.1016/j.polymdegradstab.2008.05.030}.

\bibitem[Ekelund et~al.(2010)Ekelund, Azhdar, and Gedde]{Ekelund2010}
M.~Ekelund, B.~Azhdar, and U.W. Gedde.
\newblock Evaporative loss kinetics of di(2-ethylhexyl)phthalate ({DEHP}) from pristine {DEHP} and plasticized {PVC}.
\newblock \emph{Polymer Degradation and Stability}, 95\penalty0 (9):\penalty0 1789--1793, 2010.
\newblock ISSN 01413910.
\newblock \doi{10.1016/j.polymdegradstab.2010.05.007}.

\bibitem[Smith et~al.(2004)Smith, Skidmore, Howe, and Majewski]{Smith2004}
Gregory~S. Smith, Cary~B. Skidmore, Philip~M. Howe, and Jaroslaw Majewski.
\newblock Diffusion, evaporation, and surface enrichment of a plasticizing additive in an annealed polymer thin film.
\newblock \emph{Journal of Polymer Science Part B: Polymer Physics}, 42\penalty0 (17):\penalty0 3258--3266, 2004.
\newblock ISSN 0887-6266, 1099-0488.
\newblock \doi{10.1002/polb.20171}.

\bibitem[Nouman et~al.(2017)Nouman, Saunier, Jubeli, and Yagoubi]{Nouman2017}
Micheal Nouman, Johanna Saunier, Emile Jubeli, and Najet Yagoubi.
\newblock Additive blooming in polymer materials: {Consequences} in the pharmaceutical and medical field.
\newblock \emph{Polymer Degradation and Stability}, 143:\penalty0 239--252, 2017.
\newblock ISSN 01413910.
\newblock \doi{10.1016/j.polymdegradstab.2017.07.021}.

\bibitem[Treybal(2004)]{Treybal2004}
Robert~Ewald Treybal.
\newblock \emph{Mass-Transfer Operations}.
\newblock {McGraw-Hill} Chemical Engineering Series. McGraw-Hill, New York, 3. ed., reissued edition, 2004.
\newblock ISBN 978-0-07-066615-3 978-0-07-065176-0.

\bibitem[Bellobono et~al.(1984)Bellobono, Marcandalli, Selli, Polissi, and Leidi]{Bellobono1984}
Ignazio~Renato Bellobono, Bruno Marcandalli, Elena Selli, Alessandra Polissi, and Giorgio Leidi.
\newblock A model study for release of plasticizers from polymer films through vapor phase.
\newblock \emph{Journal of Applied Polymer Science}, 29\penalty0 (10):\penalty0 3185--3195, 1984.
\newblock ISSN 00218995, 10974628.
\newblock \doi{10.1002/app.1984.070291020}.

\bibitem[Clausen et~al.(2007)Clausen, Xu, {Kofoed-S{\o}rensen}, Little, and Wolkoff]{Clausen2007}
Per~Axel Clausen, Ying Xu, Vivi {Kofoed-S{\o}rensen}, John~C. Little, and Peder Wolkoff.
\newblock The influence of humidity on the emission of di-(2-ethylhexyl) phthalate ({DEHP}) from vinyl flooring in the emission cell ``{FLEC}''.
\newblock \emph{Atmospheric Environment}, 41\penalty0 (15):\penalty0 3217--3224, 2007.
\newblock ISSN 13522310.
\newblock \doi{10.1016/j.atmosenv.2006.06.063}.

\bibitem[Begley(1997)]{Begley1997}
Timothy~H. Begley.
\newblock Methods and approaches used by {FDA} to evaluate the safety of food packaging materials.
\newblock \emph{Food Additives and Contaminants}, 14\penalty0 (6-7):\penalty0 545--553, 1997.
\newblock ISSN 0265-203X.
\newblock \doi{10.1080/02652039709374566}.

\bibitem[Baner et~al.(1996)Baner, Brandsch, Franz, and Piringer]{Baner1996}
A.~Baner, J.~Brandsch, R.~Franz, and O.~Piringer.
\newblock The application of a predictive migration model for evaluating the compliance of plastic materials with {European} food regulations\textdagger.
\newblock \emph{Food Additives and Contaminants}, 13\penalty0 (5):\penalty0 587--601, 1996.
\newblock ISSN 0265-203X.
\newblock \doi{10.1080/02652039609374443}.

\bibitem[Audouin et~al.(1992)Audouin, Dalle, Metzger, and Verdu]{Audouin1992}
L.~Audouin, B.~Dalle, G.~Metzger, and J.~Verdu.
\newblock Thermal aging of plasticized pvc. i. weight loss kinetics in the pvc \textemdash didecylphtalate system.
\newblock \emph{Journal of Applied Polymer Science}, 45\penalty0 (12):\penalty0 2091--2096, 1992.
\newblock ISSN 00218995, 10974628.
\newblock \doi{10.1002/app.1992.070451204}.

\end{thebibliography}






\end{document}